\documentclass[
preprint,  
superscriptaddress,
 amsmath,amssymb,
prl
]{revtex4-1}
\usepackage{xcolor}
\usepackage{color}
\usepackage[normalem]{ulem}

\usepackage{graphicx}
\usepackage{subfigure}
\usepackage{dcolumn}
\usepackage{bm}
\usepackage{xcolor}%
\usepackage{amsmath}
\usepackage{verbatim}
\usepackage{mathtools}
\usepackage{float}
\usepackage{lineno}

\graphicspath{ {images/} }

\begin{document}

\title{
{\bf 
Connecting packing efficiency of binary hard sphere systems to their intermediate range structure\\
}
}

\author{Houfei Yuan}
\affiliation{School of Physics and Astronomy, Shanghai Jiao Tong University, Shanghai 200240, China}

\author{Zhen Zhang}
\affiliation{Center for Alloy Innovation and Design, State Key Laboratory for Mechanical
Behavior of Materials, Xi’an Jiaotong University, Xi’an 710049, China}

\author{Walter Kob}
\email[Corresponding author: ]{walter.kob@umontpellier.fr}
\affiliation{Laboratoire Charles Coulomb,
University of Montpellier and CNRS, F-34095 Montpellier, France}
\affiliation{School of Physics and Astronomy, Shanghai Jiao Tong University, Shanghai 200240, China}

\author{Yujie Wang}
\email[Corresponding author: ]{yujiewang@sjtu.edu.cn}
\affiliation{School of Physics and Astronomy, Shanghai Jiao Tong University, Shanghai 200240, China}

\begin{abstract}
Using computed x-ray tomography we determine the
three dimensional (3d) structure of binary hard sphere mixtures as a function of composition and size ratio of the particles, $q$. Using a recently introduced four-point correlation function we reveal that this 3d structure has on intermediate and large length scales a surprisingly regular order, the symmetry of which depends on $q$. The related structural correlation length has a minimum at the composition at which the packing fraction is highest. At this composition also the number of different local particle arrangements has a maximum, indicating that efficient packing of particles is associated with a structure that is locally maximally disordered. 
\end{abstract}

\maketitle

\vspace*{10mm}

Hard sphere (HS) systems play a paramount role in statistical mechanics and material science since they are important models to study the behavior of many-particle systems like liquids, colloids, and metallic glasses~\cite{parisi_10,binder_11,hansen_13,torquato_02}. Also in the domain of granular materials many experimental and theoretical studies have used HS-like systems since granular particles usually have a high modulus and the simplicity of the particle shape allows to probe the influence of friction and roughness on the properties of granular assemblies~\cite{song_08,yuan_21}. Since for HS the interaction energy between the particles is trivial, the properties of such systems is encoded in their structure, i.e. the way the particles are arranged with respect to each other, and hence many previous studies have aimed to characterize this structure on the various length scales~\cite{watanabe_08,torquato_02,parisi_10,song_08,torquato_10}. Many of these investigations focused on one-component systems since this choice facilitates the analysis of the packing structure and also the theoretical calculations~\cite{watanabe_08,parisi_10}. However, in the context of the glass transition one also often uses a weakly polydisperse sample or a slightly asymmetric binary mixture, since this will efficiently suppress the crystallization of the liquid while keeping its structure close to the well understood one-component case, a feature that is shared by systems of  particles with short range interactions such as Lennard-Jones or soft spheres~\cite{kob_95,tong_18}. These studies have shown that, depending on the composition and the packing fraction, the local structure, i.e., the first nearest neighbor shell of a particle, shows a surprisingly rich variety, the details of which are at present still not well understood~\cite{clusel_09,tong_18,coslovich_07,royall_15,farr_09,ogarko_13}. In contrast to these almost one-component systems, the case of mixtures in which the particles have a size ratio that differs significantly from unity has been studied much less~\cite{dodds_75,statt_16,danisch_10,biazzo_09,yuan_18}.
Understanding this structure is, however, important since such highly asymmetric systems are relevant for describing the behavior of real granular materials, which are usually highly polydisperse, as well for the glass-forming ability of multi-component systems like metallic glasses in which the atoms can have very different radii~\cite{graf_03,desmond_14,williams_01,imhof_95,lazaro_19,bosse_95,wang_04}. In addition it has recently been found that such strongly asymmetric mixtures can have local structures with unexpected symmetries which can, e.g., be used to create novel materials via self-assembly~\cite{wang_21}. The main reason for our lack of understanding of these systems is that their theoretical description is significantly more complex and experiments on colloidal systems are hampered by the precise control of size ratio, while granular systems are prone to the phenomenon of phase separation~\cite{seiden_11,monteux_11}. A further problem that one faces with these systems is the difficulty to characterize their structure on intermediate length scales since the presence of two particle sizes gives rise to a highly complex distance dependency of the partial radial distribution functions, i.e., the standard quantities that are used to characterize the structure of many-body systems~\cite{hansen_86,grodon_04,grodon_05,statt_16}. Due to these difficulties there is at present little insight on how the size ratio or the composition affects the packing density or the structure of asymmetric HS systems. 

In this work we use the computed tomography (CT) technique to probe how composition and size ratio affect the packing fraction of a binary granular system. Using a recently developed method to characterize the structure of disordered systems at intermediate and large length scales we are able to show that a high packing fraction is intimately related to the presence of a short correlation length of the structure which in turn is caused by a large variety of possible local structural motifs. 

Our system is a binary mixture of particles with size ratio $q=d_b/d_s$, where $d_b$ and $d_s=3.0$~mm are the diameter of the big and small particles, respectively. The concentration of the small particles will be denoted by $x$. These particles are made of Acrylonitrile Butadiene Styrene plastic and the friction coefficient is 0.29. We consider two size ratios: $q=1.33$, which corresponds to a mixture that has particles with similar size, and $q=2.0$, i.e.,~the particle sizes are significantly different. The particles are thoroughly mixed and then dropped into a cylinder of diameters 180~mm, the walls of which have been covered randomly with half-spheres of diameter 4~mm to prevent layering of the particles. Once the cylinder is filled to a height of 200~mm it is placed in a medial CT scanner and the particle positions are obtained with a precision of about 0.01$d_s$. More details on the experimental procedure are given in the Supplementary Material (SM). Depending on the concentration $x$ we have between 22,000 and 169,000 particles in the cylinder and to obtain the following results we average over 3-5 independent samples.

\begin{figure}[tht]
\centering
\includegraphics[width=15cm]{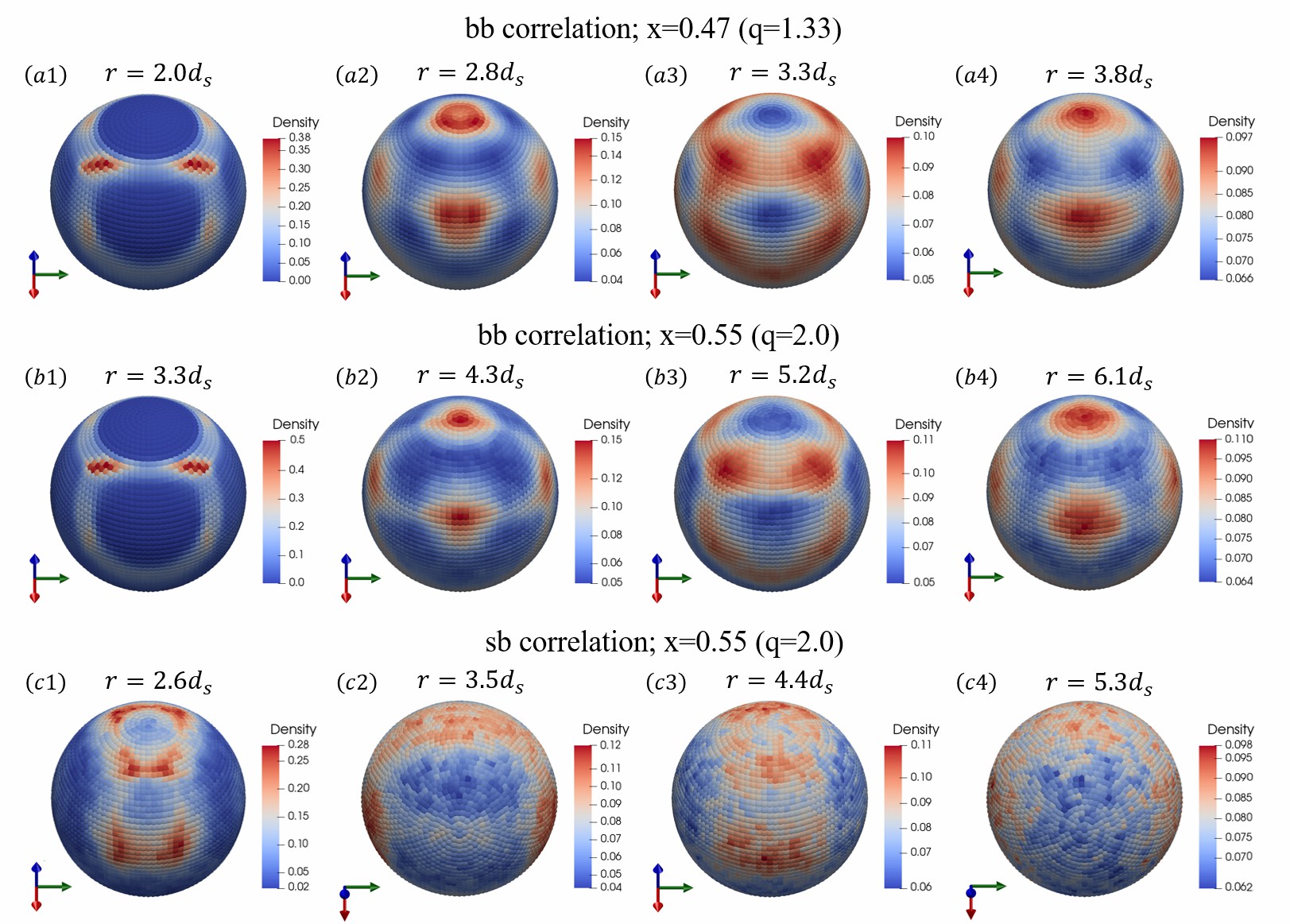}
\caption{Distribution of particles in three dimensions in a shell of thickness 0.5~$d_s$. a1)-a4): $bb-$correlation for the $x=0.47$ mixture ($q=1.33$ system); b1)-b4): $bb$-correlation for the $x=0.55$ mixture ($q=2.0$ system); c1)-c4) $sb$-correlation for the $x=0.55$ mixture ($q=2.0$ system).
}
\label{fig_1}
\end{figure}

Following Ref.~\cite{zhang_20} we analyze the 3d structure of the samples by introducing a local coordinate system. For this one picks any three particles of the same type that touch each other, i.e., they form a triangle with sides that are less than the distance of the first minimum in the radial distribution function $g_{\alpha\alpha}(r)$, with $\alpha \in \{b,s\}$ (see~SM, Fig.~\ref{fig_s2}a). We define the position of particle \#1 as the origin, the direction from particle \#1 to \#2 as the $z$-axis, the plane containing the three particles as the $x$-$z$~plane and the $y$-axis is orthogonal to it. This local reference frame allows therefore to determine the 3d distribution of the particles that have a distance $r$ from the central particle, i.e., a four-point correlation function.

In Fig.~\ref{fig_1}(a) we show for different values of $r$ the distribution $\rho_{bb}(\vec{r})$, i.e.~the 3d distribution of the density of the $b$-particles if at the center we also have a $b$-particle ($q=1.33$ and $x=0.47$). The snapshots show that $\rho_{bb}(\vec{r})$ is highly anisotropic in that it shows, at a given $r$, spots of high intensity that are arranged in a regular manner on the surface of the sphere. Depending on $r$ the geometrical arrangement of these spots have an icosahedral symmetry or a dodecahedral one, a result that is reasonable since the cavities at a given $r$ will have a high density at the distance of the subsequent layer, i.e.,~the two symmetries are dual to each other~\cite{zhang_20}. In Ref.~\cite{zhang_20} it was shown that at intermediate and large $r$ these two arrangements correspond to the location of the minima and maxima in the radial distribution functions $g_{bb}(r)$. Our finding regarding this regular alternation between these two platonic symmetries is thus the first experimental evidence that the structural order found in Ref.~\cite{zhang_20} does indeed exist in real systems. We also mention that the same alternating sequence is found for other compositions and other types of density fields (e.g., $\rho_{sb}(\vec{r})$, in which one has a $s$-particle in the center), although the symmetries of the structure might be a bit more smeared out.

For the $q=2.0$ system the field $\rho_{bb}(\vec{r})$ is shown in Fig.~\ref{fig_1}(b) and one sees that the spatial distribution is similar to the one of the $q=1.33$ system, i.e.~two alternating symmetries that are dual to each other. Figure~\ref{fig_1}(c) show that if the central particle is a $s$-particle, the density field has no longer an icosahedral/dodecahedral symmetry, although one still recognizes the presence of two alternating symmetries with increasing $r$. The first pattern has two zones of high intensities on the left and right of the sphere surface, connected by a broad band at the top of the sphere (panels c2 and c4) while the second pattern has three prominent bands orthogonal to the $x$-$z$ plane (panels c1 and c3). Thus these two spatial distributions have much less symmetry than the ones found for $q=1.33$, giving a first hint that increasing $q$ results in a decrease of the order at intermediate and large $r$. Movie~1 shows a 3d representation of the density field.

\begin{figure}[htbp]
\centering
\subfigure{
\includegraphics[width=7cm]{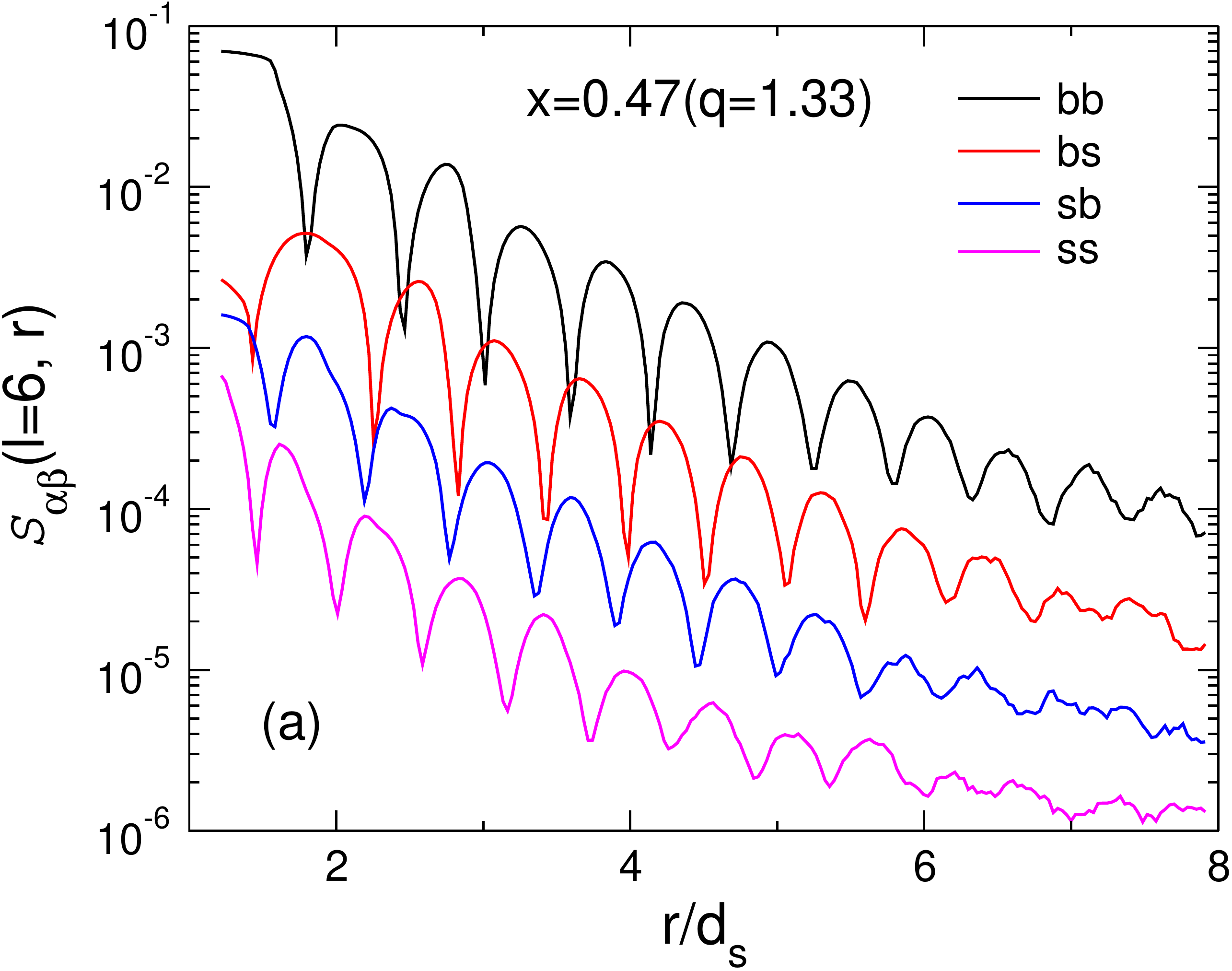}
}
\quad
\subfigure{
\includegraphics[width=7cm]{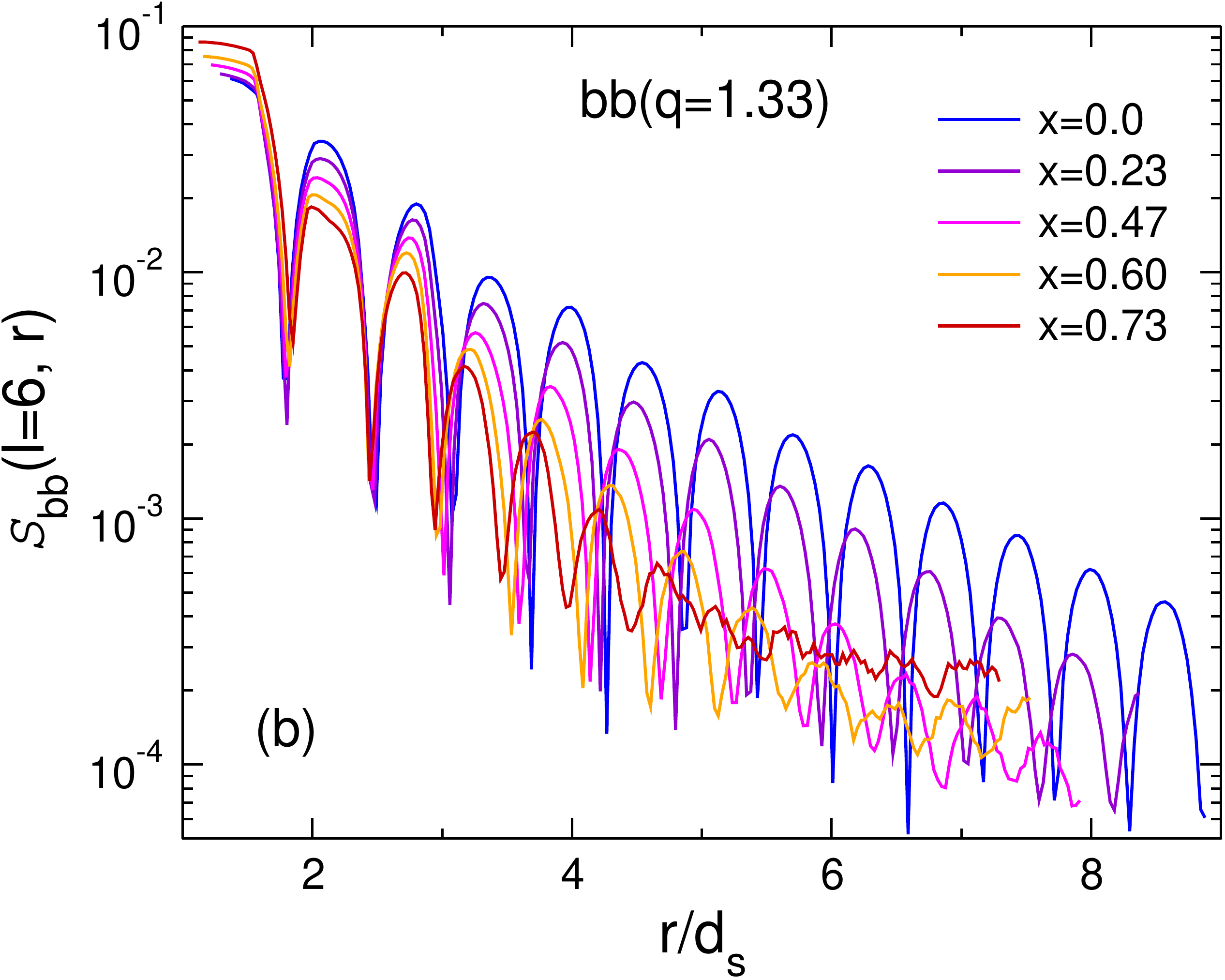}
}
\quad
\subfigure{
\includegraphics[width=7cm]{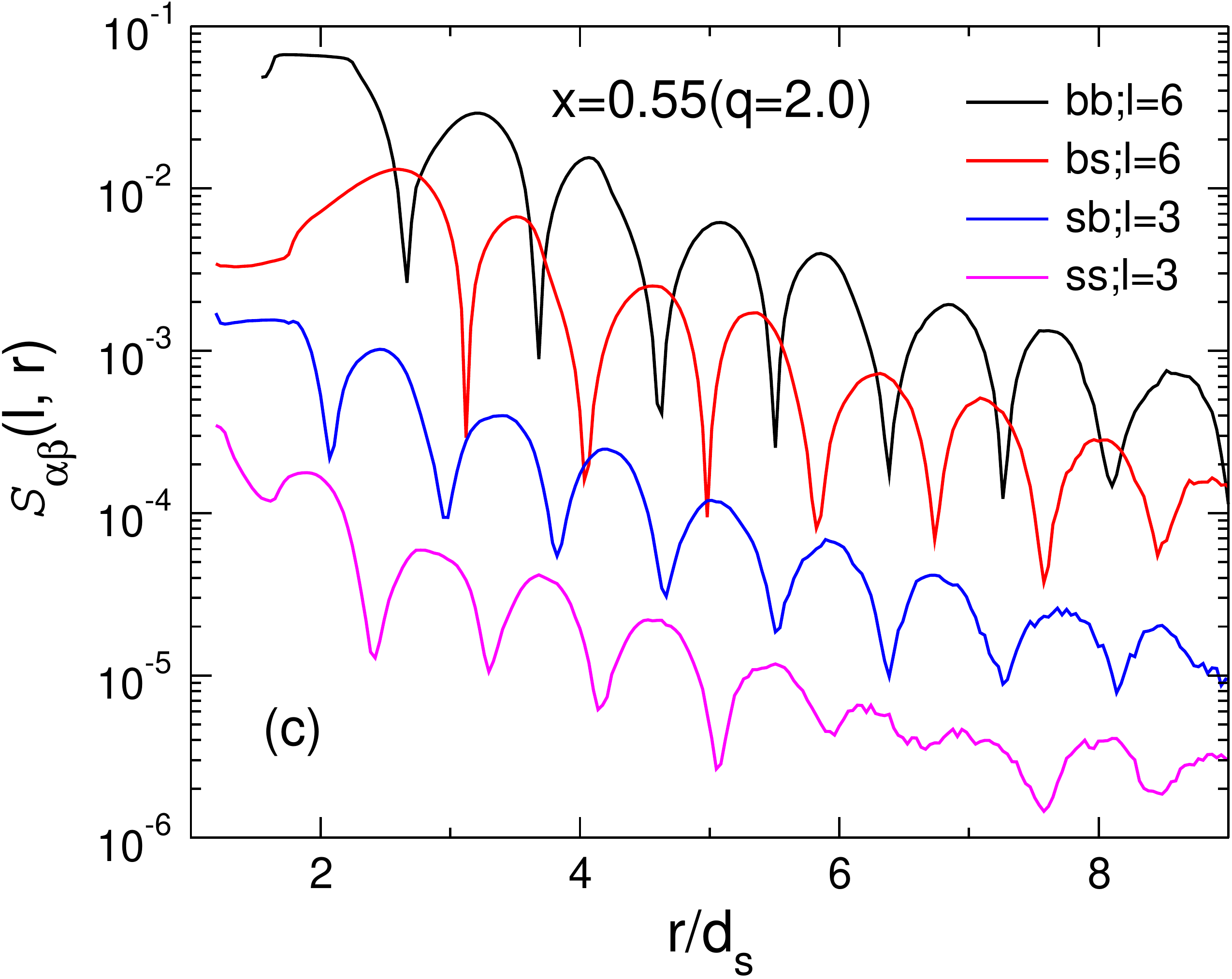}
}
\quad
\subfigure{
\includegraphics[width=7cm]{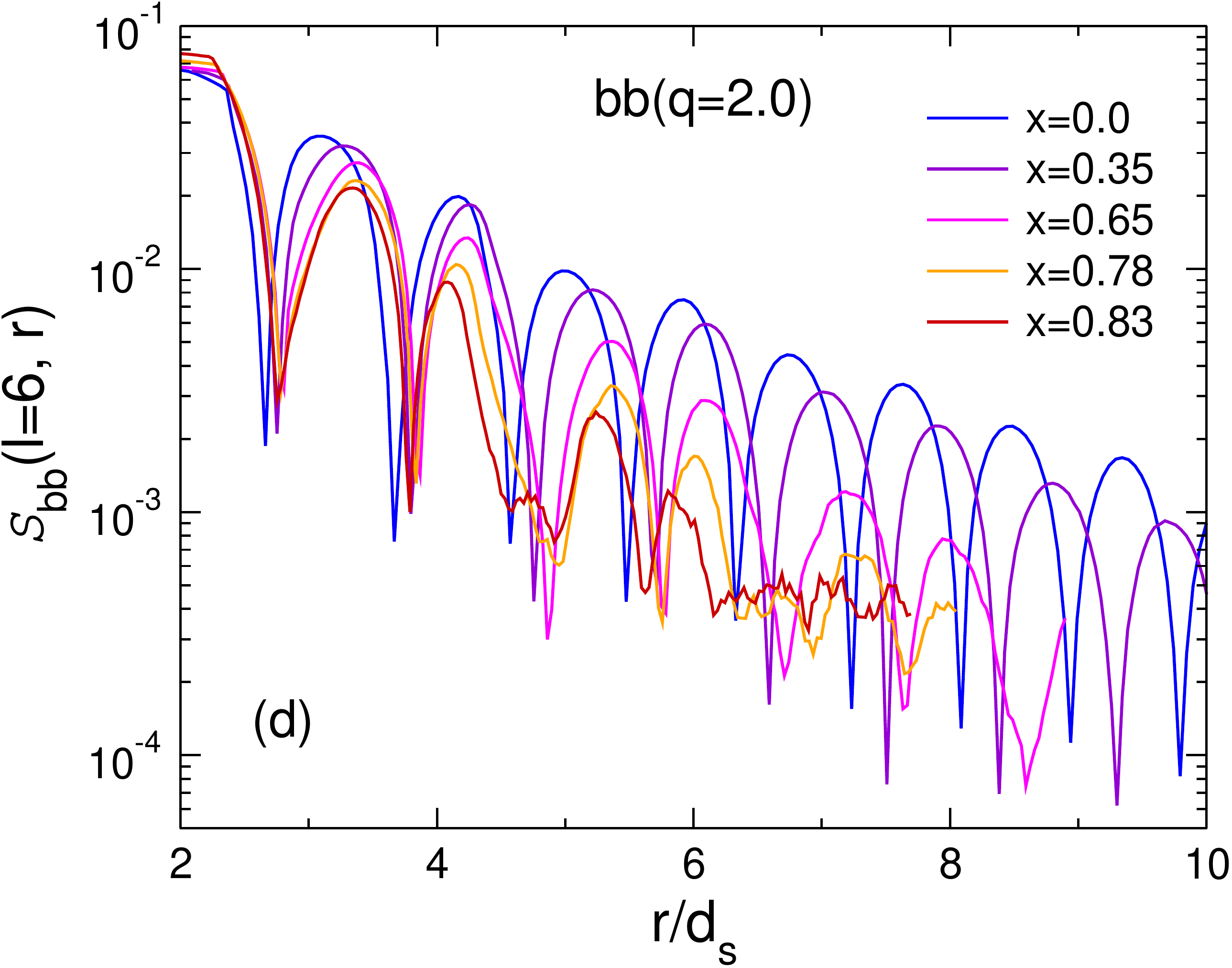}
}
\caption{$\mathcal{S}_{\alpha\beta}(l,r)$ for $bb$, $bs$, $sb$, and $ss$ correlations. Curves in panels (a) and (c) are shifted vertically by powers of $10^{-1}$ (top to bottom). (a) and (c): Different correlations for the $q=1.33$ and $q=2.0$ systems, respectively. (b) and (d): $bb$-correlation for different concentrations $x$ for the $q=1.33$ and $q=2.0$ systems, respectively.
}
\label{fig_2}
\end{figure}

The standard way to characterize the structure of many-body systems is by means of the partial radial distribution functions $g_{\alpha\beta}(r)$ or, equivalently, the partial static structure factors~\cite{hansen_13}. Figures~\ref{fig_s1} and \ref{fig_s3} of the SM show that for the case of $q=2.0$ it is difficult to interpret the $r$-dependence of these functions due to the presence of a multitude of peaks, the height of which show a complicated $r$-dependence, in agreement with the conclusion from previous studies~\cite{grodon_04,grodon_05,statt_16}.
Since also the dependence of $g_{\alpha\beta}(r)$ on composition is complex, Fig.~\ref{fig_s3} in the SM, it is not possible to extract from these functions in a non-arbitrary manner a physically meaningful decay length that could be used to describe the range of the structural order. Hence one concludes that the projection of the complex 3d structure found in Fig.~\ref{fig_1} on the one-dimensional quantity $g_{\alpha\beta}(r)$ leads to a severe loss of information and thus it is necessary to study directly the four-point correlation function.

To quantify the anisotropic density distribution of Fig.~\ref{fig_1}, we decompose $\rho_{\alpha\beta}(\vec{r})$ into spherical harmonics $Y_l^m(\theta,\phi,r)$:

\begin{equation}
\rho_{\alpha\beta}(\theta,\phi,r) =
\sum_{l=0}^\infty \sum_{m=-l}^{l}\rho_{\alpha\beta,l}^m(r) Y_l^m(\theta,\phi) \quad ,
\label{eq_1}
\end{equation}

\noindent
where $\theta$ and $\phi$ are the angular variables and the expansion coefficients $\rho_{\alpha\beta,l}^m$ are given in the SM. We then use the square root of the angular power spectrum, $\mathcal{S}_{\alpha\beta}(l,r)= [(2l+1)^{-1}\sum_{m=-l}^{l}|\rho_{\alpha\beta,}l^m(r)|^2]^{1/2}$, to characterize the anisotropy of the density distribution.

Figure~\ref{fig_2}(a) presents for $q=1.33$ and $x=0.47$ the $r$-dependence of $\mathcal{S}_{\alpha\beta}$ for the four possible combinations of the field. These curves are for $l=6$, since 
$\mathcal{S}_{\alpha\beta}(l=6,r)$ has the largest signal (see SM), due to the large number of angles around $60^\circ$ in the icosahedron~(dodecahedron)-like structure of the density field, see Fig.~\ref{fig_1}(a). We see that these functions show an oscillatory behavior and that their envelope decays in an exponential manner with $r$, in agreement with the results from Ref.~\cite{zhang_20} where the origin of these oscillations have been discussed. Thus the slope of the envelope can be used to define a structural length scale $\xi$ which will be studied in the following. Since this slope is independent of the function considered, we can average over the corresponding four values of $\xi$ and hence improve the accuracy of the estimate (see SM for details).

In Fig.~\ref{fig_2}(b) we show $\mathcal{S}_{bb}(r)$ for the $q=1.33$ system at different concentrations $x$ and one sees that the position of the peaks depends only mildly on $x$ but that their intensity changes rapidly with increasing $x$. This indicates that the geometry of the structure, {\it i.e.}, the icosahedron-dodecahedron sequence, is present up to relatively high concentration of the small particles, but that the structures are less pronounced.

\begin{figure}[tbp]
\centering
\subfigure{
\includegraphics[width=7cm]{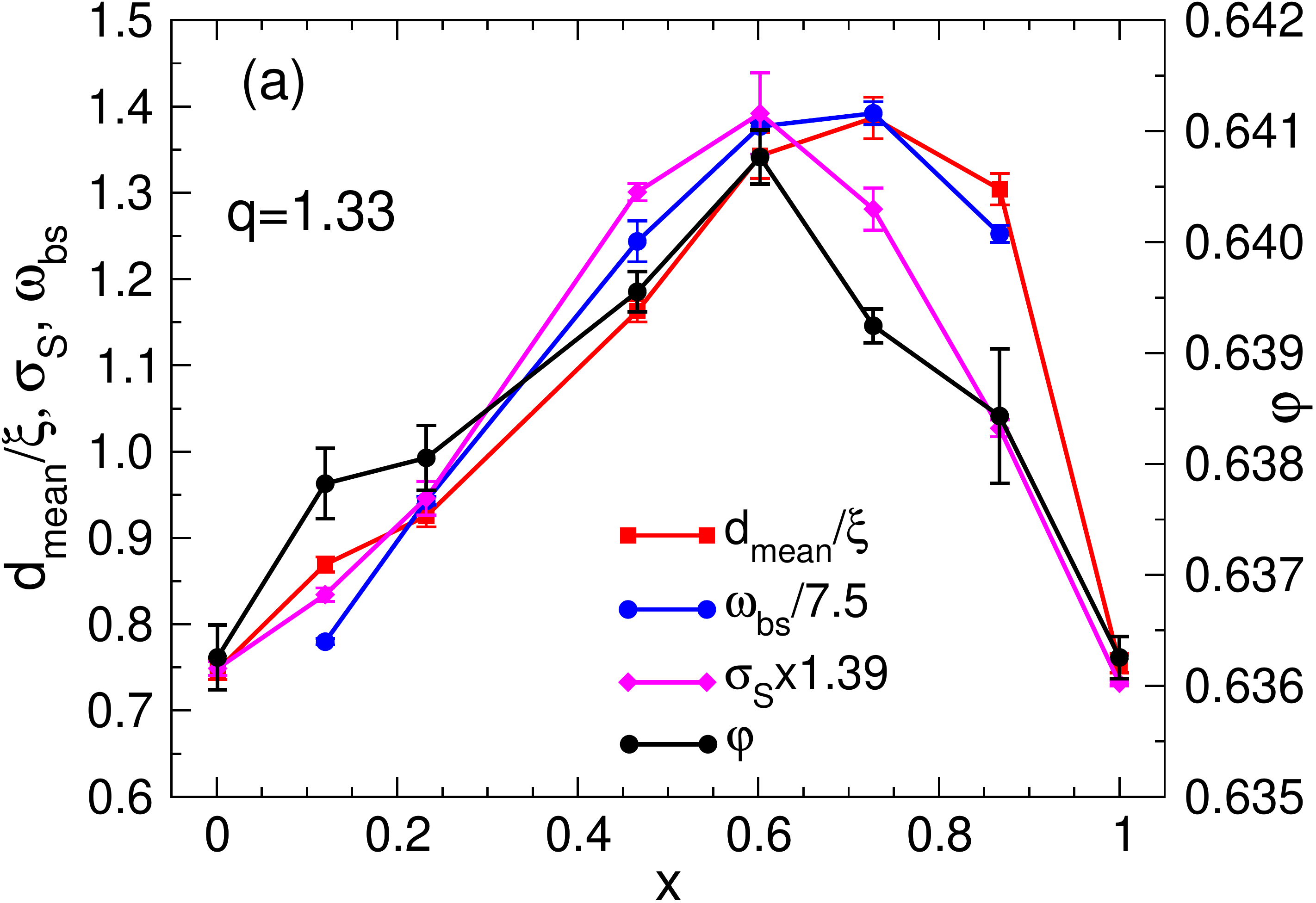}
}
\quad
\subfigure{
\includegraphics[width=7.1cm]{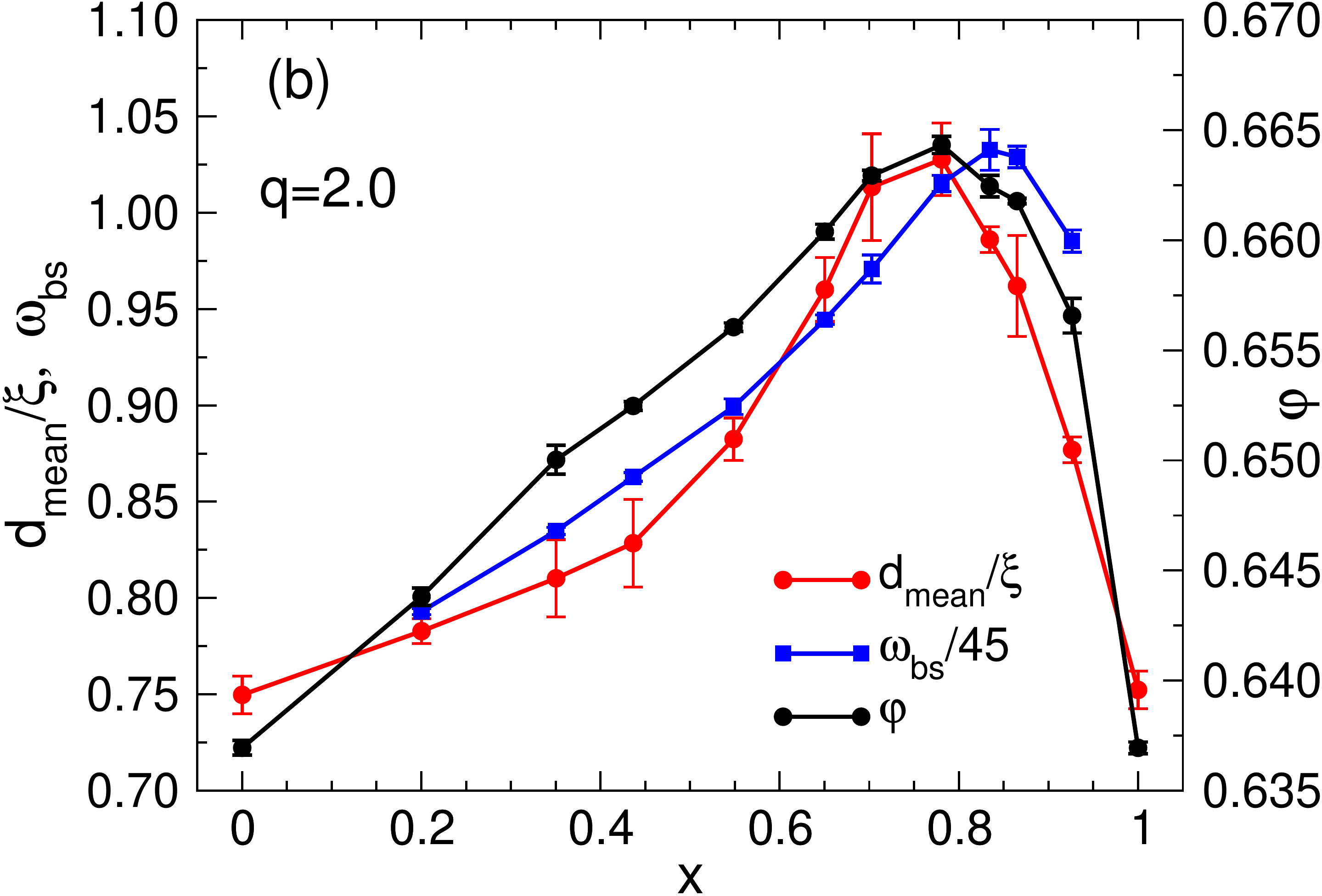}
}
\quad
\caption{Composition dependence of the length scale $\xi$ for the $q=1.33$, panel (a), and $q=2.0$, panel (b), systems. Also included is the packing fraction $\varphi$, the standard deviation of the coordination number, $\omega_{bs}$, and in panel (a) the full width at half maximum of the static structure factor, $\sigma_S$.
} 
\label{fig_3}
\end{figure}

Figure~\ref{fig_3}(c) shows $\mathcal{S}_{bb}(r)$ for the $q=2.0$ system at $x=0.55$. The value of $l$ is still 6, since the signal is again the strongest. However, if the coordinate system is centered on a $s$-particle we find that the functions $\mathcal{S}_{s\alpha}(r)$ have the largest signal for $l=3$, a result that is consistent with the fact that the density field is no longer given by a icosahedra/dodecahedral symmetry, see Fig.~\ref{fig_1}(c). Despite the different values for $l$, we see that the decay length of the signal is independent of the field considered and thus can be used as a robust indicator for the structural order.

Finally we show in Fig.~\ref{fig_2}(d) $\mathcal{S}_{bb}(r)$ for the $q=2.0$ system for different concentrations of the $s$-particles. We recognize that the $x$-dependence of the curves is smooth and that the main effect of changing composition is that the slope of the envelope changes (and below we will show that the slope, i.e.~the structural length scale, is non-monotonic in $x$). Thus we can conclude that, in contrast to the partial radial distribution functions $g_{\alpha\beta}(r)$, the quantity $\mathcal{S}_{\alpha\beta}(l,r)$ allows to quantify in a direct manner how the structure on intermediate and large scale changes as a function of composition.

Figure~\ref{fig_3}(a) shows the $x$-dependence of the length scale $\xi$ for the $q=1.33$ system and one recognizes that $\xi^{-1}$ shows a maximum at around $x_{\rm max}=0.7$, i.e.~at this concentration the structure is maximally disordered.. (Note that here we express $\xi$ in terms of the mean particle diameter $d_{\rm mean}=(1-x)d_b+xd_s$ so that the $x=0$ and $x=1$ systems have the same normalized length scale.) For (quasi-)one component systems it is common to determine the structural coherence length from $\sigma_S$, the width of the first peak in the static structure factor~\cite{binder_11}. In the SM we show that this is indeed feasible for the $q=1.33$ system while for larger $q$ this is not possible since there is no main peak, confirming once more the difficulty to characterize the structure for such systems. We have included in Fig.~\ref{fig_3}(a) also the $x$-dependence of $\sigma_S$ and one recognizes that this data is qualitatively very similar to the one of $\xi^{-1}$ giving evidence that these two quantities are indeed closely related to each other. Also shown in the graph is the packing fraction $\varphi$ defined by 
$\varphi=\sum_i v_i/\sum_i v_{\rm voro, i}$, where $v_i$ and $v_{\rm voro,i}$ are, respectively, the volume and Voronoi volume of particle $i$~\footnote{Since the CT scanner has only a finite spatial resolution, the packing density of the system with only big spheres is not quite the same as the one for a system with the small spheres. In the SM we discuss how this minor effect has been corrected.}. Since $\varphi(x)$ shows a maximum close to $\varphi_{\rm max}$ we have a first indication that a small correlation is associated with a high packing fraction.

For the $q=2.0$ system, Fig.~\ref{fig_3}(b), we find that $\xi^{-1}$ has a maximum which is shifted to larger $x$, i.e.~$x_{\rm max} \approx 0.8$.
(Note that for this value of $q$ it is not possible to extract $\sigma_S$, see SM, and hence we do not show this quantity.) Also included in the graph is the packing fraction and we find that the maximum is also shifted to larger $x$ and its height is increased, consistent with theoretical predictions~\cite{biazzo_09,danisch_10}. In agreement with the results for the $q=1.33$ system, the location of the maximum in $\varphi(x)$ coincides with the one of $\xi^{-1}$, giving further evidence that these two quantities are closely related.

\begin{figure}[ht]
\centering
\subfigure{
\includegraphics[width=7cm]{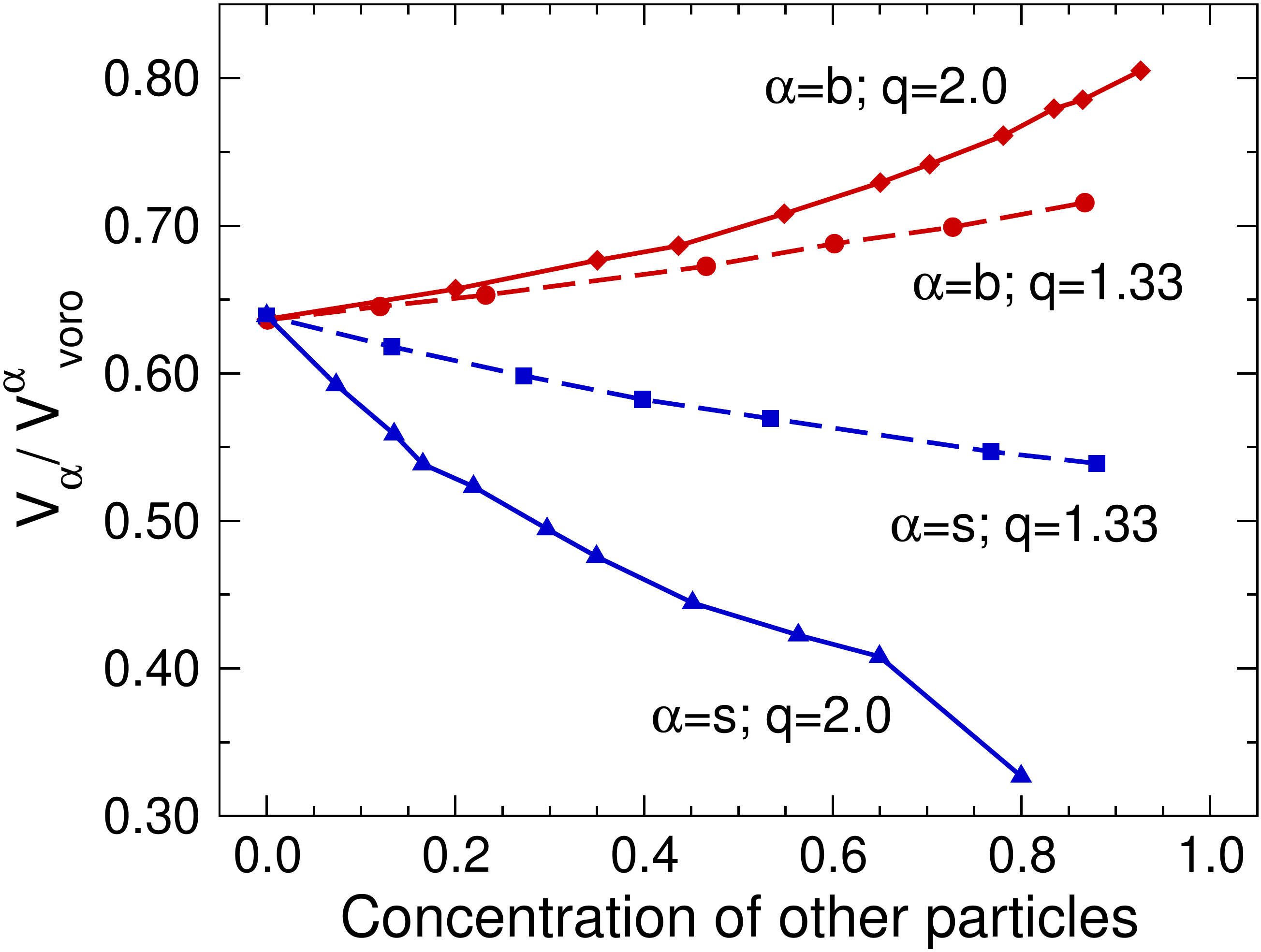}
}
\quad

\caption{Average local packing fraction $\varphi_b$ and $\varphi_s$. Full and dashed lines are for the $q=2.0$ and $q=1.33$ systems, respectively. Error bars are smaller than the size of the symbols.
}
\label{fig_4}
\end{figure}

To understand the connection between $\varphi$ and $\xi$ one has to recall that one-component HS-like systems can have a {\it local} packing fraction that is significantly higher than the {\it global} packing fraction in that, e.g., the central particle of an icosahedron has $\varphi_{\rm ico}=0.74$, thus well above the density of random close packing, $\varphi_{\rm rcp}\approx 0.64$~\cite{binder_11}. The reason for this difference is that icosahedra cannot tile space and hence the densely packed local structures have to be supplemented with structures that are packed less densely, resulting in an overall packing fraction less than $\varphi_{\rm ico}$~\cite{parisi_10,binder_11}. The presence of a second kind of particle has two effects on the packing: 1) Big particles can increase their local packing density by having small particles as nearest neighbors since the minimum distance between a big and small particle is less than the one between two big particles, thus resulting in a decrease of the Voronoi volume of the central particle. This trend is shown in Fig.~\ref{fig_4} for the two systems in that we find a slow rise of $\varphi_b$ with $x$ and the increase is faster for large $q$. 2) For small particles the presence of the big particles has the opposite effect, {\it i.e.}, the local packing fraction $\varphi_s(x)$ decreases with the addition of the big particles because a big particle that is close to a small one will strongly increase the Voronoi volume of the latter, see Fig.~\ref{fig_4}. The total packing fraction is given by

\begin{equation}
\varphi(x)=\frac{N_b S_b+N_s S_s}{N_b V_{\rm voro}^b+ N_sV_{\rm voro}^s}=\frac{\varphi_b(x) \varphi_s(x) [q^3 (1-x)+x]}{\varphi_s(x) q^3(1-x)+\varphi_b(x)x}    \quad ,
\label{eq_2}
\end{equation}

\noindent
where $N_\alpha$ and $S_\alpha$ are the number and volume of particles of type $\alpha$, respectively. Since $\varphi_s$ shows a stronger $x$-dependence than $\varphi_b$, one has a maximum in $\varphi(x)$ at $x_{\rm max}>0.5$, in agreement with the data from Fig.~\ref{fig_3}. 

Note that $\varphi_b$ (and $\varphi_s$) is an average packing fraction, i.e.~it is the weighted average over particles that have different local environments. In order to maximize $\varphi_b$ one thus needs that the relevant local environments, i.e., those with non-negligible weight, do i) have a high packing fraction and ii) can be joined together without the need to have loosely packed interfaces between them. This latter condition is met most easily if there is a large number of possible local environments that have a high packing density since this allows for a greater flexibility in the assembly of the total packing. To estimate the number of such local environments at a given $x$ we probe $C_{\alpha\beta}(k)$, the probability that a particle of type $\alpha$ has exactly $k$ nearest neighbors of type $\beta$. Figure~\ref{fig_s9} shows that these distributions are Gaussian-like and thus their standard deviation, $\omega_{\alpha\beta}$, can be taken as a measure for the variety of the local environments of a given particle type. The $x$-dependence of $\omega_{bb}$ is included in Fig.~\ref{fig_3} as well and one sees that this curve does indeed match very well the $x$-dependence of $\varphi$. (Figure~\ref{fig_s10} demonstrates that the shape of $\omega_{\alpha\beta}(x)$ is independent of the choice of $\alpha$ and $\beta$ and, Fig.~\ref{fig_s11}, that the statistically relevant local environments do indeed have a high packing fraction.) Thus this supports the view that in order to have a high packing fraction one needs indeed a large variety of local structures. Since these local structures are by definition different, the resulting global structural correlation length will be small. Hence this rationalizes why $\varphi(x)$ and $\xi^{-1}$ peak at the same concentration $x_{\rm max}$, see Fig.~\ref{fig_3}.

In this work we have used a CT scanner to determine the structural properties of granular packings on intermediate and large lengths scales. Due to the preparation protocol and the existence of friction, these structures are not exactly the same as the frictionless packings studied in theoretical works~\cite{biazzo_09,danisch_10,dodds_75}, but they do correspond to packings that occur in real granular systems. Using a novel method to characterize the structure in 3d, we are able to determine for the first time a static length scale $\xi$ even for strongly asymmetric mixtures. The fact that $\xi^{-1}(x)$ shows a maximum at the same concentration at which the maximum packing fraction occurs indicates that in multi-component disordered systems efficient packing is related to a {\it short} correlation length. (We emphasize that this result is not in contradiction with the expectation that {\it at fixed concentration $x$} the correlation length grows if the packing fraction is increased~\cite{binder_11}.) Surprisingly, the presence of a short correlation length does not exclude the possibility that the structure shows order even on intermediate length scales, see Fig.~\ref{fig_1}(b)/(c). The nature of this order depends on the size ratio and its classification is not straightforward because the structure has low symmetry, calling for further studies to clarify this dependence. Note that the presence of the symmetry shown in Fig.~\ref{fig_1}(c) has not been documented before, demonstrating that probing the structure in 3d is a powerful approach which allows to characterize the structure of many-body systems that so far could not be described in an insightful manner. 
Exploiting this approach to multi-component systems will thus allow not only to advance our understanding of disordered systems but also facilitate the creation of novel self-assembled materials and glass-forming systems~\cite{wang_21}.

The presence of a multitude of local structures close to $x_{\rm max}$ invites to speculate about the meaning of eutectic points in the equilibrium phase diagrams of multi-component systems and their glass-forming ability~\cite{zhang_14}. The high number of such structures for compositions close to $x_{\rm max}$ will have the consequence that the system has difficulty to crystallize because of entropic contribution due to these structures will lower the free energy of the disordered phase~\cite{zhang_14,russo_18}. 
Thus we can expect that $x_{\rm max}$ will be close to an eutectic point. (Note that in non-HS systems the energetic contributions can play a role as well.) Since at $x_{\rm max}$ the packing density is high, one expects that close to this concentration  the dynamics of the system is slow, indicating that close to $x_{\rm max}$ the glass-forming ability of the system is high, a result that will be useful for the identification of novel glass-formers.

Finally we mention that our result on the existence of a maximum in $\varphi(x)$, directly related to the competition between opposite trends for the local packing density of the big and small particles, calls for studies that connect this observation to the formalism of the Edwards measure for granular systems~\cite{edwards_89,yuan_21}. Since this measure is based on the entropy related to the global packing density, it is important to probe how the mentioned opposite trends are reflected in this theoretical framework, i.e., to see whether it is possible to rationalize our results within a thermodynamic approach.

Acknowledgments: We thank H.~Tanaka and F.~Zamponi for useful discussions. W.K. is senior member of the Institut Universitaire de France. The work was supported by the National Natural Science Foundation of China (No. 11974240), the Shanghai Science and Technology Committee (No. 19XD1402100), and the China Scholarship Council Grant 201606050112.\\

H.Y. and Z.Z. contributed equally to this work.

\vfill


\begin{thebibliography}{40}%
\makeatletter
\providecommand \@ifxundefined [1]{%
 \@ifx{#1\undefined}
}%
\providecommand \@ifnum [1]{%
 \ifnum #1\expandafter \@firstoftwo
 \else \expandafter \@secondoftwo
 \fi
}%
\providecommand \@ifx [1]{%
 \ifx #1\expandafter \@firstoftwo
 \else \expandafter \@secondoftwo
 \fi
}%
\providecommand \natexlab [1]{#1}%
\providecommand \enquote  [1]{``#1''}%
\providecommand \bibnamefont  [1]{#1}%
\providecommand \bibfnamefont [1]{#1}%
\providecommand \citenamefont [1]{#1}%
\providecommand \href@noop [0]{\@secondoftwo}%
\providecommand \href [0]{\begingroup \@sanitize@url \@href}%
\providecommand \@href[1]{\@@startlink{#1}\@@href}%
\providecommand \@@href[1]{\endgroup#1\@@endlink}%
\providecommand \@sanitize@url [0]{\catcode `\\12\catcode `\$12\catcode
  `\&12\catcode `\#12\catcode `\^12\catcode `\_12\catcode `\%12\relax}%
\providecommand \@@startlink[1]{}%
\providecommand \@@endlink[0]{}%
\providecommand \url  [0]{\begingroup\@sanitize@url \@url }%
\providecommand \@url [1]{\endgroup\@href {#1}{\urlprefix }}%
\providecommand \urlprefix  [0]{URL }%
\providecommand \Eprint [0]{\href }%
\providecommand \doibase [0]{http://dx.doi.org/}%
\providecommand \selectlanguage [0]{\@gobble}%
\providecommand \bibinfo  [0]{\@secondoftwo}%
\providecommand \bibfield  [0]{\@secondoftwo}%
\providecommand \translation [1]{[#1]}%
\providecommand \BibitemOpen [0]{}%
\providecommand \bibitemStop [0]{}%
\providecommand \bibitemNoStop [0]{.\EOS\space}%
\providecommand \EOS [0]{\spacefactor3000\relax}%
\providecommand \BibitemShut  [1]{\csname bibitem#1\endcsname}%
\let\auto@bib@innerbib\@empty
\bibitem [{\citenamefont {Parisi}\ and\ \citenamefont
  {Zamponi}(2010)}]{parisi_10}%
  \BibitemOpen
  \bibfield  {author} {\bibinfo {author} {\bibfnamefont {G.}~\bibnamefont
  {Parisi}}\ and\ \bibinfo {author} {\bibfnamefont {F.}~\bibnamefont
  {Zamponi}},\ }\href@noop {} {\bibfield  {journal} {\bibinfo  {journal} {Rev.
  Mod. Phys.}\ }\textbf {\bibinfo {volume} {82}},\ \bibinfo {pages} {789}
  (\bibinfo {year} {2010})}\BibitemShut {NoStop}%
\bibitem [{\citenamefont {Binder}\ and\ \citenamefont {Kob}(2011)}]{binder_11}%
  \BibitemOpen
  \bibfield  {author} {\bibinfo {author} {\bibfnamefont {K.}~\bibnamefont
  {Binder}}\ and\ \bibinfo {author} {\bibfnamefont {W.}~\bibnamefont {Kob}},\
  }\href@noop {} {\emph {\bibinfo {title} {Glassy materials and disordered
  solids: An introduction to their statistical mechanics}}}\ (\bibinfo
  {publisher} {World Scientific},\ \bibinfo {year} {2011})\BibitemShut
  {NoStop}%
\bibitem [{\citenamefont {Hansen}\ and\ \citenamefont
  {McDonald}(2013)}]{hansen_13}%
  \BibitemOpen
  \bibfield  {author} {\bibinfo {author} {\bibfnamefont {J.-P.}\ \bibnamefont
  {Hansen}}\ and\ \bibinfo {author} {\bibfnamefont {I.~R.}\ \bibnamefont
  {McDonald}},\ }\href@noop {} {\emph {\bibinfo {title} {Theory of simple
  liquids}}}\ (\bibinfo  {publisher} {Academic Press},\ \bibinfo {year}
  {2013})\BibitemShut {NoStop}%
\bibitem [{\citenamefont {Torquato}(2002)}]{torquato_02}%
  \BibitemOpen
  \bibfield  {author} {\bibinfo {author} {\bibfnamefont {S.}~\bibnamefont
  {Torquato}},\ }\href@noop {} {\emph {\bibinfo {title} {Random Heterogeneous
  Materials: Microstructure and Macroscopic Properties}}}\ (\bibinfo
  {publisher} {Springer, New York},\ \bibinfo {year} {2002})\BibitemShut
  {NoStop}%
\bibitem [{\citenamefont {Song}\ \emph {et~al.}(2008)\citenamefont {Song},
  \citenamefont {Wang},\ and\ \citenamefont {Makse}}]{song_08}%
  \BibitemOpen
  \bibfield  {author} {\bibinfo {author} {\bibfnamefont {C.}~\bibnamefont
  {Song}}, \bibinfo {author} {\bibfnamefont {P.}~\bibnamefont {Wang}}, \ and\
  \bibinfo {author} {\bibfnamefont {H.~A.}\ \bibnamefont {Makse}},\ }\href@noop
  {} {\bibfield  {journal} {\bibinfo  {journal} {Nature}\ }\textbf {\bibinfo
  {volume} {453}},\ \bibinfo {pages} {629} (\bibinfo {year}
  {2008})}\BibitemShut {NoStop}%
\bibitem [{\citenamefont {Yuan}\ \emph {et~al.}(2021)\citenamefont {Yuan},
  \citenamefont {Xing}, \citenamefont {Zheng}, \citenamefont {Li},
  \citenamefont {Yuan}, \citenamefont {Zhang}, \citenamefont {Zeng},
  \citenamefont {Xia}, \citenamefont {Tong}, \citenamefont {Kob}, \citenamefont
  {Zhang},\ and\ \citenamefont {Wang}}]{yuan_21}%
  \BibitemOpen
  \bibfield  {author} {\bibinfo {author} {\bibfnamefont {Y.}~\bibnamefont
  {Yuan}}, \bibinfo {author} {\bibfnamefont {Y.}~\bibnamefont {Xing}}, \bibinfo
  {author} {\bibfnamefont {J.}~\bibnamefont {Zheng}}, \bibinfo {author}
  {\bibfnamefont {Z.}~\bibnamefont {Li}}, \bibinfo {author} {\bibfnamefont
  {H.}~\bibnamefont {Yuan}}, \bibinfo {author} {\bibfnamefont {S.}~\bibnamefont
  {Zhang}}, \bibinfo {author} {\bibfnamefont {Z.}~\bibnamefont {Zeng}},
  \bibinfo {author} {\bibfnamefont {C.}~\bibnamefont {Xia}}, \bibinfo {author}
  {\bibfnamefont {H.}~\bibnamefont {Tong}}, \bibinfo {author} {\bibfnamefont
  {W.}~\bibnamefont {Kob}}, \bibinfo {author} {\bibfnamefont {J.}~\bibnamefont
  {Zhang}}, \ and\ \bibinfo {author} {\bibfnamefont {Y.}~\bibnamefont {Wang}},\
  }\href@noop {} {\bibfield  {journal} {\bibinfo  {journal} {Phys. Rev. Lett.}\
  }\textbf {\bibinfo {volume} {127}},\ \bibinfo {pages} {018002} (\bibinfo
  {year} {2021})}\BibitemShut {NoStop}%
\bibitem [{\citenamefont {Watanabe}\ and\ \citenamefont
  {Tanaka}(2008)}]{watanabe_08}%
  \BibitemOpen
  \bibfield  {author} {\bibinfo {author} {\bibfnamefont {K.}~\bibnamefont
  {Watanabe}}\ and\ \bibinfo {author} {\bibfnamefont {H.}~\bibnamefont
  {Tanaka}},\ }\href@noop {} {\bibfield  {journal} {\bibinfo  {journal} {Phys.
  Rev. Lett.}\ }\textbf {\bibinfo {volume} {100}},\ \bibinfo {pages} {158002}
  (\bibinfo {year} {2008})}\BibitemShut {NoStop}%
\bibitem [{\citenamefont {Torquato}\ and\ \citenamefont
  {Stillinger}(2010)}]{torquato_10}%
  \BibitemOpen
  \bibfield  {author} {\bibinfo {author} {\bibfnamefont {S.}~\bibnamefont
  {Torquato}}\ and\ \bibinfo {author} {\bibfnamefont {F.~H.}\ \bibnamefont
  {Stillinger}},\ }\href@noop {} {\bibfield  {journal} {\bibinfo  {journal}
  {Rev. Mod. Phys.}\ }\textbf {\bibinfo {volume} {82}},\ \bibinfo {pages}
  {2633} (\bibinfo {year} {2010})}\BibitemShut {NoStop}%
\bibitem [{\citenamefont {Kob}\ and\ \citenamefont {Andersen}(1995)}]{kob_95}%
  \BibitemOpen
  \bibfield  {author} {\bibinfo {author} {\bibfnamefont {W.}~\bibnamefont
  {Kob}}\ and\ \bibinfo {author} {\bibfnamefont {H.~C.}\ \bibnamefont
  {Andersen}},\ }\href@noop {} {\bibfield  {journal} {\bibinfo  {journal}
  {Phys. Rev. E}\ }\textbf {\bibinfo {volume} {51}},\ \bibinfo {pages} {4626}
  (\bibinfo {year} {1995})}\BibitemShut {NoStop}%
\bibitem [{\citenamefont {Tong}\ and\ \citenamefont {Tanaka}(2018)}]{tong_18}%
  \BibitemOpen
  \bibfield  {author} {\bibinfo {author} {\bibfnamefont {H.}~\bibnamefont
  {Tong}}\ and\ \bibinfo {author} {\bibfnamefont {H.}~\bibnamefont {Tanaka}},\
  }\href@noop {} {\bibfield  {journal} {\bibinfo  {journal} {Phys. Rev. X}\
  }\textbf {\bibinfo {volume} {8}},\ \bibinfo {pages} {011041} (\bibinfo {year}
  {2018})}\BibitemShut {NoStop}%
\bibitem [{\citenamefont {Clusel}\ \emph {et~al.}(2009)\citenamefont {Clusel},
  \citenamefont {Corwin}, \citenamefont {Siemens},\ and\ \citenamefont
  {Bruji{\'c}}}]{clusel_09}%
  \BibitemOpen
  \bibfield  {author} {\bibinfo {author} {\bibfnamefont {M.}~\bibnamefont
  {Clusel}}, \bibinfo {author} {\bibfnamefont {E.~I.}\ \bibnamefont {Corwin}},
  \bibinfo {author} {\bibfnamefont {A.~O.}\ \bibnamefont {Siemens}}, \ and\
  \bibinfo {author} {\bibfnamefont {J.}~\bibnamefont {Bruji{\'c}}},\
  }\href@noop {} {\bibfield  {journal} {\bibinfo  {journal} {Nature}\ }\textbf
  {\bibinfo {volume} {460}},\ \bibinfo {pages} {611} (\bibinfo {year}
  {2009})}\BibitemShut {NoStop}%
\bibitem [{\citenamefont {Coslovich}\ and\ \citenamefont
  {Pastore}(2007)}]{coslovich_07}%
  \BibitemOpen
  \bibfield  {author} {\bibinfo {author} {\bibfnamefont {D.}~\bibnamefont
  {Coslovich}}\ and\ \bibinfo {author} {\bibfnamefont {G.}~\bibnamefont
  {Pastore}},\ }\href@noop {} {\bibfield  {journal} {\bibinfo  {journal} {J.
  Chem. Phys.}\ }\textbf {\bibinfo {volume} {127}},\ \bibinfo {pages} {124504}
  (\bibinfo {year} {2007})}\BibitemShut {NoStop}%
\bibitem [{\citenamefont {Royall}\ and\ \citenamefont
  {Williams}(2015)}]{royall_15}%
  \BibitemOpen
  \bibfield  {author} {\bibinfo {author} {\bibfnamefont {C.~P.}\ \bibnamefont
  {Royall}}\ and\ \bibinfo {author} {\bibfnamefont {S.~R.}\ \bibnamefont
  {Williams}},\ }\href@noop {} {\bibfield  {journal} {\bibinfo  {journal}
  {Phys. Rep.}\ }\textbf {\bibinfo {volume} {560}},\ \bibinfo {pages} {1}
  (\bibinfo {year} {2015})}\BibitemShut {NoStop}%
\bibitem [{\citenamefont {Farr}\ and\ \citenamefont {Groot}(2009)}]{farr_09}%
  \BibitemOpen
  \bibfield  {author} {\bibinfo {author} {\bibfnamefont {R.~S.}\ \bibnamefont
  {Farr}}\ and\ \bibinfo {author} {\bibfnamefont {R.~D.}\ \bibnamefont
  {Groot}},\ }\href@noop {} {\bibfield  {journal} {\bibinfo  {journal} {J.
  Chem. Phys.}\ }\textbf {\bibinfo {volume} {131}},\ \bibinfo {pages} {244104}
  (\bibinfo {year} {2009})}\BibitemShut {NoStop}%
\bibitem [{\citenamefont {Ogarko}\ and\ \citenamefont
  {Luding}(2013)}]{ogarko_13}%
  \BibitemOpen
  \bibfield  {author} {\bibinfo {author} {\bibfnamefont {V.}~\bibnamefont
  {Ogarko}}\ and\ \bibinfo {author} {\bibfnamefont {S.}~\bibnamefont
  {Luding}},\ }\href@noop {} {\bibfield  {journal} {\bibinfo  {journal} {Soft
  Matter}\ }\textbf {\bibinfo {volume} {9}},\ \bibinfo {pages} {9530} (\bibinfo
  {year} {2013})}\BibitemShut {NoStop}%
\bibitem [{\citenamefont {Dodds}(1975)}]{dodds_75}%
  \BibitemOpen
  \bibfield  {author} {\bibinfo {author} {\bibfnamefont {J.}~\bibnamefont
  {Dodds}},\ }\href@noop {} {\bibfield  {journal} {\bibinfo  {journal}
  {Nature}\ }\textbf {\bibinfo {volume} {256}},\ \bibinfo {pages} {187}
  (\bibinfo {year} {1975})}\BibitemShut {NoStop}%
\bibitem [{\citenamefont {Statt}\ \emph {et~al.}(2016)\citenamefont {Statt},
  \citenamefont {Pinchaipat}, \citenamefont {Turci}, \citenamefont {Evans},\
  and\ \citenamefont {Royall}}]{statt_16}%
  \BibitemOpen
  \bibfield  {author} {\bibinfo {author} {\bibfnamefont {A.}~\bibnamefont
  {Statt}}, \bibinfo {author} {\bibfnamefont {R.}~\bibnamefont {Pinchaipat}},
  \bibinfo {author} {\bibfnamefont {F.}~\bibnamefont {Turci}}, \bibinfo
  {author} {\bibfnamefont {R.}~\bibnamefont {Evans}}, \ and\ \bibinfo {author}
  {\bibfnamefont {C.~P.}\ \bibnamefont {Royall}},\ }\href@noop {} {\bibfield
  {journal} {\bibinfo  {journal} {J. Chem. Phys.}\ }\textbf {\bibinfo {volume}
  {144}},\ \bibinfo {pages} {144506} (\bibinfo {year} {2016})}\BibitemShut
  {NoStop}%
\bibitem [{\citenamefont {Danisch}\ \emph {et~al.}(2010)\citenamefont
  {Danisch}, \citenamefont {Jin},\ and\ \citenamefont {Makse}}]{danisch_10}%
  \BibitemOpen
  \bibfield  {author} {\bibinfo {author} {\bibfnamefont {M.}~\bibnamefont
  {Danisch}}, \bibinfo {author} {\bibfnamefont {Y.}~\bibnamefont {Jin}}, \ and\
  \bibinfo {author} {\bibfnamefont {H.~A.}\ \bibnamefont {Makse}},\ }\href@noop
  {} {\bibfield  {journal} {\bibinfo  {journal} {Phys. Rev. E}\ }\textbf
  {\bibinfo {volume} {81}},\ \bibinfo {pages} {051303} (\bibinfo {year}
  {2010})}\BibitemShut {NoStop}%
\bibitem [{\citenamefont {Biazzo}\ \emph {et~al.}(2009)\citenamefont {Biazzo},
  \citenamefont {Caltagirone}, \citenamefont {Parisi},\ and\ \citenamefont
  {Zamponi}}]{biazzo_09}%
  \BibitemOpen
  \bibfield  {author} {\bibinfo {author} {\bibfnamefont {I.}~\bibnamefont
  {Biazzo}}, \bibinfo {author} {\bibfnamefont {F.}~\bibnamefont {Caltagirone}},
  \bibinfo {author} {\bibfnamefont {G.}~\bibnamefont {Parisi}}, \ and\ \bibinfo
  {author} {\bibfnamefont {F.}~\bibnamefont {Zamponi}},\ }\href@noop {}
  {\bibfield  {journal} {\bibinfo  {journal} {Phys. Rev. Lett.}\ }\textbf
  {\bibinfo {volume} {102}},\ \bibinfo {pages} {195701} (\bibinfo {year}
  {2009})}\BibitemShut {NoStop}%
\bibitem [{\citenamefont {Yuan}\ \emph {et~al.}(2018)\citenamefont {Yuan},
  \citenamefont {Liu}, \citenamefont {Zhuang}, \citenamefont {Jin},\ and\
  \citenamefont {Li}}]{yuan_18}%
  \BibitemOpen
  \bibfield  {author} {\bibinfo {author} {\bibfnamefont {Y.}~\bibnamefont
  {Yuan}}, \bibinfo {author} {\bibfnamefont {L.}~\bibnamefont {Liu}}, \bibinfo
  {author} {\bibfnamefont {Y.}~\bibnamefont {Zhuang}}, \bibinfo {author}
  {\bibfnamefont {W.}~\bibnamefont {Jin}}, \ and\ \bibinfo {author}
  {\bibfnamefont {S.}~\bibnamefont {Li}},\ }\href@noop {} {\bibfield  {journal}
  {\bibinfo  {journal} {Phys. Rev. E}\ }\textbf {\bibinfo {volume} {98}},\
  \bibinfo {pages} {042903} (\bibinfo {year} {2018})}\BibitemShut {NoStop}%
\bibitem [{\citenamefont {Graf}\ \emph {et~al.}(2003)\citenamefont {Graf},
  \citenamefont {Vossen}, \citenamefont {Imhof},\ and\ \citenamefont {van
  Blaaderen}}]{graf_03}%
  \BibitemOpen
  \bibfield  {author} {\bibinfo {author} {\bibfnamefont {C.}~\bibnamefont
  {Graf}}, \bibinfo {author} {\bibfnamefont {D.~L.}\ \bibnamefont {Vossen}},
  \bibinfo {author} {\bibfnamefont {A.}~\bibnamefont {Imhof}}, \ and\ \bibinfo
  {author} {\bibfnamefont {A.}~\bibnamefont {van Blaaderen}},\ }\href@noop {}
  {\bibfield  {journal} {\bibinfo  {journal} {Langmuir}\ }\textbf {\bibinfo
  {volume} {19}},\ \bibinfo {pages} {6693} (\bibinfo {year}
  {2003})}\BibitemShut {NoStop}%
\bibitem [{\citenamefont {Desmond}\ and\ \citenamefont
  {Weeks}(2014)}]{desmond_14}%
  \BibitemOpen
  \bibfield  {author} {\bibinfo {author} {\bibfnamefont {K.~W.}\ \bibnamefont
  {Desmond}}\ and\ \bibinfo {author} {\bibfnamefont {E.~R.}\ \bibnamefont
  {Weeks}},\ }\href@noop {} {\bibfield  {journal} {\bibinfo  {journal} {Phys.
  Rev. E}\ }\textbf {\bibinfo {volume} {90}},\ \bibinfo {pages} {022204}
  (\bibinfo {year} {2014})}\BibitemShut {NoStop}%
\bibitem [{\citenamefont {Williams}\ and\ \citenamefont {van
  Megen}(2001)}]{williams_01}%
  \BibitemOpen
  \bibfield  {author} {\bibinfo {author} {\bibfnamefont {S.}~\bibnamefont
  {Williams}}\ and\ \bibinfo {author} {\bibfnamefont {W.}~\bibnamefont {van
  Megen}},\ }\href@noop {} {\bibfield  {journal} {\bibinfo  {journal} {Phys.
  Rev. E}\ }\textbf {\bibinfo {volume} {64}},\ \bibinfo {pages} {041502}
  (\bibinfo {year} {2001})}\BibitemShut {NoStop}%
\bibitem [{\citenamefont {Imhof}\ and\ \citenamefont {Dhont}(1995)}]{imhof_95}%
  \BibitemOpen
  \bibfield  {author} {\bibinfo {author} {\bibfnamefont {A.}~\bibnamefont
  {Imhof}}\ and\ \bibinfo {author} {\bibfnamefont {J.}~\bibnamefont {Dhont}},\
  }\href@noop {} {\bibfield  {journal} {\bibinfo  {journal} {Phys. Rev. Lett.}\
  }\textbf {\bibinfo {volume} {75}},\ \bibinfo {pages} {1662} (\bibinfo {year}
  {1995})}\BibitemShut {NoStop}%
\bibitem [{\citenamefont {L{\'a}zaro-L{\'a}zaro}\ \emph
  {et~al.}(2019)\citenamefont {L{\'a}zaro-L{\'a}zaro}, \citenamefont
  {Perera-Burgos}, \citenamefont {Laermann}, \citenamefont {Sentjabrskaja},
  \citenamefont {P{\'e}rez-{\'A}ngel}, \citenamefont {Laurati}, \citenamefont
  {Egelhaaf}, \citenamefont {Medina-Noyola}, \citenamefont {Voigtmann},
  \citenamefont {Casta{\~n}eda-Priego} \emph {et~al.}}]{lazaro_19}%
  \BibitemOpen
  \bibfield  {author} {\bibinfo {author} {\bibfnamefont {E.}~\bibnamefont
  {L{\'a}zaro-L{\'a}zaro}}, \bibinfo {author} {\bibfnamefont {J.~A.}\
  \bibnamefont {Perera-Burgos}}, \bibinfo {author} {\bibfnamefont
  {P.}~\bibnamefont {Laermann}}, \bibinfo {author} {\bibfnamefont
  {T.}~\bibnamefont {Sentjabrskaja}}, \bibinfo {author} {\bibfnamefont
  {G.}~\bibnamefont {P{\'e}rez-{\'A}ngel}}, \bibinfo {author} {\bibfnamefont
  {M.}~\bibnamefont {Laurati}}, \bibinfo {author} {\bibfnamefont {S.~U.}\
  \bibnamefont {Egelhaaf}}, \bibinfo {author} {\bibfnamefont {M.}~\bibnamefont
  {Medina-Noyola}}, \bibinfo {author} {\bibfnamefont {T.}~\bibnamefont
  {Voigtmann}}, \bibinfo {author} {\bibfnamefont {R.}~\bibnamefont
  {Casta{\~n}eda-Priego}},  \emph {et~al.},\ }\href@noop {} {\bibfield
  {journal} {\bibinfo  {journal} {Phys. Rev. E}\ }\textbf {\bibinfo {volume}
  {99}},\ \bibinfo {pages} {042603} (\bibinfo {year} {2019})}\BibitemShut
  {NoStop}%
\bibitem [{\citenamefont {Bosse}\ and\ \citenamefont
  {Kaneko}(1995)}]{bosse_95}%
  \BibitemOpen
  \bibfield  {author} {\bibinfo {author} {\bibfnamefont {J.}~\bibnamefont
  {Bosse}}\ and\ \bibinfo {author} {\bibfnamefont {Y.}~\bibnamefont {Kaneko}},\
  }\href@noop {} {\bibfield  {journal} {\bibinfo  {journal} {Phys. Rev. Lett.}\
  }\textbf {\bibinfo {volume} {74}},\ \bibinfo {pages} {4023} (\bibinfo {year}
  {1995})}\BibitemShut {NoStop}%
\bibitem [{\citenamefont {Wang}\ \emph {et~al.}(2004)\citenamefont {Wang},
  \citenamefont {Dong},\ and\ \citenamefont {Shek}}]{wang_04}%
  \BibitemOpen
  \bibfield  {author} {\bibinfo {author} {\bibfnamefont {W.}~\bibnamefont
  {Wang}}, \bibinfo {author} {\bibfnamefont {C.}~\bibnamefont {Dong}}, \ and\
  \bibinfo {author} {\bibfnamefont {C.}~\bibnamefont {Shek}},\ }\href@noop {}
  {\bibfield  {journal} {\bibinfo  {journal} {Mater. Sci. Eng., R}\ }\textbf
  {\bibinfo {volume} {44}},\ \bibinfo {pages} {45} (\bibinfo {year}
  {2004})}\BibitemShut {NoStop}%
\bibitem [{\citenamefont {Wang}\ \emph {et~al.}(2021)\citenamefont {Wang},
  \citenamefont {Dasgupta}, \citenamefont {van~der Wee}, \citenamefont
  {Zanaga}, \citenamefont {Altantzis}, \citenamefont {Wu}, \citenamefont
  {Coli}, \citenamefont {Murray}, \citenamefont {Bals}, \citenamefont
  {Dijkstra},\ and\ \citenamefont {van Blaaderen}}]{wang_21}%
  \BibitemOpen
  \bibfield  {author} {\bibinfo {author} {\bibfnamefont {D.}~\bibnamefont
  {Wang}}, \bibinfo {author} {\bibfnamefont {T.}~\bibnamefont {Dasgupta}},
  \bibinfo {author} {\bibfnamefont {E.~B.}\ \bibnamefont {van~der Wee}},
  \bibinfo {author} {\bibfnamefont {D.}~\bibnamefont {Zanaga}}, \bibinfo
  {author} {\bibfnamefont {T.}~\bibnamefont {Altantzis}}, \bibinfo {author}
  {\bibfnamefont {Y.}~\bibnamefont {Wu}}, \bibinfo {author} {\bibfnamefont
  {G.~M.}\ \bibnamefont {Coli}}, \bibinfo {author} {\bibfnamefont {C.~B.}\
  \bibnamefont {Murray}}, \bibinfo {author} {\bibfnamefont {S.}~\bibnamefont
  {Bals}}, \bibinfo {author} {\bibfnamefont {M.}~\bibnamefont {Dijkstra}}, \
  and\ \bibinfo {author} {\bibfnamefont {A.}~\bibnamefont {van Blaaderen}},\
  }\href@noop {} {\bibfield  {journal} {\bibinfo  {journal} {Nature Phys.}\
  }\textbf {\bibinfo {volume} {17}},\ \bibinfo {pages} {128} (\bibinfo {year}
  {2021})}\BibitemShut {NoStop}%
\bibitem [{\citenamefont {Seiden}\ and\ \citenamefont
  {Thomas}(2011)}]{seiden_11}%
  \BibitemOpen
  \bibfield  {author} {\bibinfo {author} {\bibfnamefont {G.}~\bibnamefont
  {Seiden}}\ and\ \bibinfo {author} {\bibfnamefont {P.~J.}\ \bibnamefont
  {Thomas}},\ }\href@noop {} {\bibfield  {journal} {\bibinfo  {journal} {Rev.
  Mod. Phys.}\ }\textbf {\bibinfo {volume} {83}},\ \bibinfo {pages} {1323}
  (\bibinfo {year} {2011})}\BibitemShut {NoStop}%
\bibitem [{\citenamefont {Monteux}\ and\ \citenamefont
  {Lequeux}(2011)}]{monteux_11}%
  \BibitemOpen
  \bibfield  {author} {\bibinfo {author} {\bibfnamefont {C.}~\bibnamefont
  {Monteux}}\ and\ \bibinfo {author} {\bibfnamefont {F.}~\bibnamefont
  {Lequeux}},\ }\href@noop {} {\bibfield  {journal} {\bibinfo  {journal}
  {Langmuir}\ }\textbf {\bibinfo {volume} {27}},\ \bibinfo {pages} {2917}
  (\bibinfo {year} {2011})}\BibitemShut {NoStop}%
\bibitem [{\citenamefont {Zerah}\ and\ \citenamefont
  {Hansen}(1986)}]{hansen_86}%
  \BibitemOpen
  \bibfield  {author} {\bibinfo {author} {\bibfnamefont {G.}~\bibnamefont
  {Zerah}}\ and\ \bibinfo {author} {\bibfnamefont {J.-P.}\ \bibnamefont
  {Hansen}},\ }\href@noop {} {\bibfield  {journal} {\bibinfo  {journal} {J.
  Chem. Phys.}\ }\textbf {\bibinfo {volume} {84}},\ \bibinfo {pages} {2336}
  (\bibinfo {year} {1986})}\BibitemShut {NoStop}%
\bibitem [{\citenamefont {Grodon}\ \emph {et~al.}(2004)\citenamefont {Grodon},
  \citenamefont {Dijkstra}, \citenamefont {Evans},\ and\ \citenamefont
  {Roth}}]{grodon_04}%
  \BibitemOpen
  \bibfield  {author} {\bibinfo {author} {\bibfnamefont {C.}~\bibnamefont
  {Grodon}}, \bibinfo {author} {\bibfnamefont {M.}~\bibnamefont {Dijkstra}},
  \bibinfo {author} {\bibfnamefont {R.}~\bibnamefont {Evans}}, \ and\ \bibinfo
  {author} {\bibfnamefont {R.}~\bibnamefont {Roth}},\ }\href@noop {} {\bibfield
   {journal} {\bibinfo  {journal} {J. Chem. Phys.}\ }\textbf {\bibinfo {volume}
  {121}},\ \bibinfo {pages} {7869} (\bibinfo {year} {2004})}\BibitemShut
  {NoStop}%
\bibitem [{\citenamefont {Grodon}\ \emph {et~al.}(2005)\citenamefont {Grodon},
  \citenamefont {Dijkstra}, \citenamefont {Evans},\ and\ \citenamefont
  {Roth}}]{grodon_05}%
  \BibitemOpen
  \bibfield  {author} {\bibinfo {author} {\bibfnamefont {C.}~\bibnamefont
  {Grodon}}, \bibinfo {author} {\bibfnamefont {M.}~\bibnamefont {Dijkstra}},
  \bibinfo {author} {\bibfnamefont {R.}~\bibnamefont {Evans}}, \ and\ \bibinfo
  {author} {\bibfnamefont {R.}~\bibnamefont {Roth}},\ }\href@noop {} {\bibfield
   {journal} {\bibinfo  {journal} {Mol. Phys.}\ }\textbf {\bibinfo {volume}
  {103}},\ \bibinfo {pages} {3009} (\bibinfo {year} {2005})}\BibitemShut
  {NoStop}%
\bibitem [{\citenamefont {Zhang}\ and\ \citenamefont {Kob}(2020)}]{zhang_20}%
  \BibitemOpen
  \bibfield  {author} {\bibinfo {author} {\bibfnamefont {Z.}~\bibnamefont
  {Zhang}}\ and\ \bibinfo {author} {\bibfnamefont {W.}~\bibnamefont {Kob}},\
  }\href@noop {} {\bibfield  {journal} {\bibinfo  {journal} {Proc. Natl. Acad.
  Sci. (U.S.A.)}\ }\textbf {\bibinfo {volume} {117}},\ \bibinfo {pages} {14032}
  (\bibinfo {year} {2020})}\BibitemShut {NoStop}%
\bibitem [{Note1()}]{Note1}%
  \BibitemOpen
  \bibinfo {note} {Since the CT scanner has only a finite spatial resolution,
  the packing density of the system with only big spheres is not quite the same
  as the one for a system with the small spheres. In the SM we discuss how this
  minor effect has been corrected.}\BibitemShut {Stop}%
\bibitem [{\citenamefont {Zhang}\ \emph {et~al.}(2014)\citenamefont {Zhang},
  \citenamefont {Smith}, \citenamefont {Wang}, \citenamefont {Liu},
  \citenamefont {Schroers}, \citenamefont {Shattuck},\ and\ \citenamefont
  {O'Hern}}]{zhang_14}%
  \BibitemOpen
  \bibfield  {author} {\bibinfo {author} {\bibfnamefont {K.}~\bibnamefont
  {Zhang}}, \bibinfo {author} {\bibfnamefont {W.~W.}\ \bibnamefont {Smith}},
  \bibinfo {author} {\bibfnamefont {M.}~\bibnamefont {Wang}}, \bibinfo {author}
  {\bibfnamefont {Y.}~\bibnamefont {Liu}}, \bibinfo {author} {\bibfnamefont
  {J.}~\bibnamefont {Schroers}}, \bibinfo {author} {\bibfnamefont {M.~D.}\
  \bibnamefont {Shattuck}}, \ and\ \bibinfo {author} {\bibfnamefont {C.~S.}\
  \bibnamefont {O'Hern}},\ }\href@noop {} {\bibfield  {journal} {\bibinfo
  {journal} {Phys. Rev. E}\ }\textbf {\bibinfo {volume} {90}},\ \bibinfo
  {pages} {032311} (\bibinfo {year} {2014})}\BibitemShut {NoStop}%
\bibitem [{\citenamefont {Russo}\ \emph {et~al.}(2018)\citenamefont {Russo},
  \citenamefont {Romano},\ and\ \citenamefont {Tanaka}}]{russo_18}%
  \BibitemOpen
  \bibfield  {author} {\bibinfo {author} {\bibfnamefont {J.}~\bibnamefont
  {Russo}}, \bibinfo {author} {\bibfnamefont {F.}~\bibnamefont {Romano}}, \
  and\ \bibinfo {author} {\bibfnamefont {H.}~\bibnamefont {Tanaka}},\
  }\href@noop {} {\bibfield  {journal} {\bibinfo  {journal} {Phys. Rev. X}\
  }\textbf {\bibinfo {volume} {8}},\ \bibinfo {pages} {021040} (\bibinfo {year}
  {2018})}\BibitemShut {NoStop}%
\bibitem [{\citenamefont {Edwards}\ and\ \citenamefont
  {Oakeshott}(1989)}]{edwards_89}%
  \BibitemOpen
  \bibfield  {author} {\bibinfo {author} {\bibfnamefont {S.~F.}\ \bibnamefont
  {Edwards}}\ and\ \bibinfo {author} {\bibfnamefont {R.}~\bibnamefont
  {Oakeshott}},\ }\href@noop {} {\bibfield  {journal} {\bibinfo  {journal}
  {Physica A}\ }\textbf {\bibinfo {volume} {157}},\ \bibinfo {pages} {1080}
  (\bibinfo {year} {1989})}\BibitemShut {NoStop}%
\bibitem [{\citenamefont {Cao}\ \emph {et~al.}(2018)\citenamefont {Cao},
  \citenamefont {Li}, \citenamefont {Kou}, \citenamefont {Xia}, \citenamefont
  {Li}, \citenamefont {Chen}, \citenamefont {Xie}, \citenamefont {Xiao},
  \citenamefont {Kob}, \citenamefont {Hong}, \citenamefont {Zhang},\ and\
  \citenamefont {Wang}}]{cao_18}%
  \BibitemOpen
  \bibfield  {author} {\bibinfo {author} {\bibfnamefont {Y.}~\bibnamefont
  {Cao}}, \bibinfo {author} {\bibfnamefont {J.}~\bibnamefont {Li}}, \bibinfo
  {author} {\bibfnamefont {B.}~\bibnamefont {Kou}}, \bibinfo {author}
  {\bibfnamefont {C.}~\bibnamefont {Xia}}, \bibinfo {author} {\bibfnamefont
  {Z.}~\bibnamefont {Li}}, \bibinfo {author} {\bibfnamefont {R.}~\bibnamefont
  {Chen}}, \bibinfo {author} {\bibfnamefont {H.}~\bibnamefont {Xie}}, \bibinfo
  {author} {\bibfnamefont {T.}~\bibnamefont {Xiao}}, \bibinfo {author}
  {\bibfnamefont {W.}~\bibnamefont {Kob}}, \bibinfo {author} {\bibfnamefont
  {L.}~\bibnamefont {Hong}}, \bibinfo {author} {\bibfnamefont {J.}~\bibnamefont
  {Zhang}}, \ and\ \bibinfo {author} {\bibfnamefont {Y.}~\bibnamefont {Wang}},\
  }\href@noop {} {\bibfield  {journal} {\bibinfo  {journal} {Nat. Commun.}\
  }\textbf {\bibinfo {volume} {9}},\ \bibinfo {pages} {1} (\bibinfo {year}
  {2018})}\BibitemShut {NoStop}%
\bibitem [{\citenamefont {Santos}\ \emph {et~al.}(2020)\citenamefont {Santos},
  \citenamefont {Yuste},\ and\ \citenamefont {L{\'o}pez~de Haro}}]{santos_20}%
  \BibitemOpen
  \bibfield  {author} {\bibinfo {author} {\bibfnamefont {A.}~\bibnamefont
  {Santos}}, \bibinfo {author} {\bibfnamefont {S.~B.}\ \bibnamefont {Yuste}}, \
  and\ \bibinfo {author} {\bibfnamefont {M.}~\bibnamefont {L{\'o}pez~de
  Haro}},\ }\href@noop {} {\bibfield  {journal} {\bibinfo  {journal} {J. Chem.
  Phys.}\ }\textbf {\bibinfo {volume} {153}},\ \bibinfo {pages} {120901}
  (\bibinfo {year} {2020})}\BibitemShut {NoStop}%
\end{thebibliography}

%

\newpage
\clearpage

\renewcommand{\figurename}{Figure}
\renewcommand{\thefigure}{S\arabic{figure}}
\setcounter{figure}{0}

\begin{center}
{\bf {\large Supplementary Material}}
\end{center}

\noindent
{\bf \large Connecting packing efficiency of binary hard sphere systems to their intermediate range structure}\\

Houfei Yuan$^1$, Zhen Zhang$^2$, Walter Kob $^{3,1}$, and Yujie Wang$^1$

$^1$ School of Physics and Astronomy, Shanghai Jiao Tong University, Shanghai 200240, China

$^2$ Center for Alloy Innovation and Design, State Key Laboratory for Mechanical
Behavior of Materials, Xi’an Jiaotong University, Xi’an 710049, China

$^3$ Laboratoire Charles Coulomb,
University of Montpellier and CNRS, F-34095 Montpellier, France\\

In this Supplementary Material we describe the details of the experiment,
the properties of the radial distribution function and the four-point
correlation function, the static structure factor, the coordination
numbers, and the local packing fractions. 

\clearpage

\section{1. System and image processing}
\vspace*{-5mm}
The particles are made of  Acrylonitrile Butadiene Styrene plastic (Ching Cheuk Precision Tooling, Shenzhen. The polydipersity is less than 1.7\%. Note that small particles are slightly more polydisperse than the big ones, making that the packing fraction of the one-component system of $s$-particles is 0.639, about 0.31\% higher than the packing fraction for the $b$-particles. To remove this trivial $x$-dependence we have subtracted from the $\varphi(x)$ curve a linear background, making that the packing fractions at $x=1$ matches the one at $x=0.$ 

\begin{figure}[tbh]
\centering
\includegraphics[width=13cm]{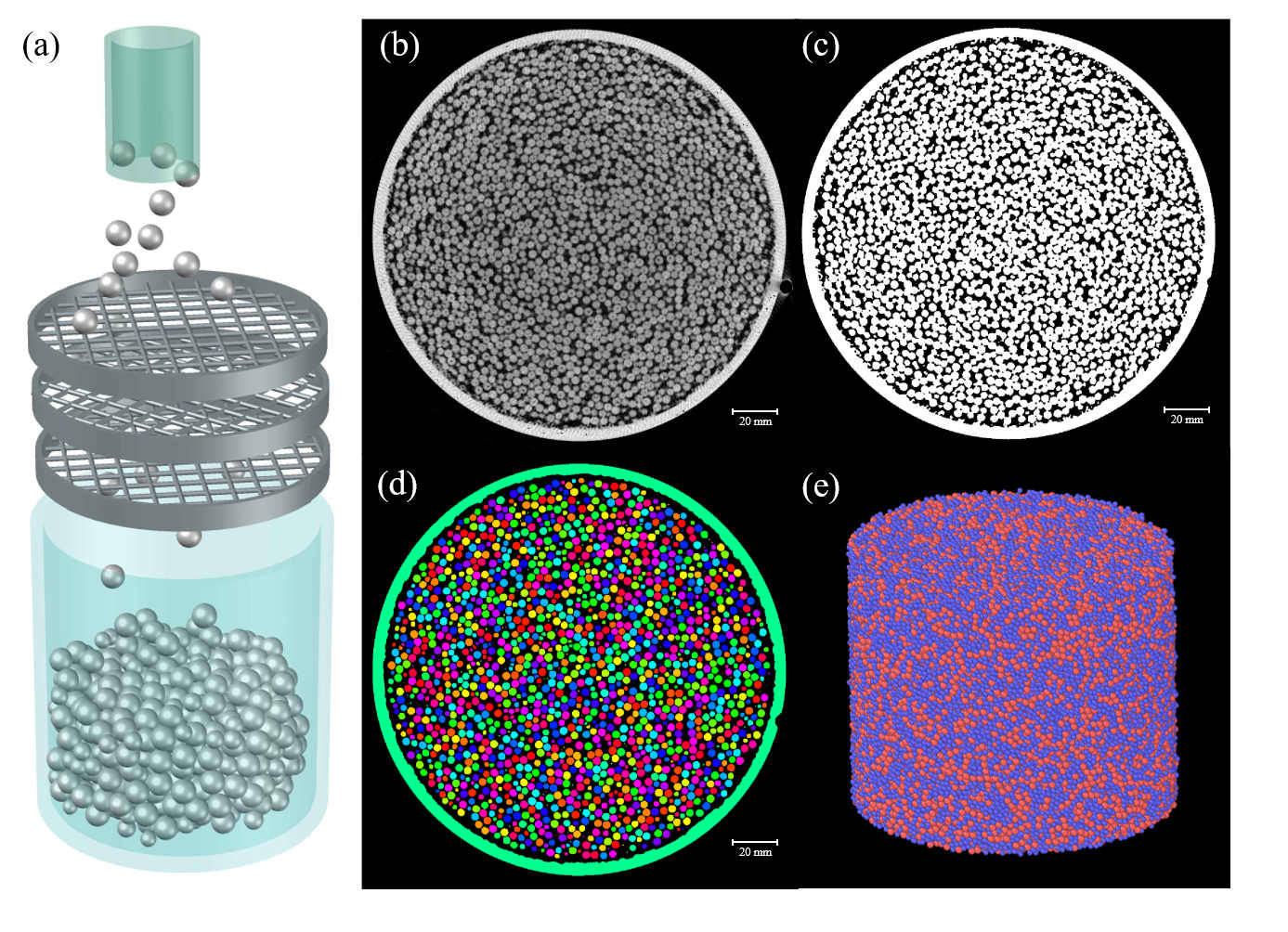}
\caption{(a) Schematic view of the setup for filling the cylinder. (b)-(e) Reconstruction of the structure by image processing: (b) Cross-section (orthogonal to the axis of the cylindrical sample) of the CT image; (c) Binarized image of (b); (d) Marker-based watershed segmented image of (c); (e)  Configuration re-constructed from the data of the 3d coordinates and particles sizes. Red and blue spheres represent the 4~mm and 3~mm particles, respectively. Scale bars are 20~mm.
}
\label{fig_s1}
\end{figure}

\begin{figure}[tbhp]
\centering
\subfigure{
\includegraphics[width=7cm]{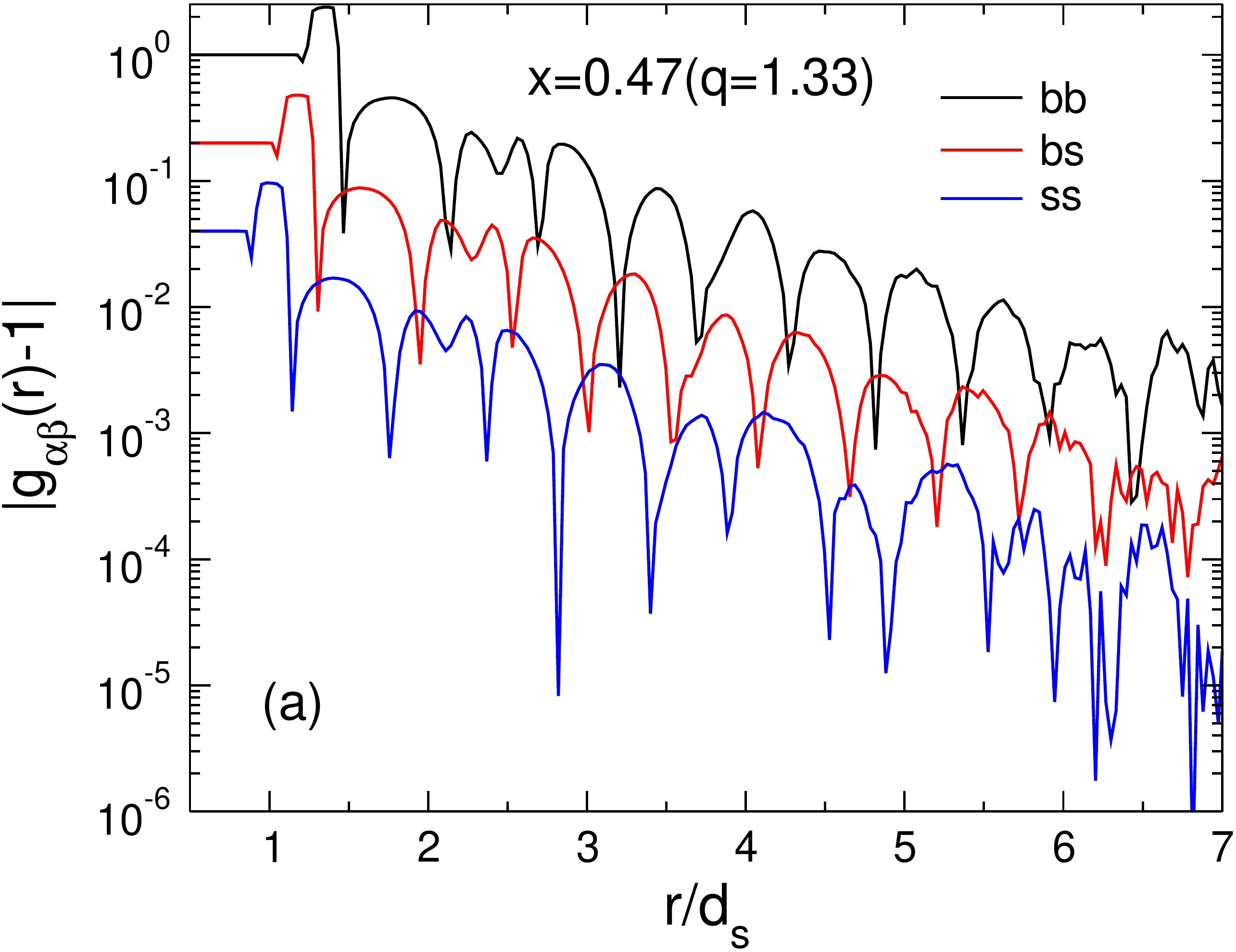}
}
\quad
\subfigure{
\includegraphics[width=7cm]{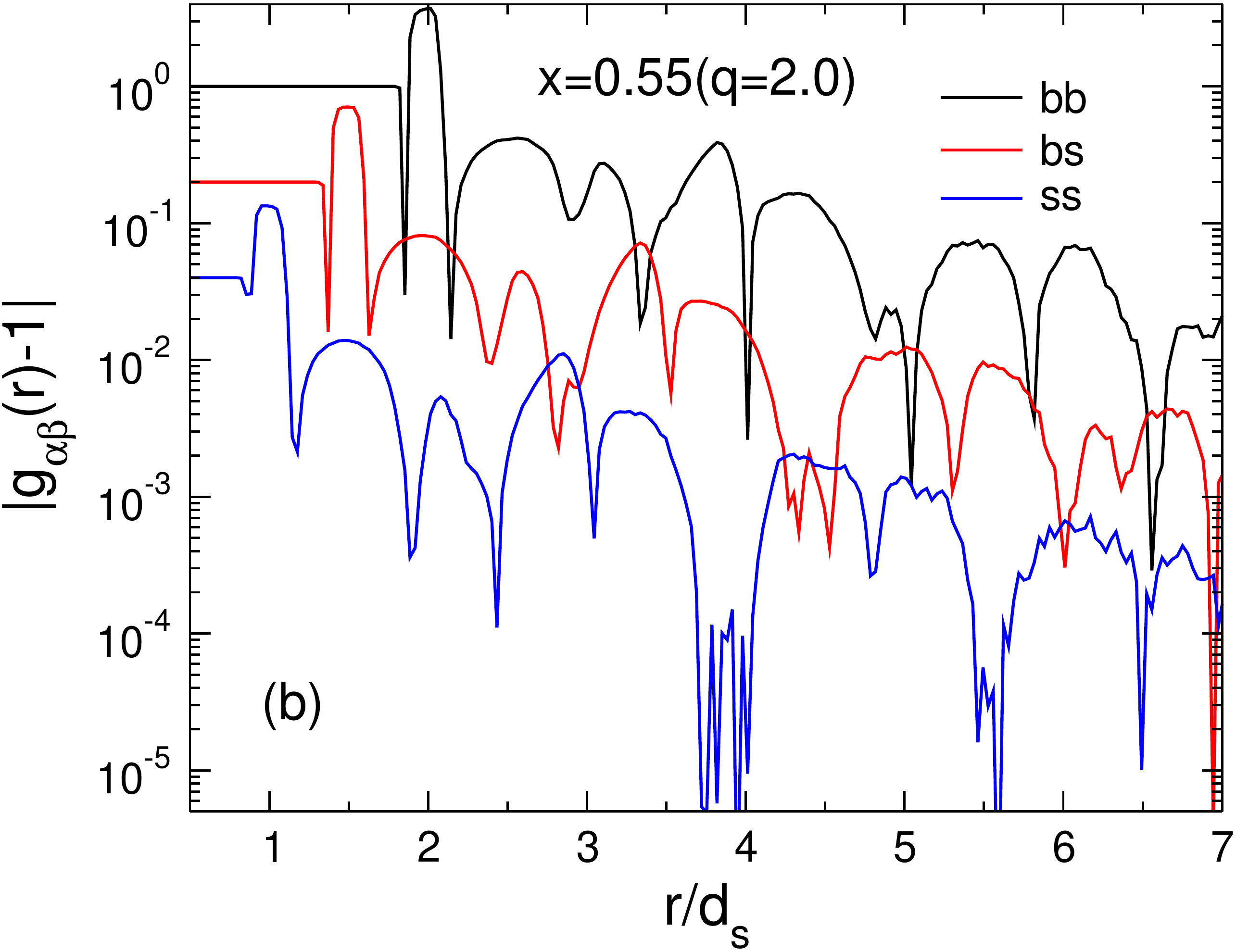}
}
\caption{$|g_{\alpha\beta}(r)-1|$ for the $bb$, $bs$, and $ss$ correlations. (a):  The $q=1.33$ system for $x=0.47$; (b): The $q=2.0$ system for $x=0.55$.
} 
\label{fig_s2}
\end{figure}

To prepare the samples we mount three sieves on top of the cylinder that serves as sample holder, see Fig.~\ref{fig_s1}(a), and then slowly pour the binary mixture of particles on the top sieve. The mesh size is 8~mm and the vertical distance between the sieves is 70~mm. The presence of these sieves makes that the particles are spread out horizontally, thus avoiding the formation of a cone in the middle of the sample. Once all the particles were poured into the cylinder the latter was placed in a CT scanner.

The scanner acquired images of the particles, Fig.~\ref{fig_s1}(b), are analyzed using MATLAB programs as documented in previous studies~\cite{cao_18}. The main steps include the image binarization procedure which results in a discrete representation of the sample, Fig.~\ref{fig_s1}(c). Subsequently we use a marker-based watershed algorithm to identify the centroid and size of all the particles, Fig.~\ref{fig_s1}(d). A snapshot of the so obtained particle representation is shown in Fig.~\ref{fig_s1}(e). Note that while the spatial resolution of the CT scan is 0.2mm, the mentioned algorithms allow to locate the particles to within 0.02~mm, i.e., the positions have an error of less than $10^{-2}d$~\cite{cao_18}.


\section{2. Radial distribution functions}
\vspace*{-5mm}
The structure of many-particles systems is usually characterized by means of the radial distribution function $g(r)$~\cite{hansen_13,binder_11}. In Fig.~\ref{fig_s2} we show the partial radial distribution function defined by~\cite{binder_11}

\begin{equation}
g_{\alpha\beta}(r)=\frac{N_\alpha+N_\beta}{4\pi r^2(\rho_\alpha+\rho_\beta)} 
\sum_{k=1}^{N_\alpha} \sum_{l=1}^{N_\beta}
\delta(r-|\mathbf{r}_k -\mathbf{r}_l|)\quad ,
\label{eq_s1}
\end{equation}

\noindent
where $\rho_\alpha$ is the number density of species $\alpha$.

For the case $q=1.33$, Fig.~\ref{fig_s2}(a), we find that the three partials have basically the same $r$-dependence and that it resembles strongly the one for the radial distribution function of a one-component HS system~\cite{hansen_13,binder_11}. In Fig.~\ref{fig_s2}(b) we show the same quantities for $q=2.0$ and one recognizes that in this case the three partials are very different from each other. Also the $r$-dependence is very complex in that the functions shows various peaks the height of which does not decrease monotonically with increasing $r$. Because of this complexity it is difficult to use these functions to gain insight into the structure of this asymmetric composition, in agreement with previous studies~\cite{statt_16,santos_20}.

\begin{figure}[htbp]
\centering
\subfigure{
\includegraphics[width=7cm]{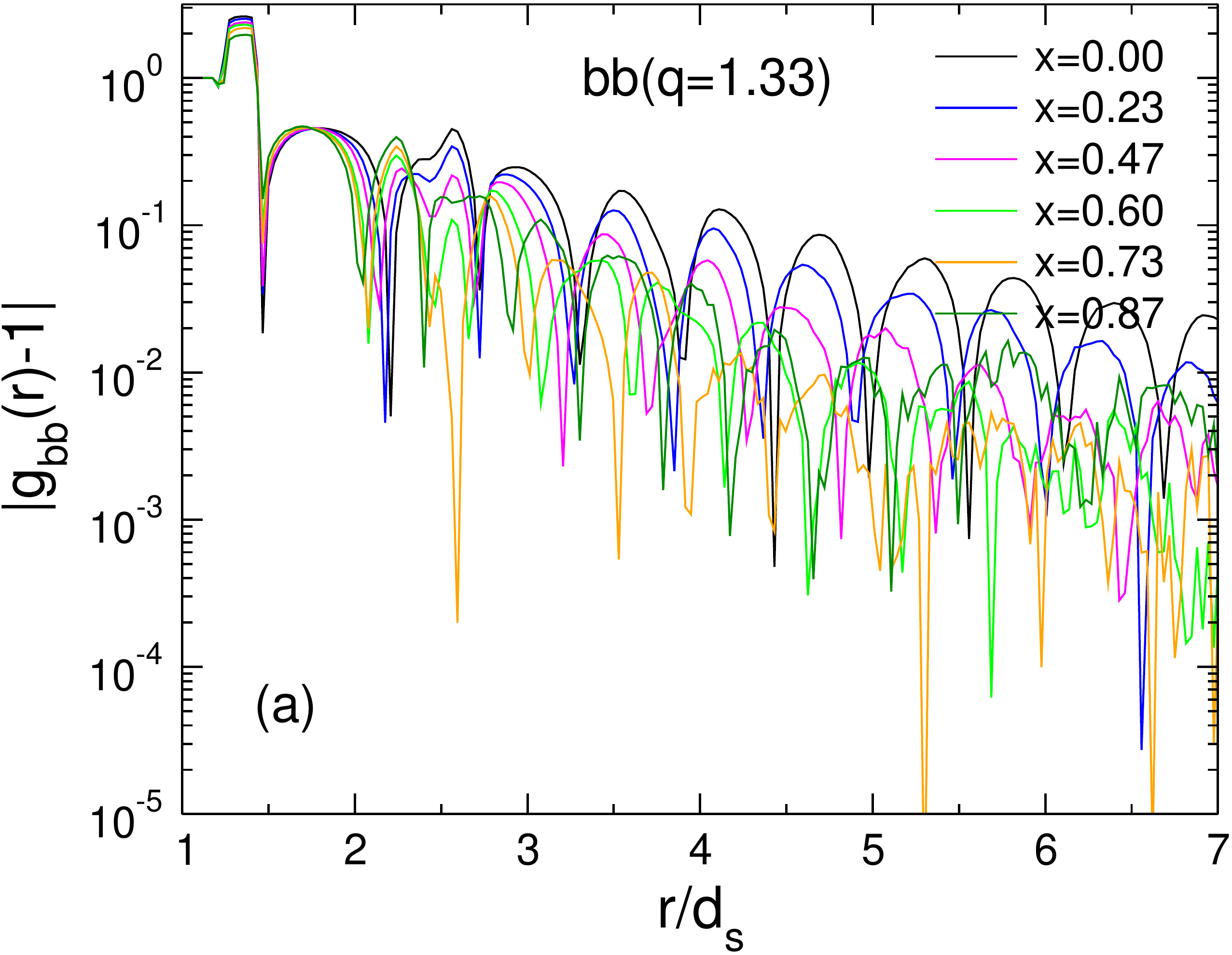}
}
\quad
\subfigure{
\includegraphics[width=7cm]{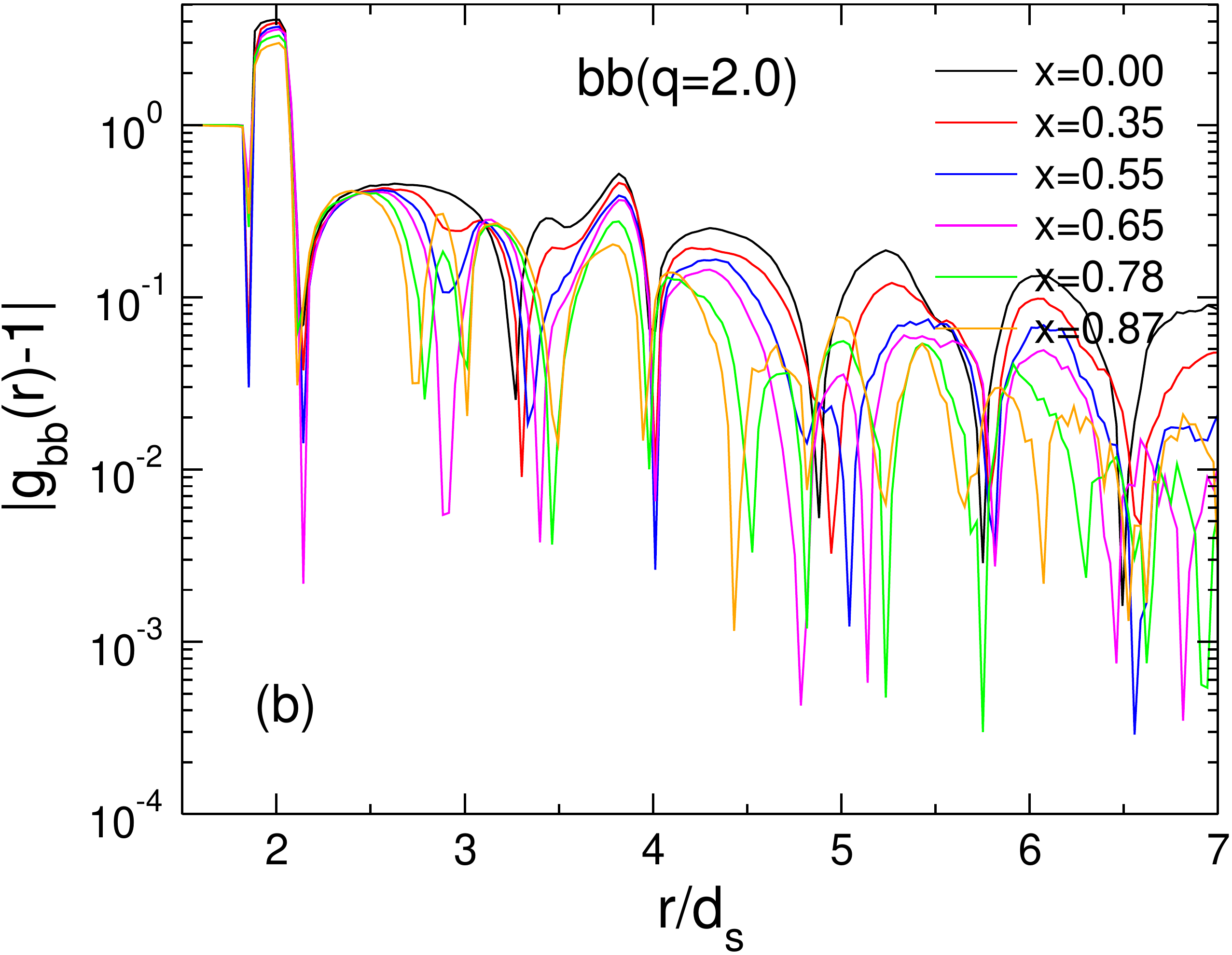}
}
\quad
\subfigure{
\includegraphics[width=7cm]{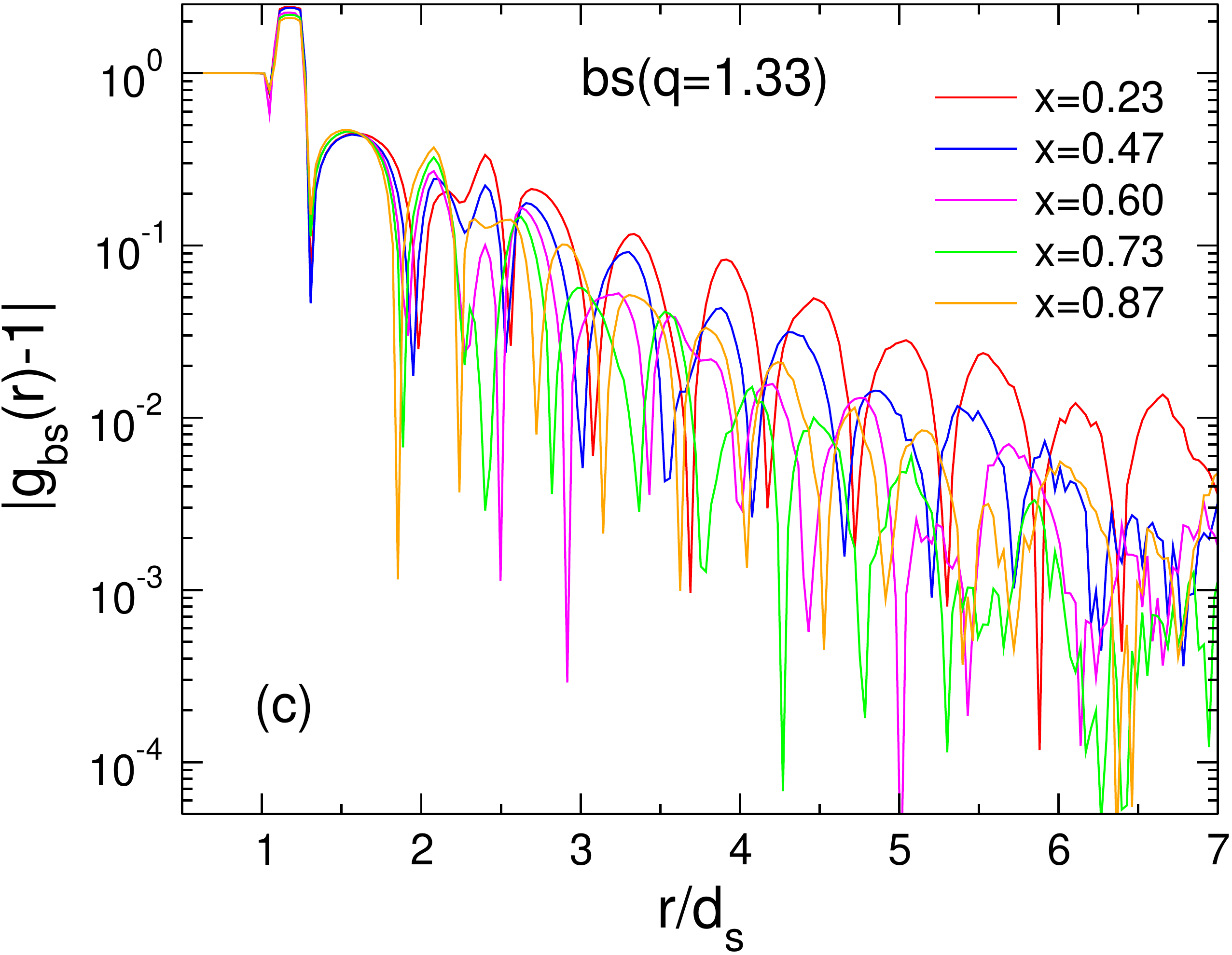}
}
\quad
\subfigure{
\includegraphics[width=7cm]{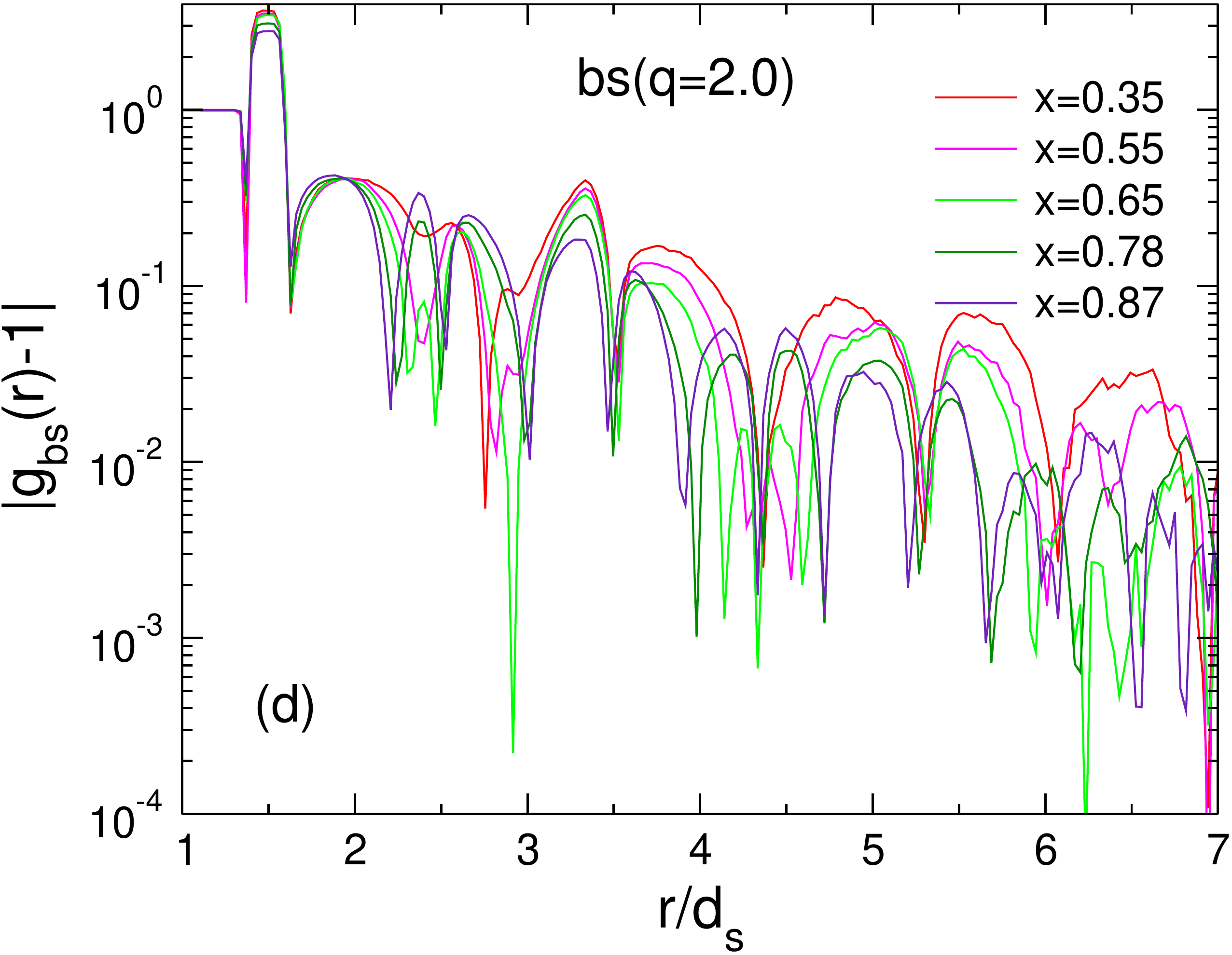}
}
\quad
\subfigure{
\includegraphics[width=7cm]{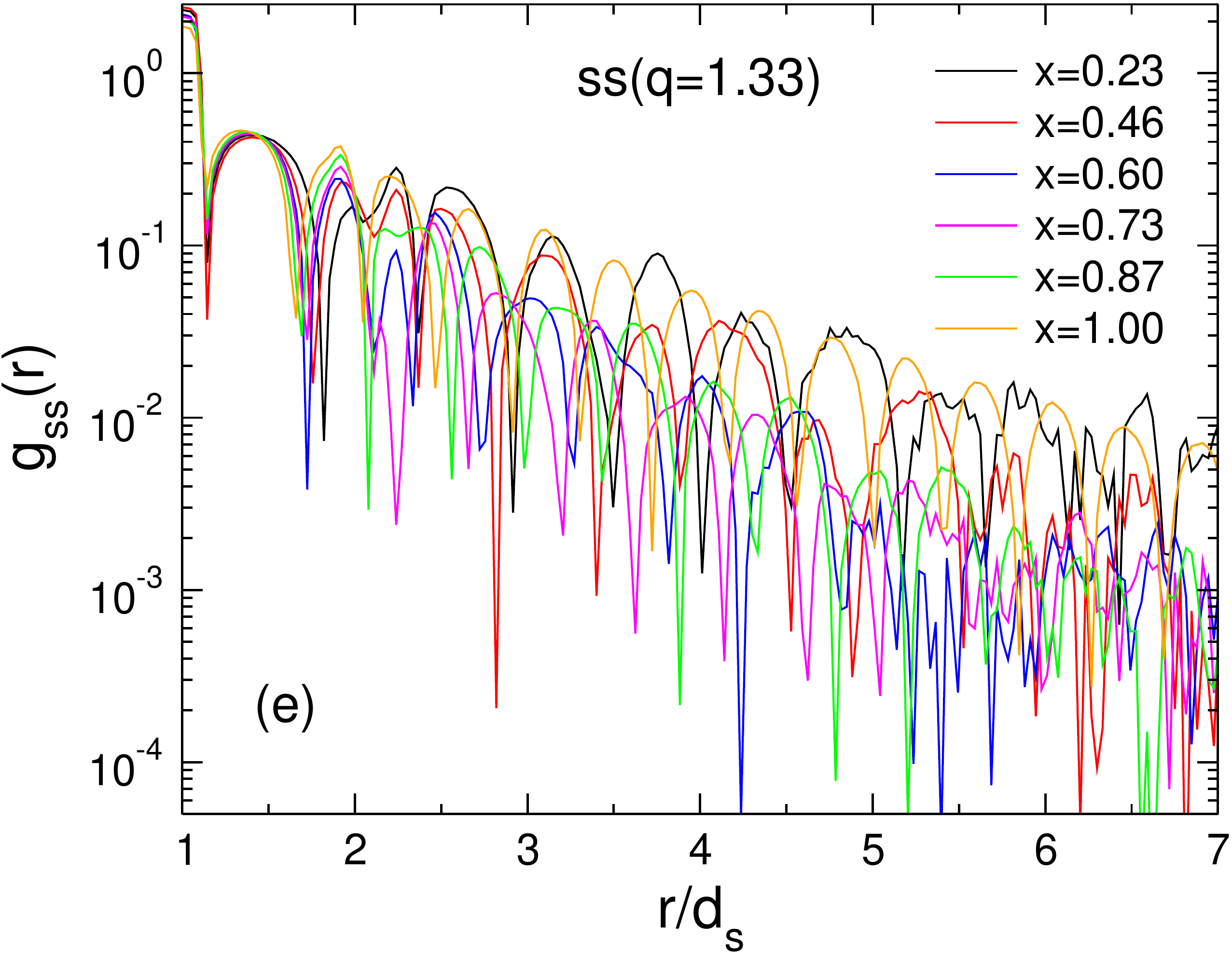}
}
\quad
\subfigure{
\includegraphics[width=7cm]{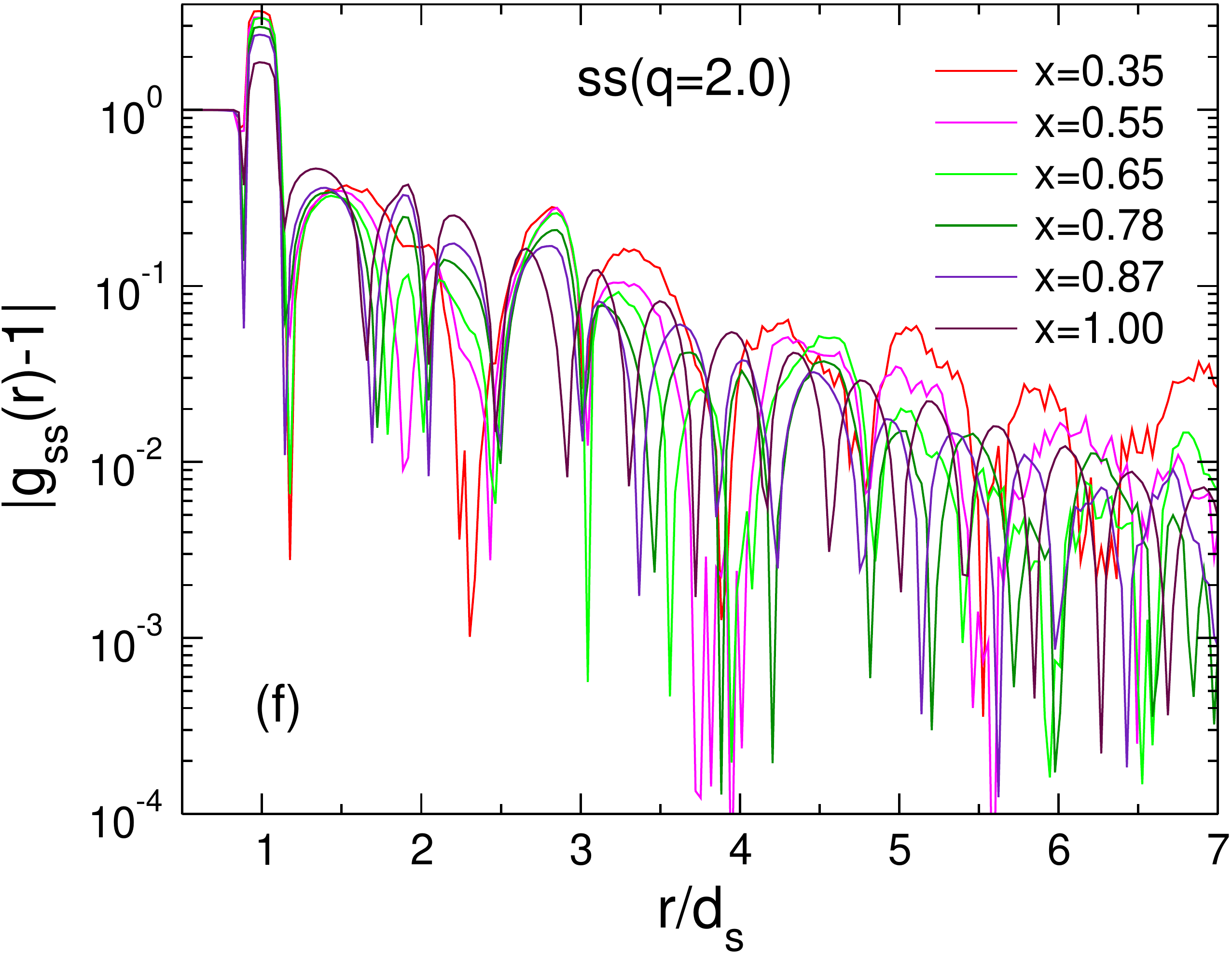}
}
\caption{$|g(r)-1|$ for $bb$, $bs$, and $ss$-correlation of different concentration. (a), (c), (e): $q=1.33$ systems; (b), (d), (f): $q=2.0$ systems.
}
\label{fig_s3}
\end{figure}

The data in Fig.~\ref{fig_s2} are for systems that have roughly an equimolar composition. In Fig.~\ref{fig_s3} we show the three partial radial distribution functions for different compositions. For the case $q=1.33$, Fig.~\ref{fig_s3}(a),(c),(e), the $x$-dependence is, as expected, relatively mild, in that the change of the position and the intensity of the various peaks is smooth. This is in contrast to the case of $q=2.0$, Fig.~\ref{fig_s2}(b),(d),(f), for which one finds that at certain compositions some of the peaks split into two sub-peaks. Hence it is evident that for a system with strongly different particle sizes it is difficult to grasp the structure from a quantity like the radial distribution functions.

Note that this figure demonstrates that all correlation functions show for small values of $r$, i.e., the first few maxima in $|g_{\alpha\beta}(r)-1|$, only a very mild and regular dependence on $x$, and this behavior is found for both values of $q$. Thus this indicates that the local particle arrangement of these packings does not reflect the strong $x$-dependence found for the structure at intermediate and large distances, supporting therefore the view that the large $r$-regime is more useful to understand the relevant structure of these packings.

\clearpage
\section{3. Four point correlation functions}
\vspace*{-5mm}
The expansion coefficients of the density field $\rho_{\alpha\beta}(\theta,\phi,r)$ in Eq.~(\ref{eq_1}) of the main text are given by

\begin{equation}
\rho_{\alpha \beta ,l}^m =\int_0^{2\pi} d\phi \int_0^\pi d\theta \sin \theta
\rho_{\alpha\beta}(\theta,\phi,r) Y_l^{m*}(\theta,\phi) \quad ,
\end{equation}

\noindent
where $Y_l^{m*}$ is the complex conjugate of the spherical harmonic function of degree $l$ and order $m$.

\begin{figure}[htbp]
\centering
\subfigure{
\includegraphics[width=6cm]{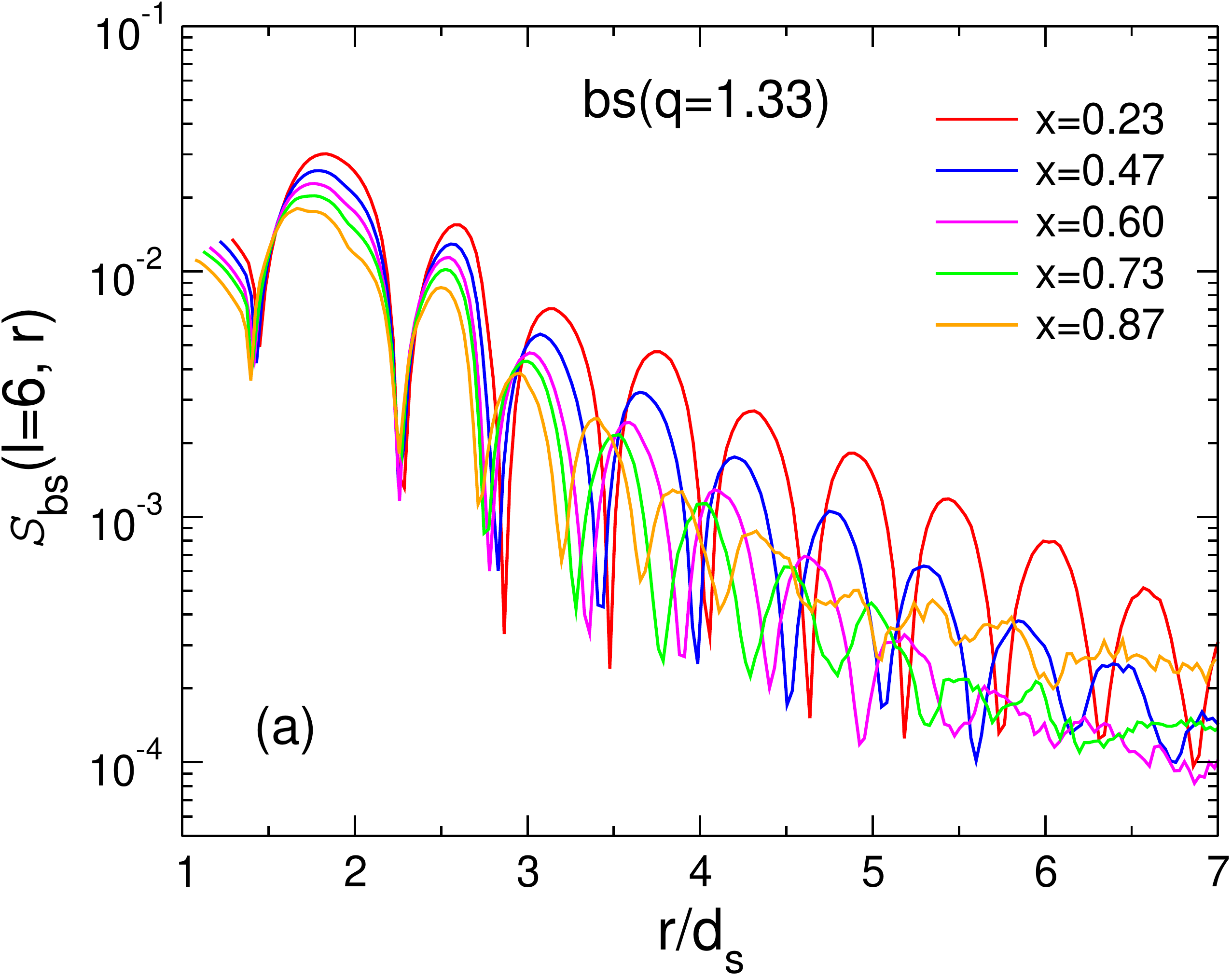}
}
\quad
\subfigure{
\includegraphics[width=6cm]{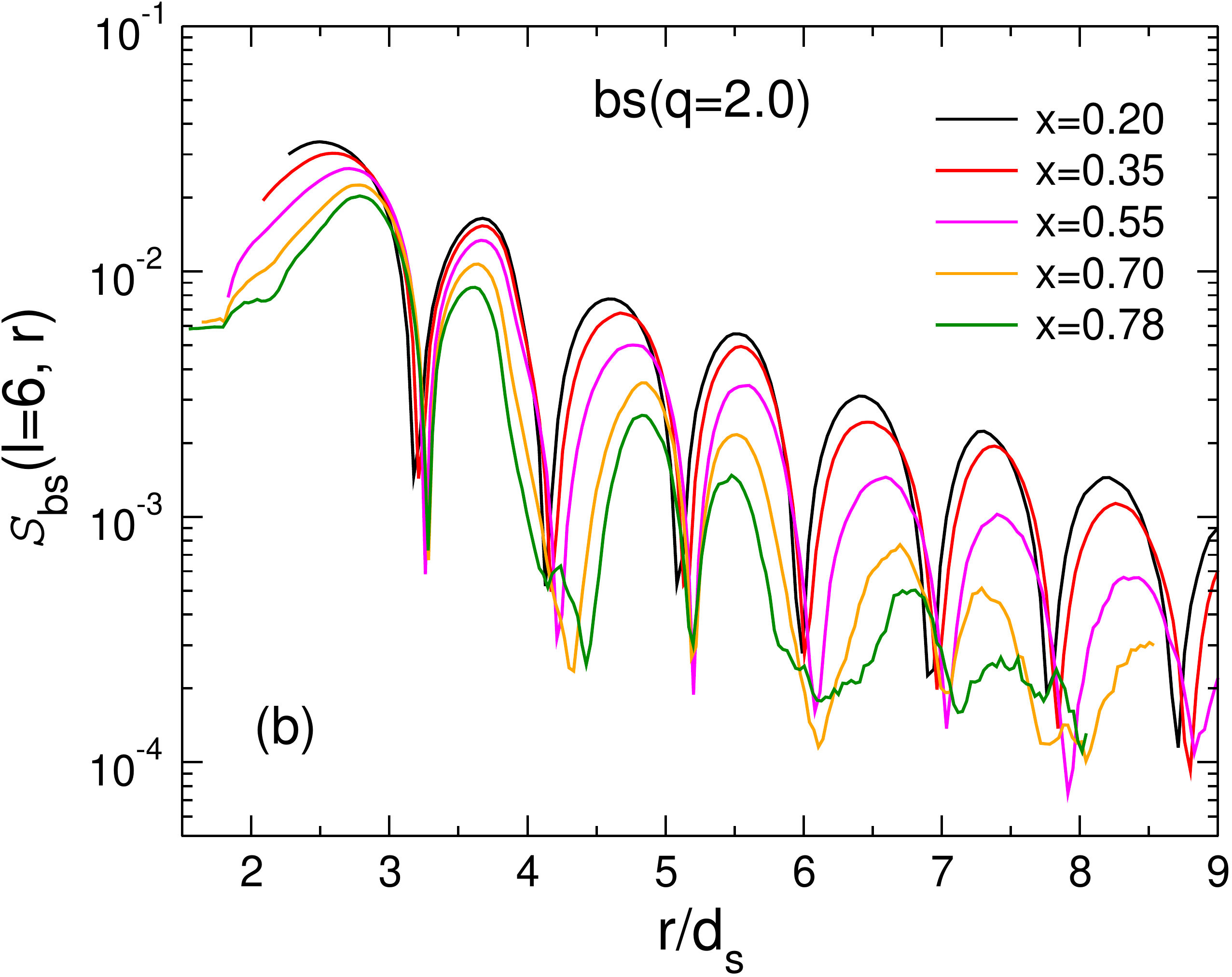}
}
\quad
\subfigure{
\includegraphics[width=6cm]{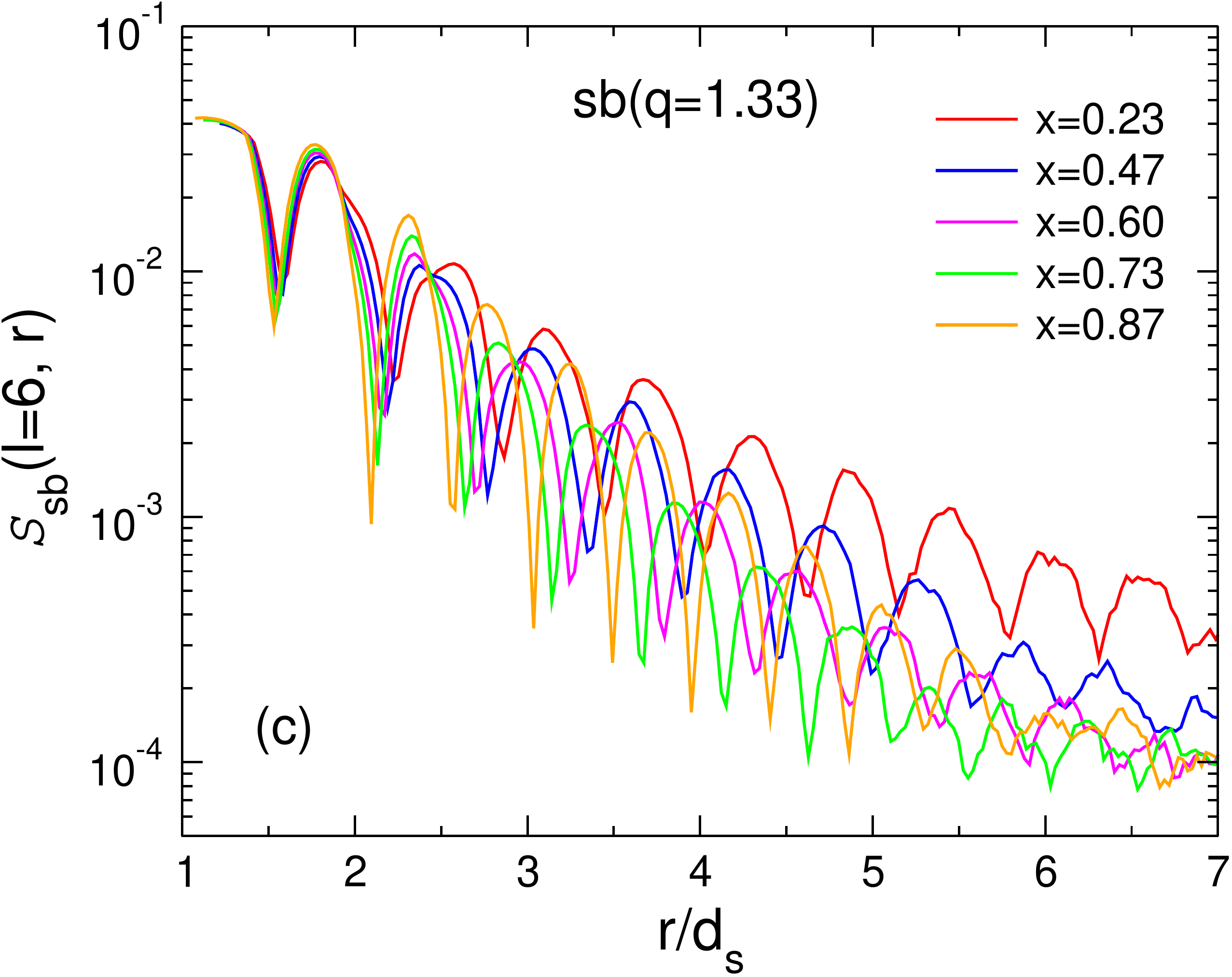}
}
\quad
\subfigure{
\includegraphics[width=6cm]{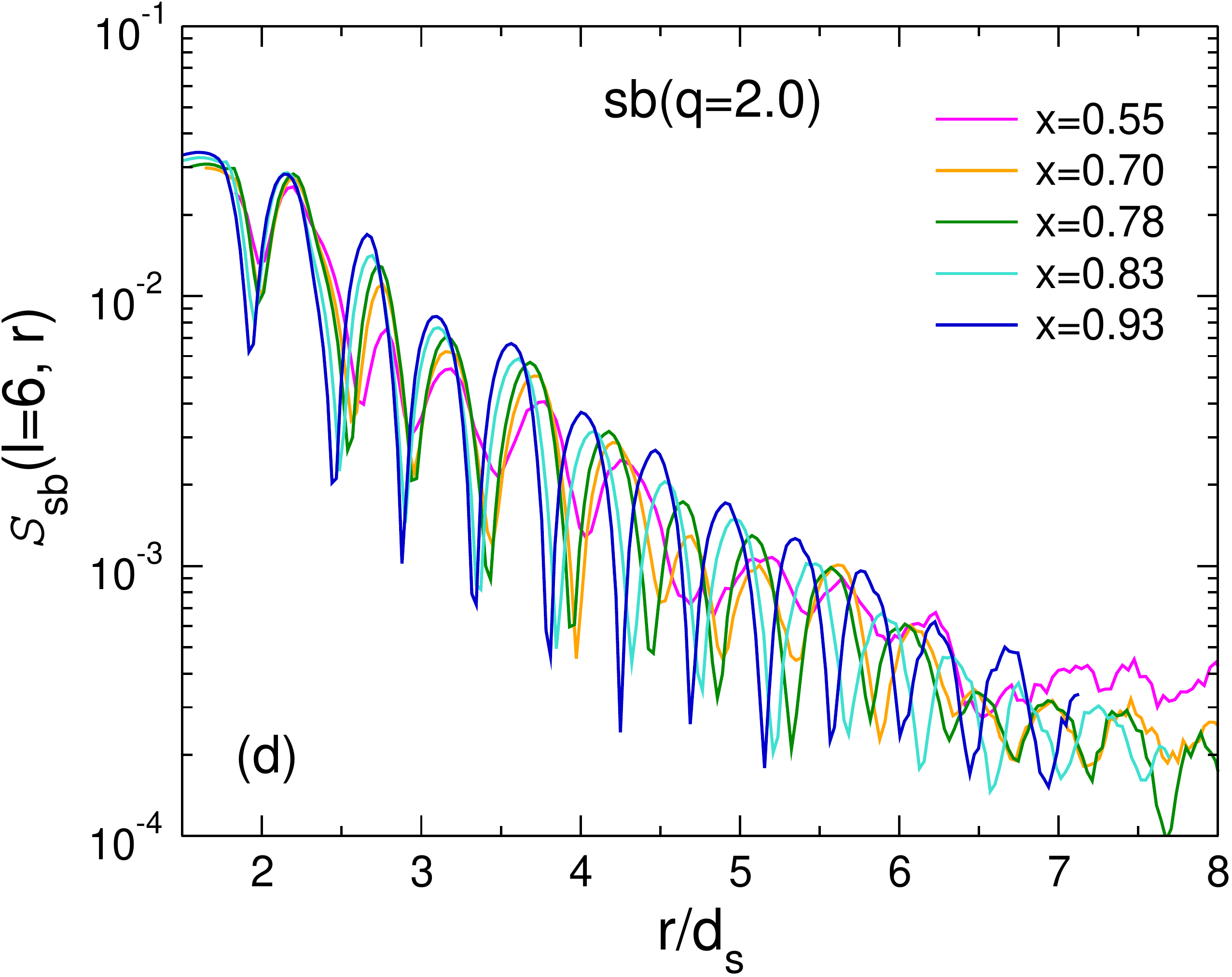}
}
\quad
\subfigure{
\includegraphics[width=6cm]{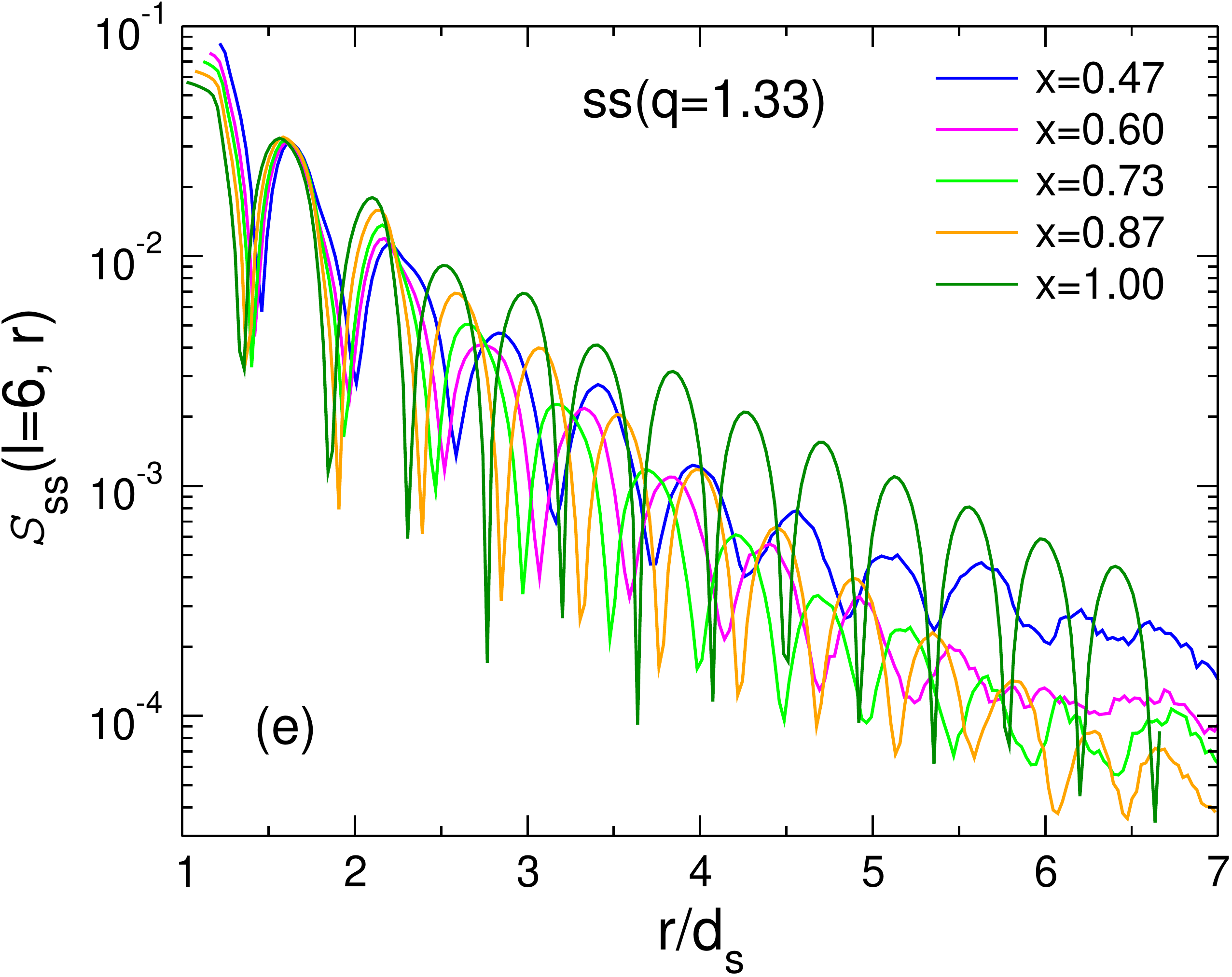}
}
\quad
\subfigure{
\includegraphics[width=6cm]{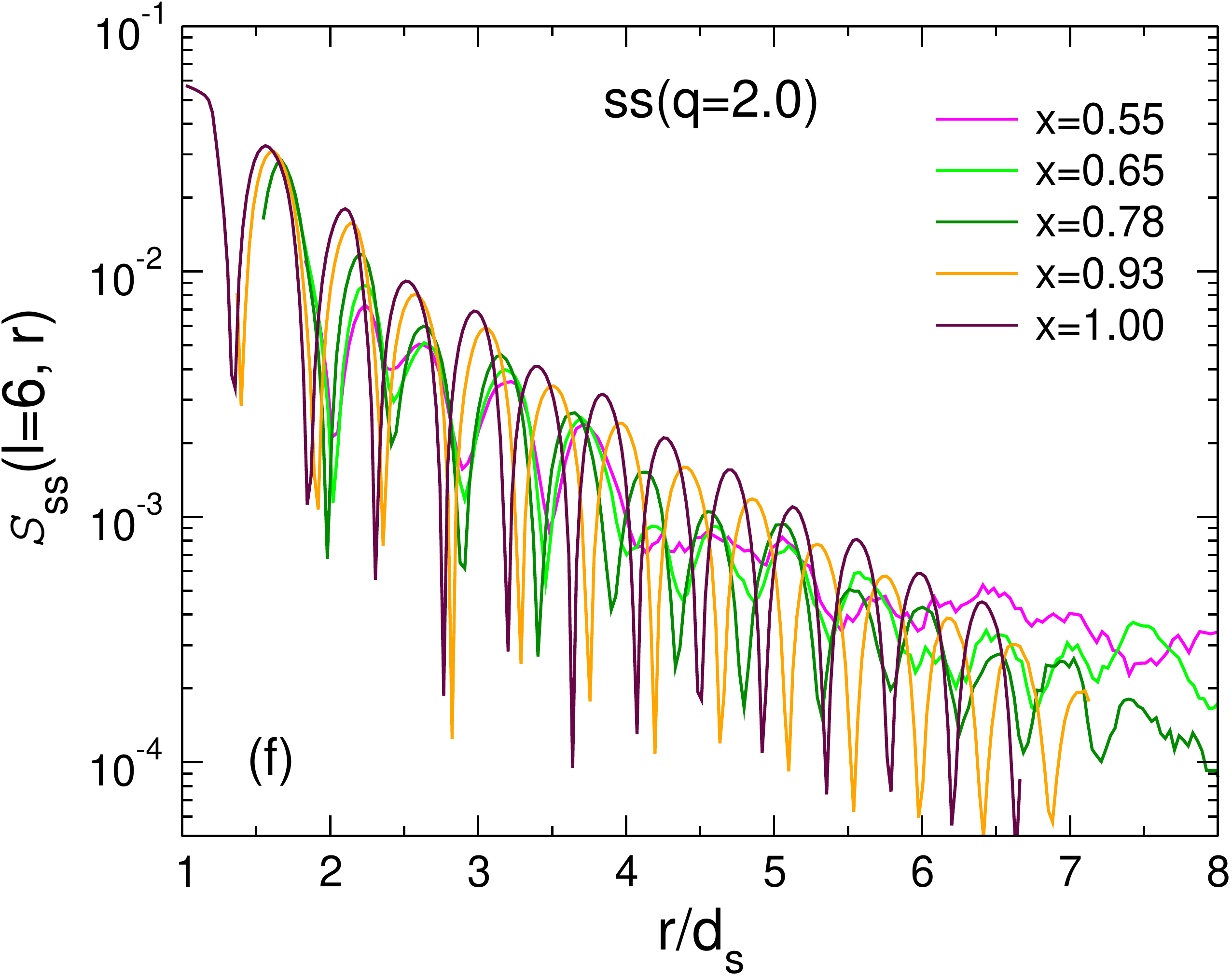}
}
\caption{$\mathcal{S}_{\alpha\beta}(l=6,r)$ for the $bs$, $sb$, and $ss$-correlation at different concentrations $x$. (a), (c), (e): $q=1.33$ systems; (b), (d), (f): $q=2.0$ systems.
}
\label{fig_s4}
\end{figure}

In Fig.~\ref{fig_s4} we show the $r$-dependence of $\mathcal{S}_{\alpha\beta}$, defined in the main text, for different concentrations $x$. (The $x$-dependence for the $bb$-correlation is shown in the main text.) In contrast to the radial distribution functions $g_{\alpha\beta}(r)$, this $x$-dependence is very smooth in that the general structure of the function depends only weakly on the concentration and it is mainly the height of the peaks that shows a dependence on $x$. These results agree thus with the observations we made in the context of Fig.~\ref{fig_2} of the main text.

\begin{figure}[htbp]
\centering
\subfigure{
\includegraphics[width=7cm]{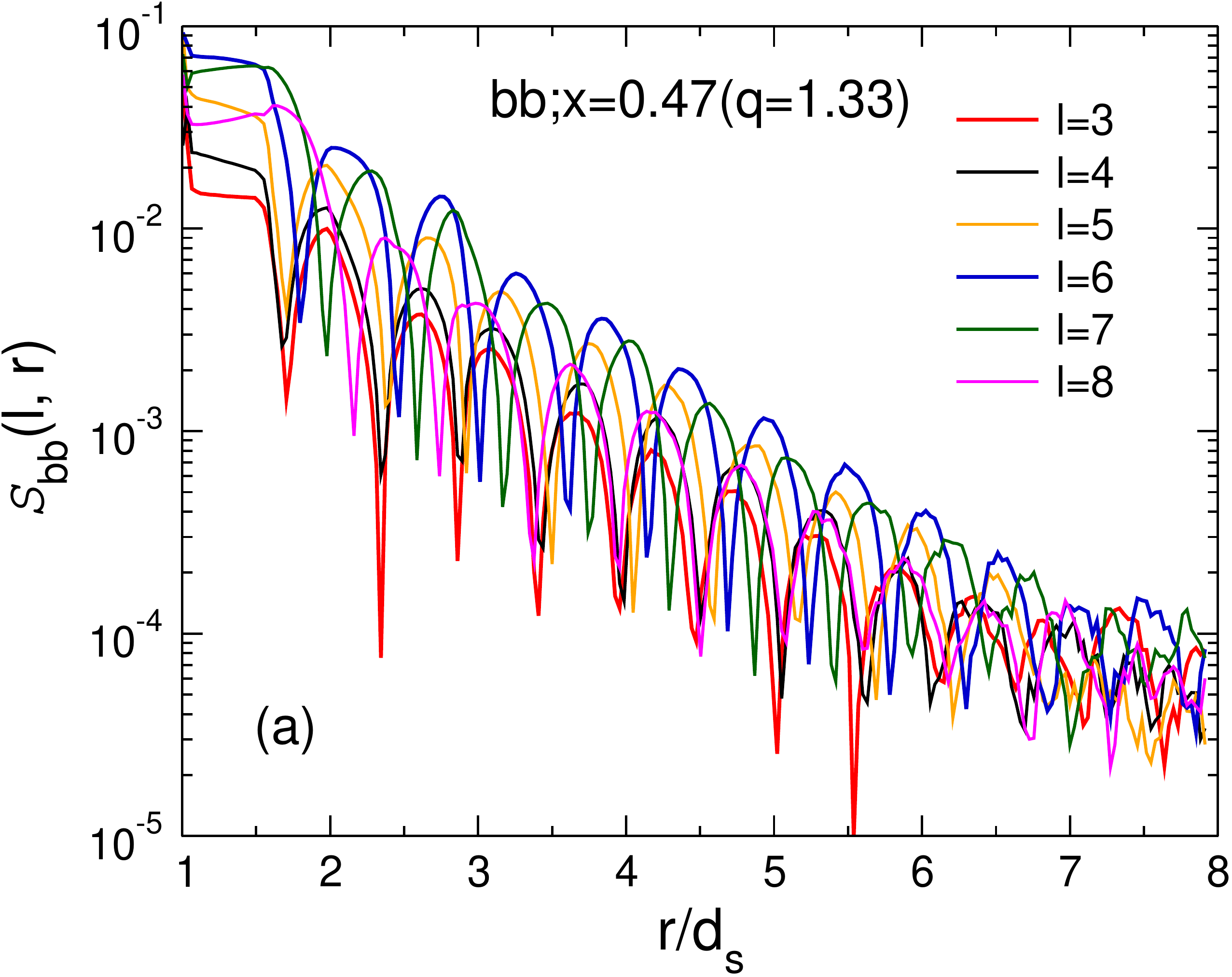}
}
\quad
\subfigure{
\includegraphics[width=7cm]{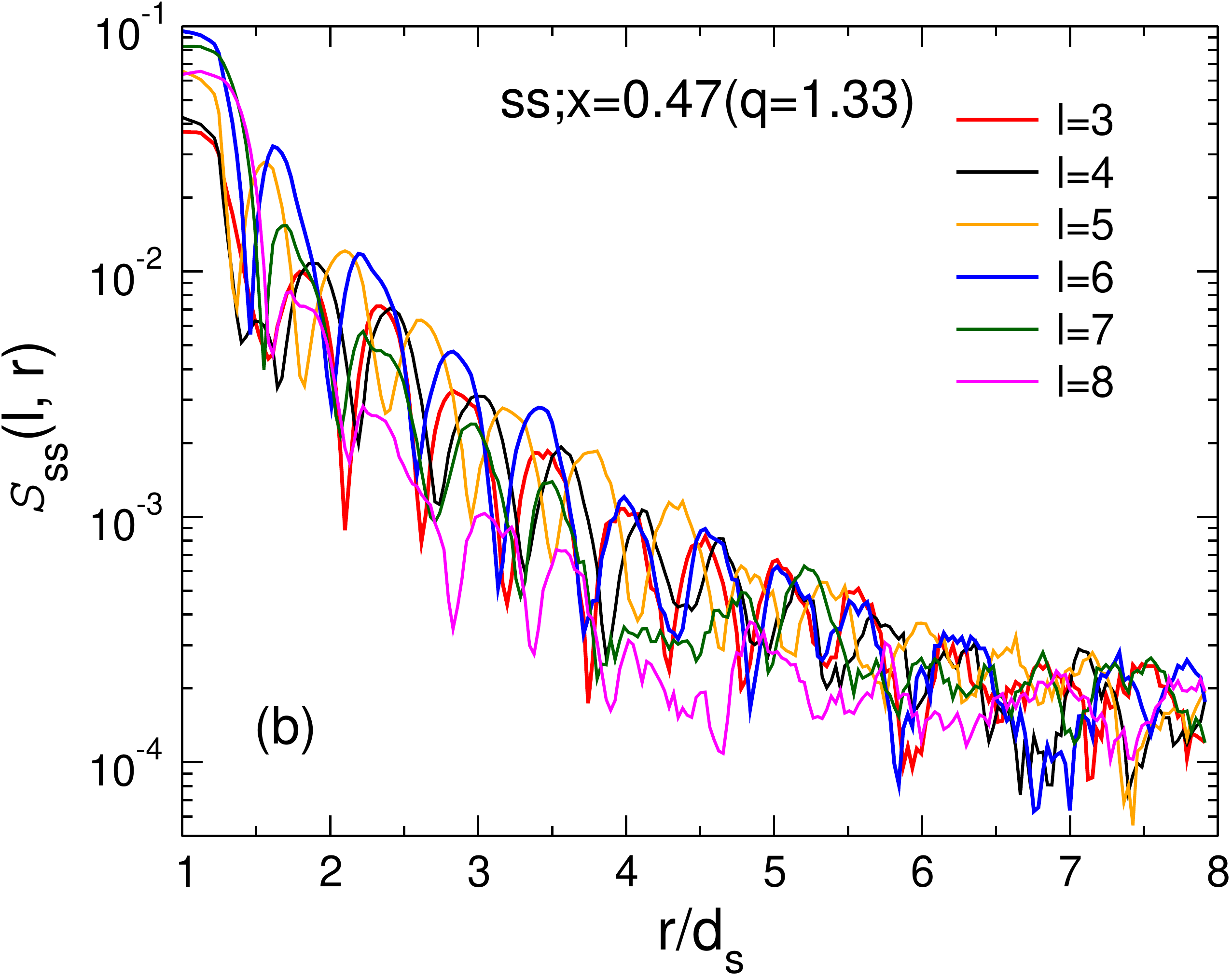}
}
\quad
\subfigure{
\includegraphics[width=7cm]{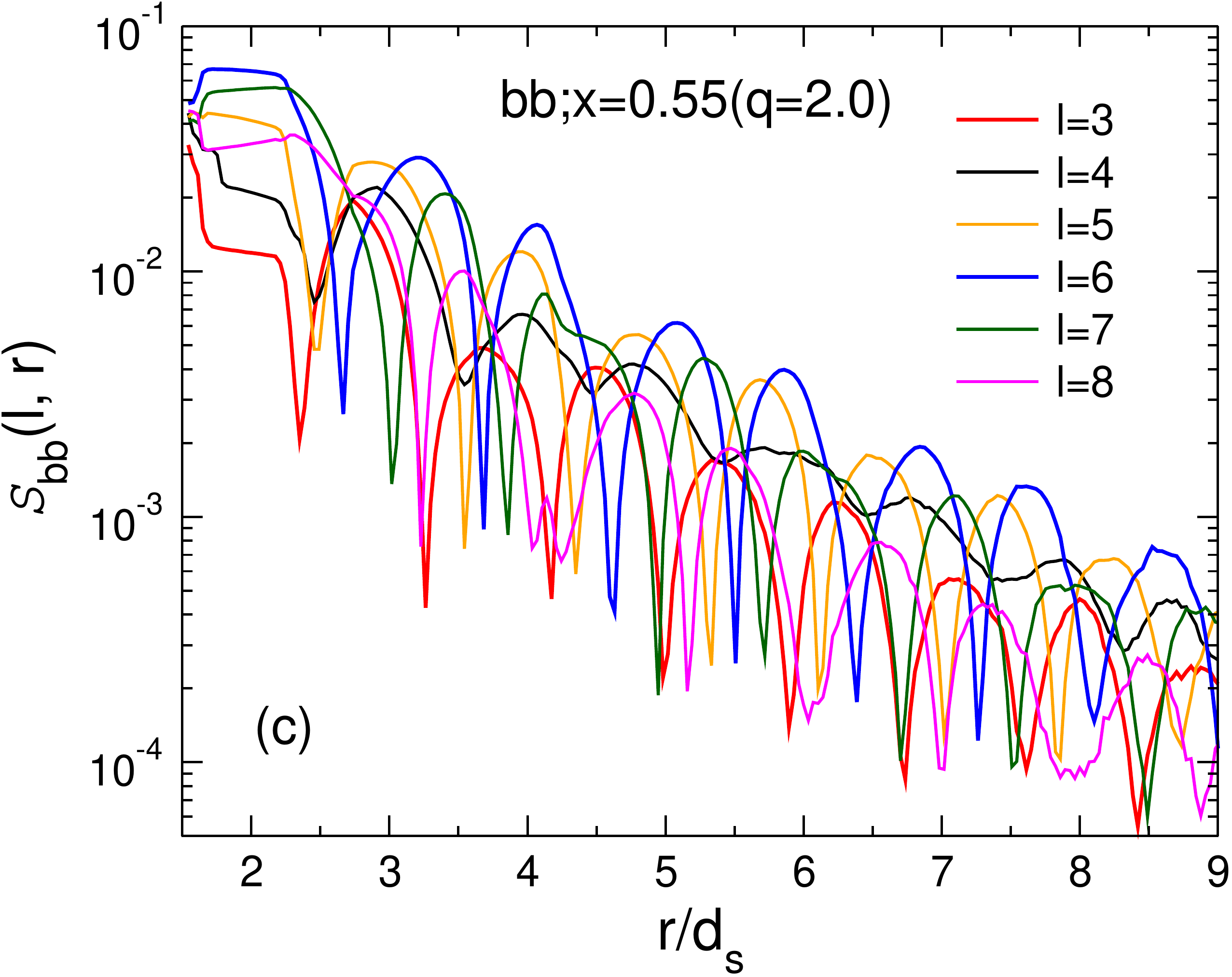}
}
\quad
\subfigure{
\includegraphics[width=7cm]{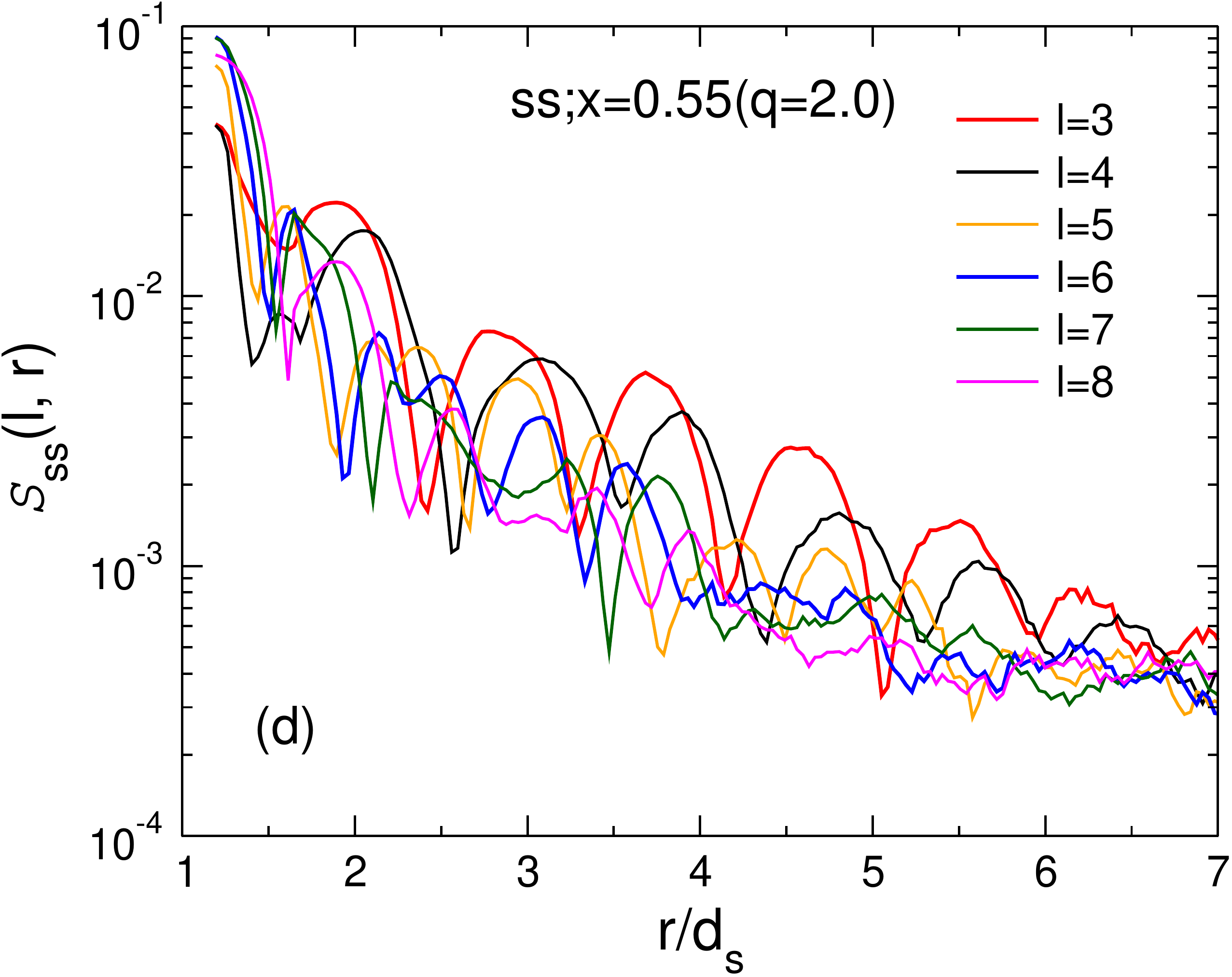}
}
\caption{$\mathcal{S}_{\alpha\beta}(l,r)$ for different values of $l$. (a) and (c) $bb$-correlation for the $q=1.33$ and $q=2.0$ systems, respectively. (b) and (d) $ss$-correlation for the $q=1.33$ and $q=2.0$ systems, respectively.
} 
\label{fig_s5}
\end{figure}

One important parameter for $\mathcal{S}_{\alpha\beta}(l,r)$ is the index $l$ of the considered spherical harmonics. For the $q=1.33$ system one finds that $l=6$ has a very strong signal for all four combinations of $\alpha$ and $\beta$, see Fig.~\ref{fig_s5}(a) and (b). For the case of the asymmetric system, $q=2.0$, the $l$-dependence is more complex: For the $bb$-correlation one has again that $l=6$ has the strongest signal, Fig.~\ref{fig_s5}(c). This result is related to the fact that for a strong size difference the structure is strongly dominated by the one of the big particles which show the icosahedral/dodecahedra sequence found in one-component systems (see Fig.~\ref{fig_1} of the main text), at least if the concentration $x$ is not very close to 1.0. However, if the central particle is a small one, it is the $l=3$ correlator that has the largest signal, see Fig.~\ref{fig_s5}(d). This result is likely related to the fact that three small particles (forming the coordinate system) that are close to a big particle will form a local structure that has many angles close to the tetrahedral angle 109$^{\circ}$, thus giving a strong contribution to $\mathcal{S}_{ss}(3,r)$ and less to $\mathcal{S}_{ss}(6,r)$ (the latter corresponding to 60$^\circ$ angles).

\begin{figure}[htbp]
\centering
\subfigure{
\includegraphics[width=6cm]{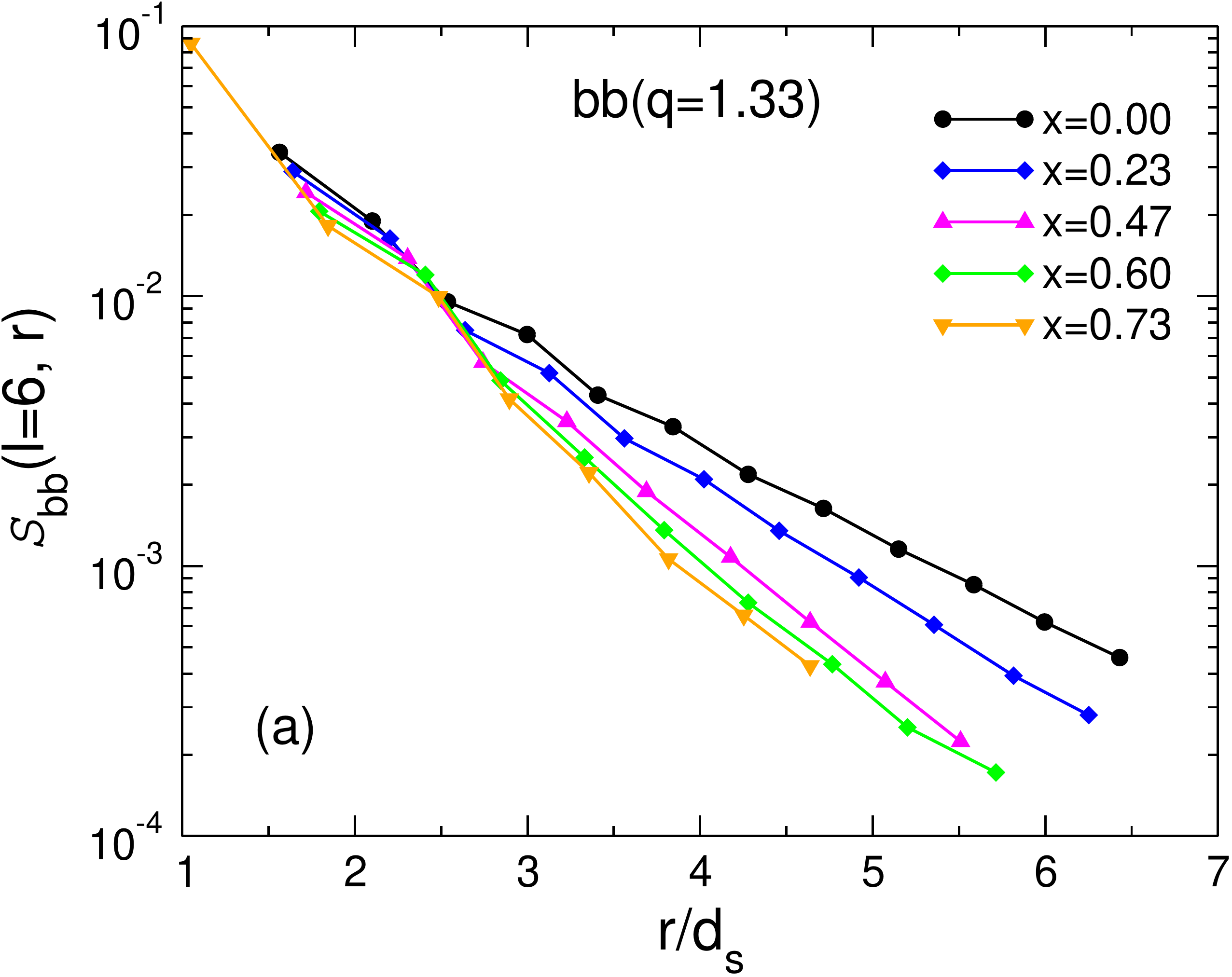}
}
\quad
\subfigure{
\includegraphics[width=6cm]{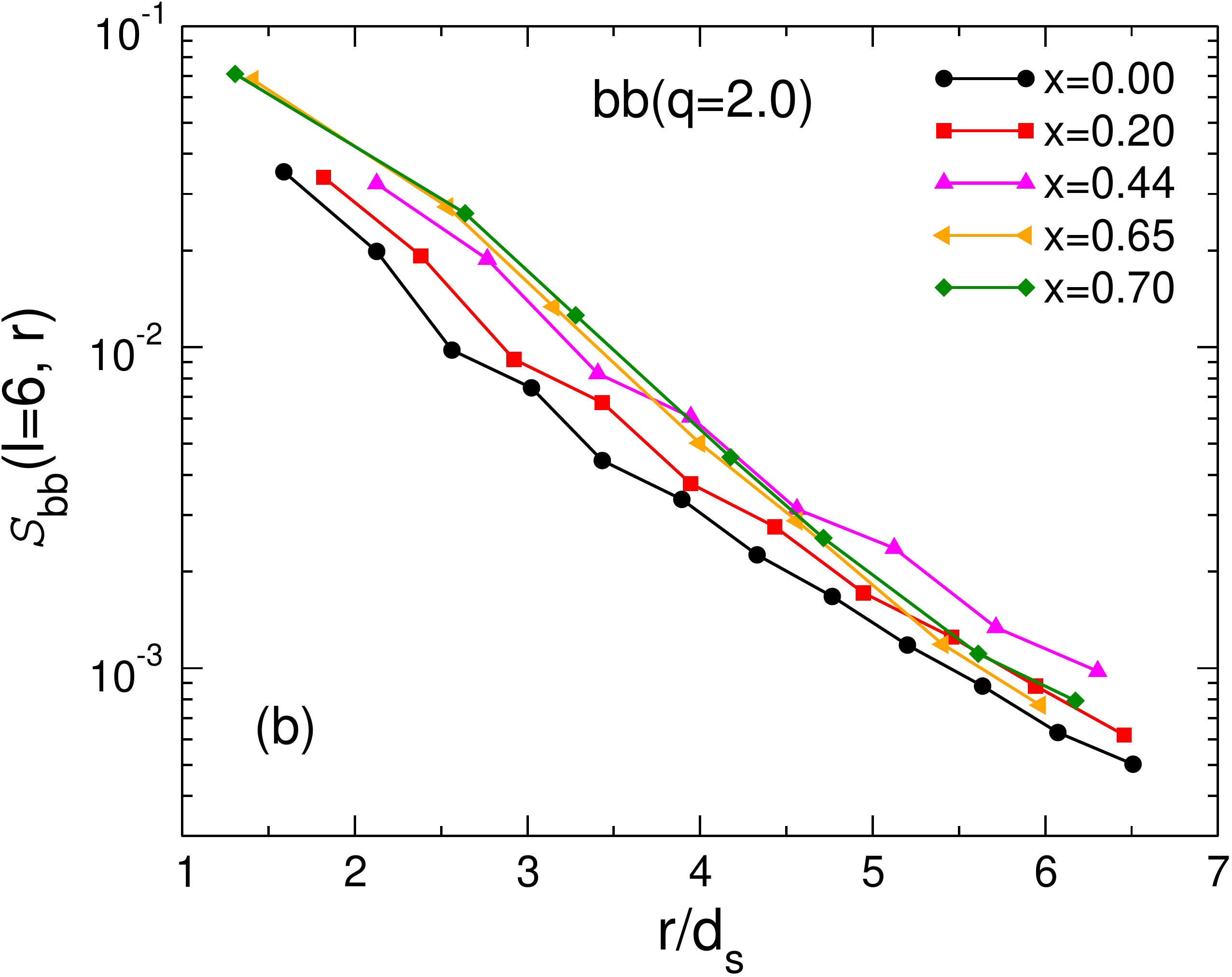}
}
\quad
\subfigure{
\includegraphics[width=6cm]{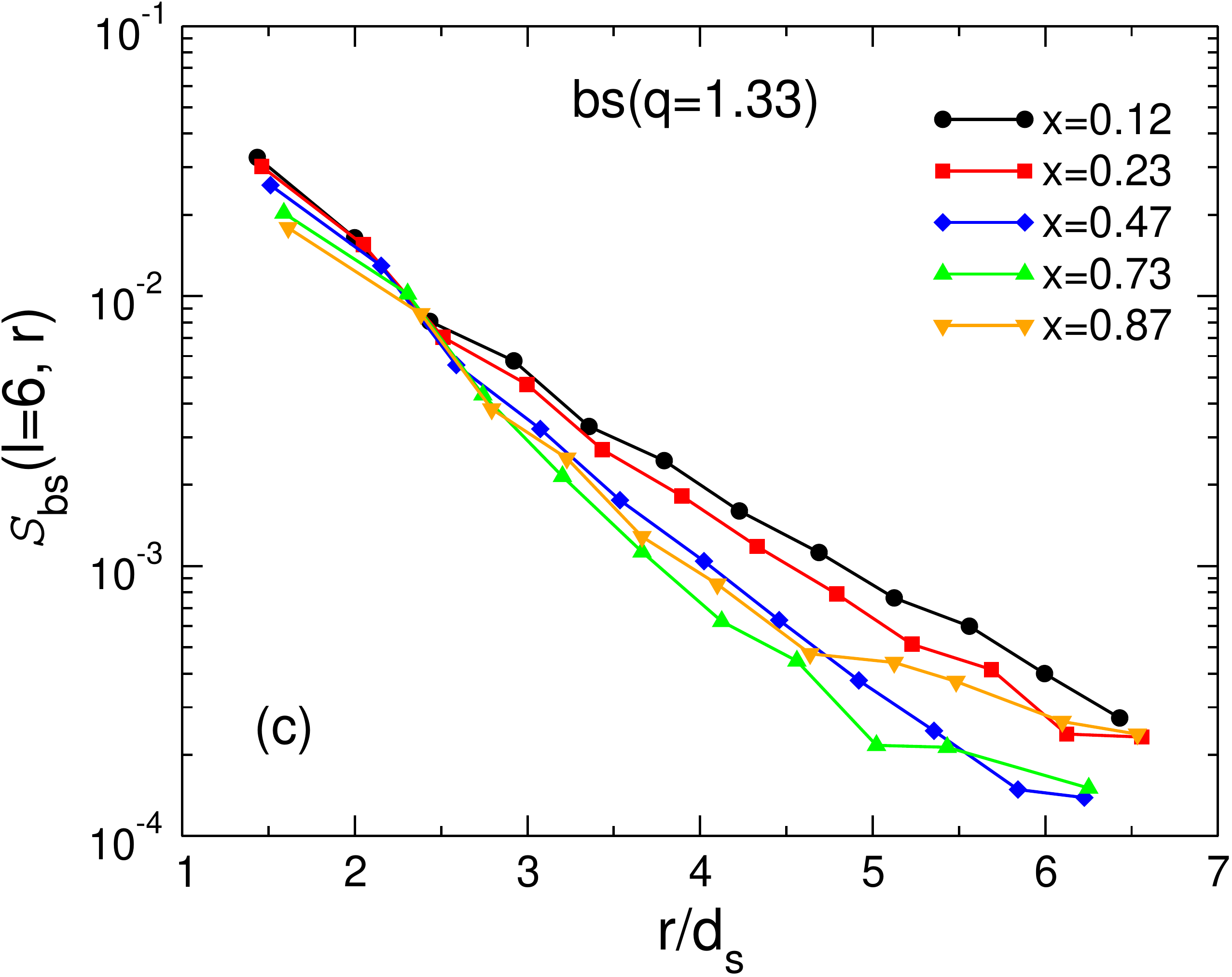}
}
\quad
\subfigure{
\includegraphics[width=6cm]{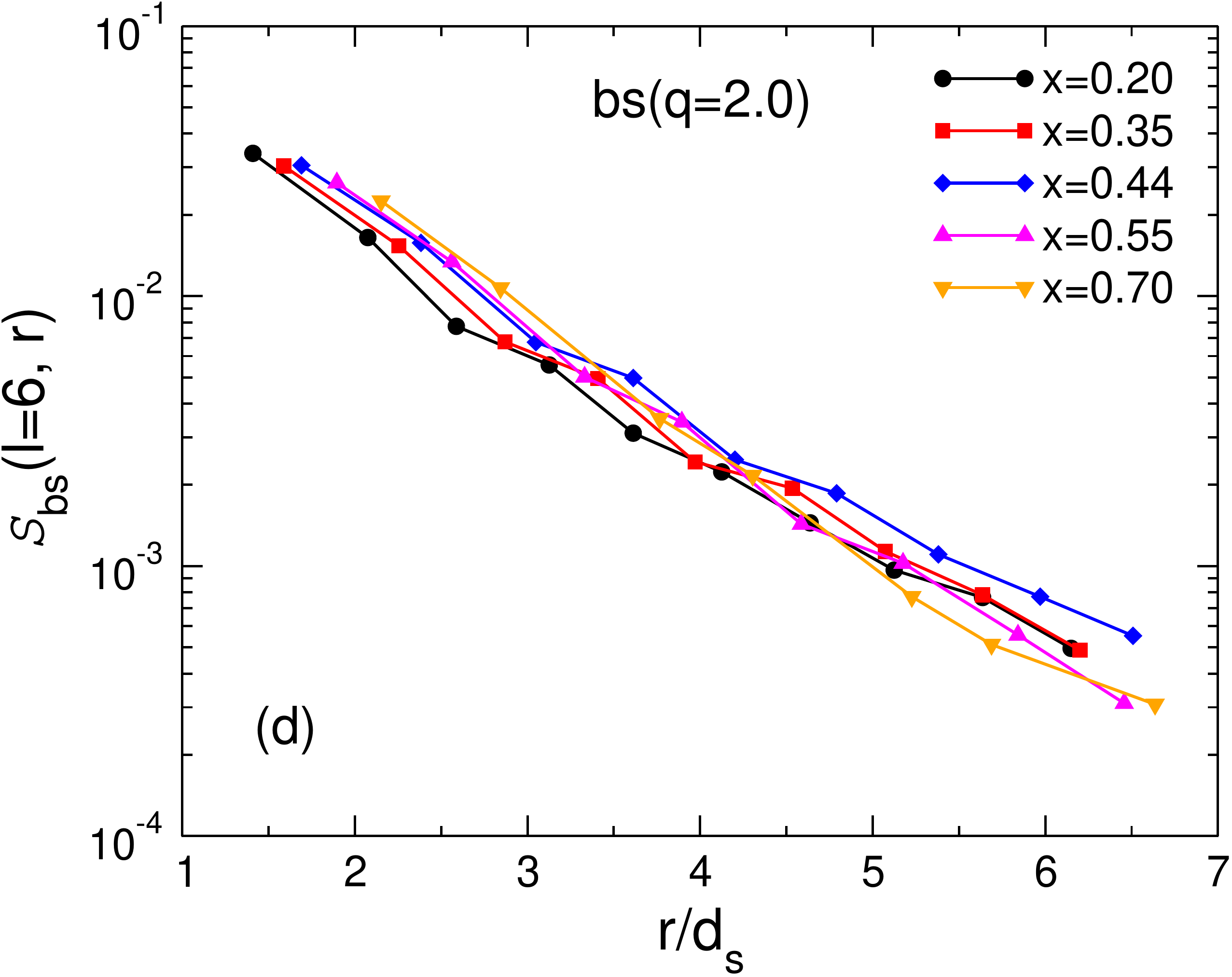}
}
\quad
\subfigure{
\includegraphics[width=6cm]{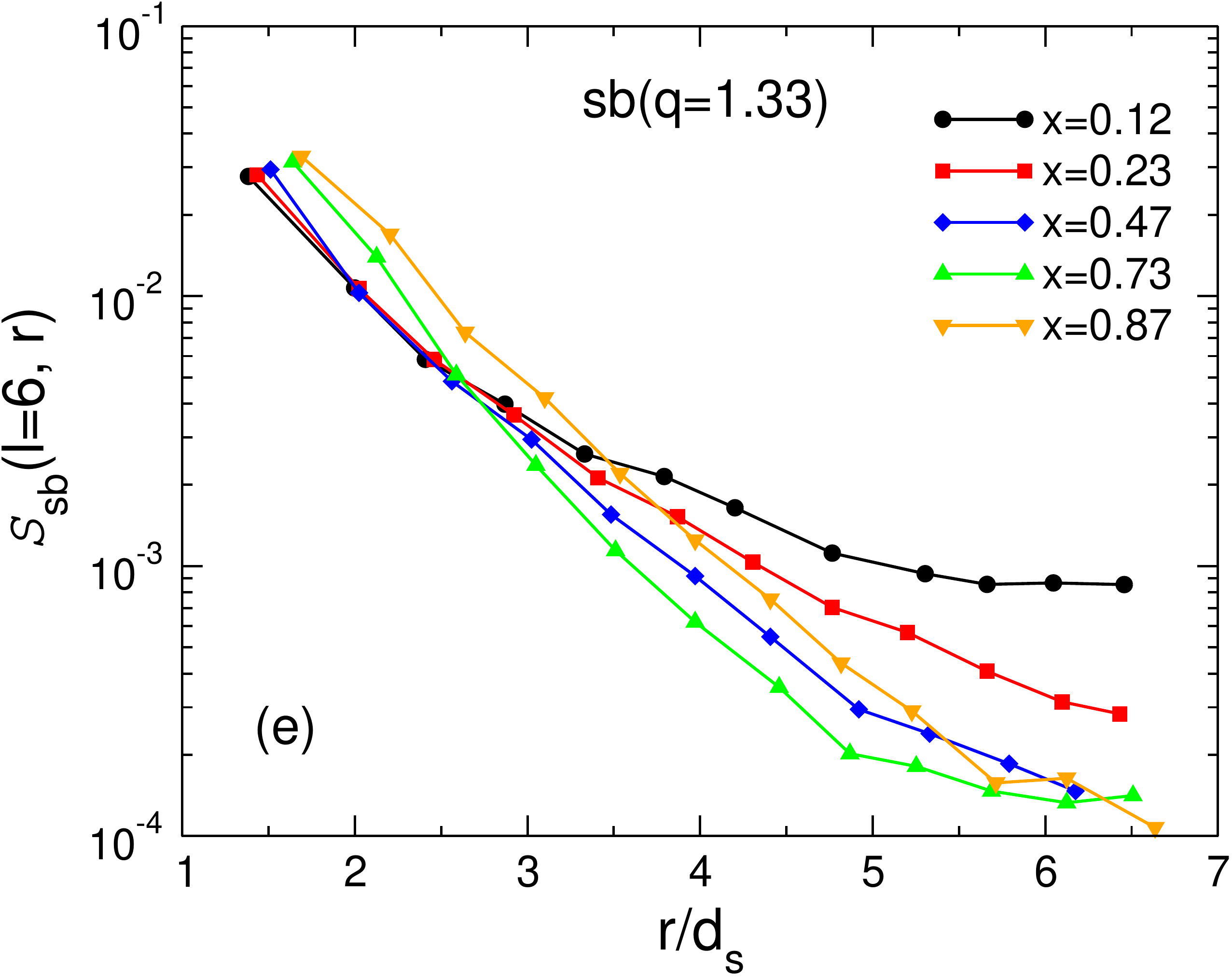}
}
\quad
\subfigure{
\includegraphics[width=6cm]{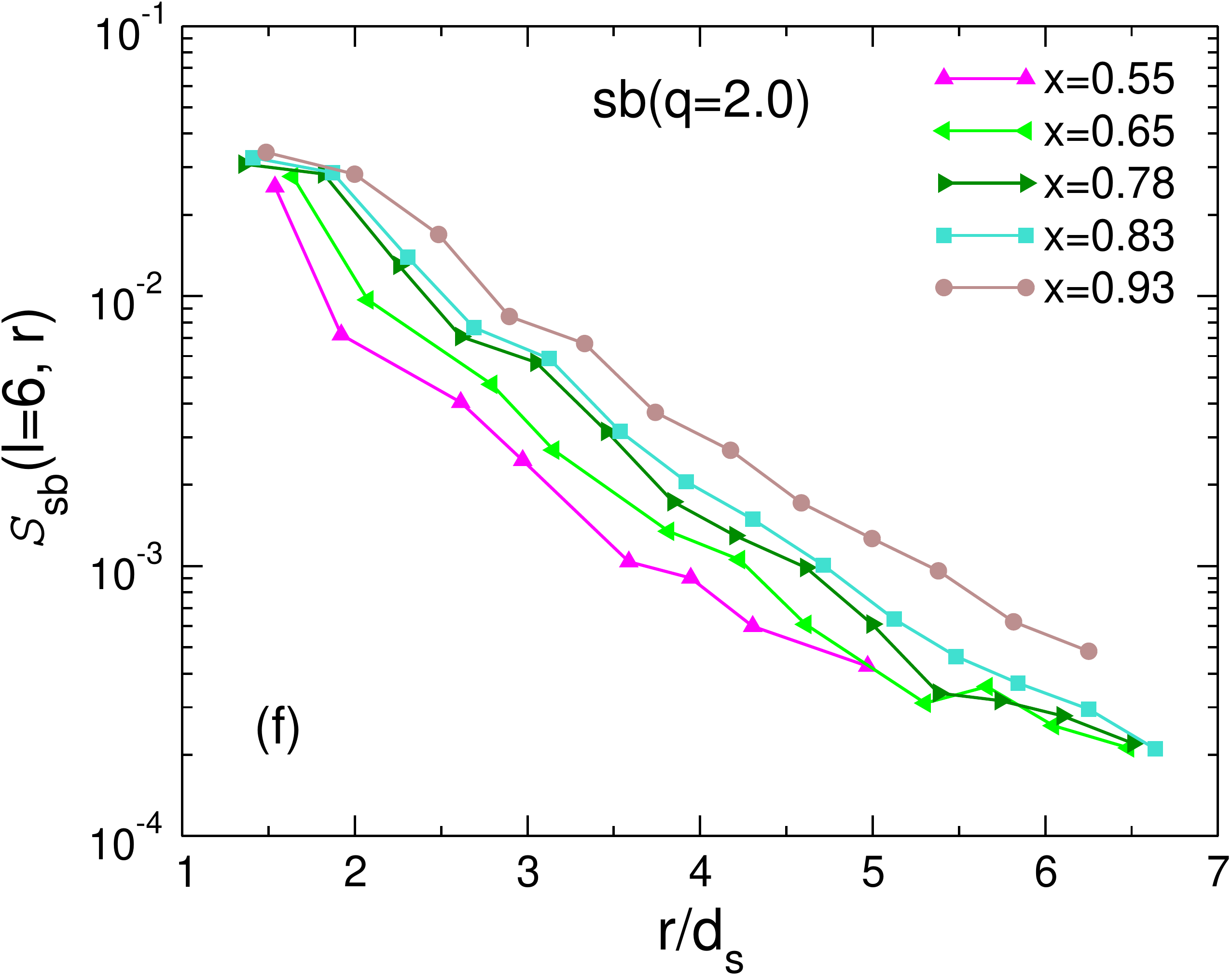}
}
\quad
\subfigure{
\includegraphics[width=6cm]{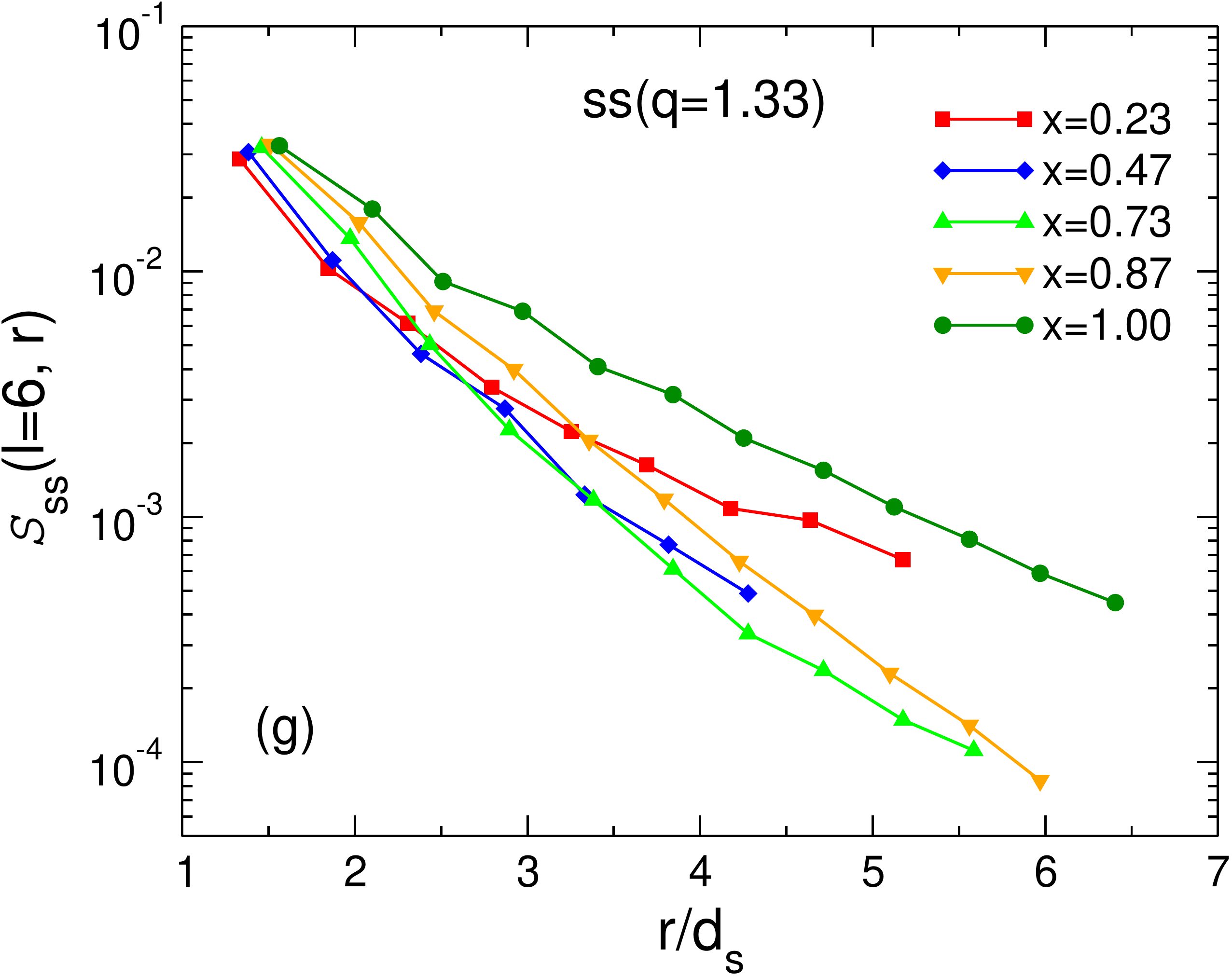}
}
\quad
\subfigure{
\includegraphics[width=6cm]{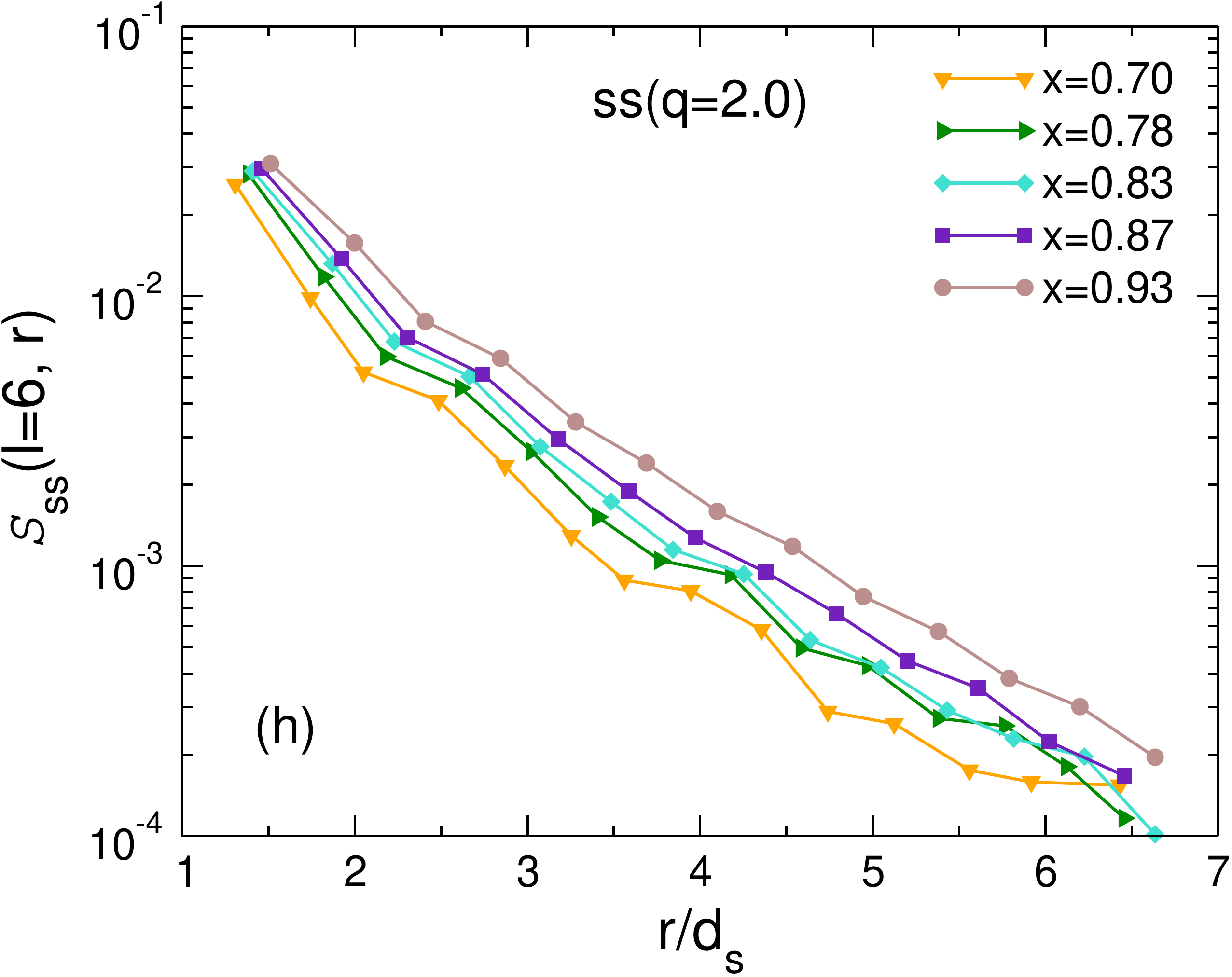}
}
\caption{Local maxima in $\mathcal{S}_{\alpha\beta}(l=6,r)$ for different concentrations $x$. (a), (c), (e), (g): The $q=1.33$ systems; (b), (d), (f), (h): The $q=2.0$ systems.
}
\label{fig_s6}
\end{figure}

Figures ~\ref{fig_s4} and \ref{fig_s5} show that the upper envelope of the curves can be used to define a slope, the inverse of which gives then a length scale. In Fig.~\ref{fig_s6} we show the location of the local maxima in $\mathcal{S}_{\alpha\beta}(l=6,r)$ and one sees that most of these curves can be approximated well by a straight line, i.e.,~a single length scale.

\begin{figure}[htbp]
\centering
\subfigure{
\includegraphics[width=7cm]{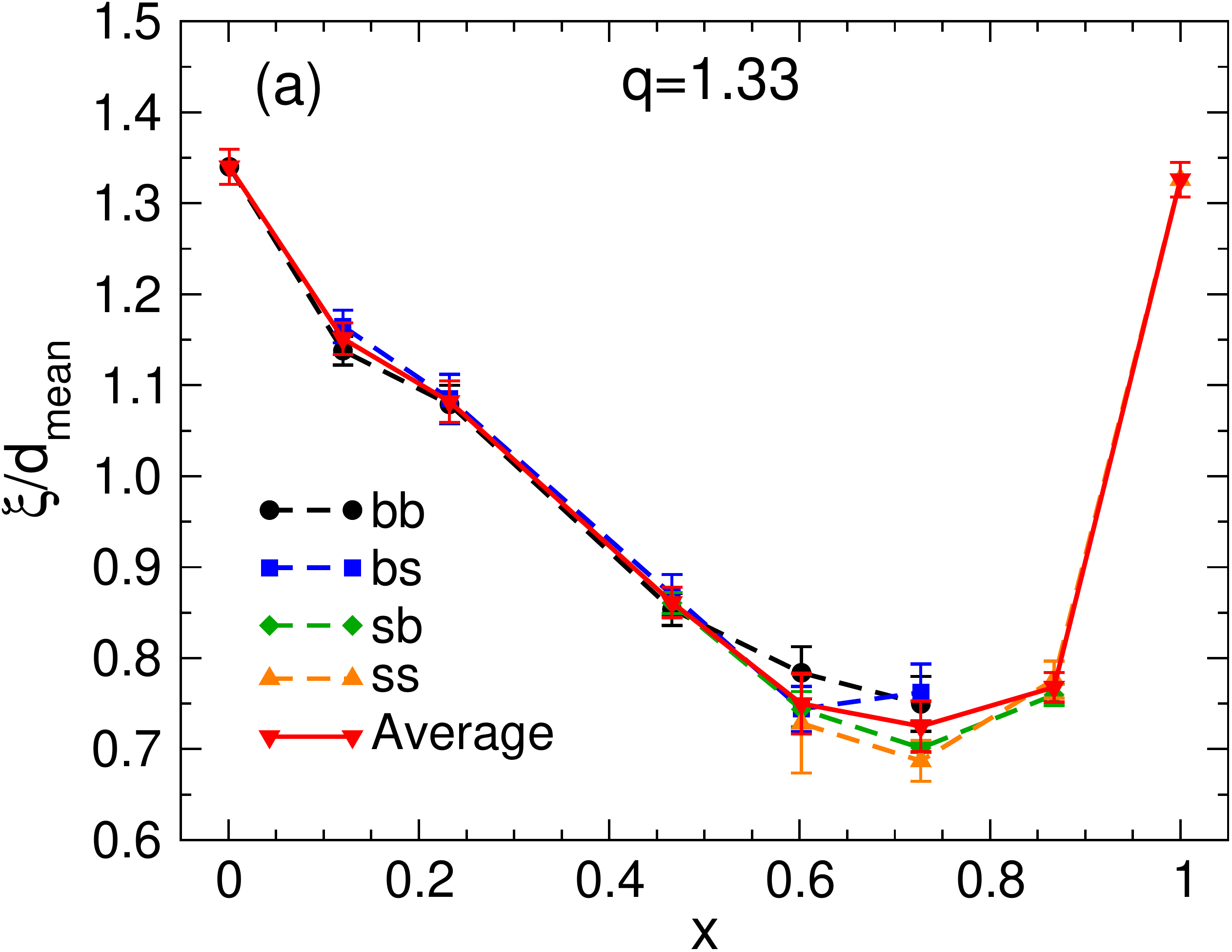}
}
\quad
\subfigure{
\includegraphics[width=7cm]{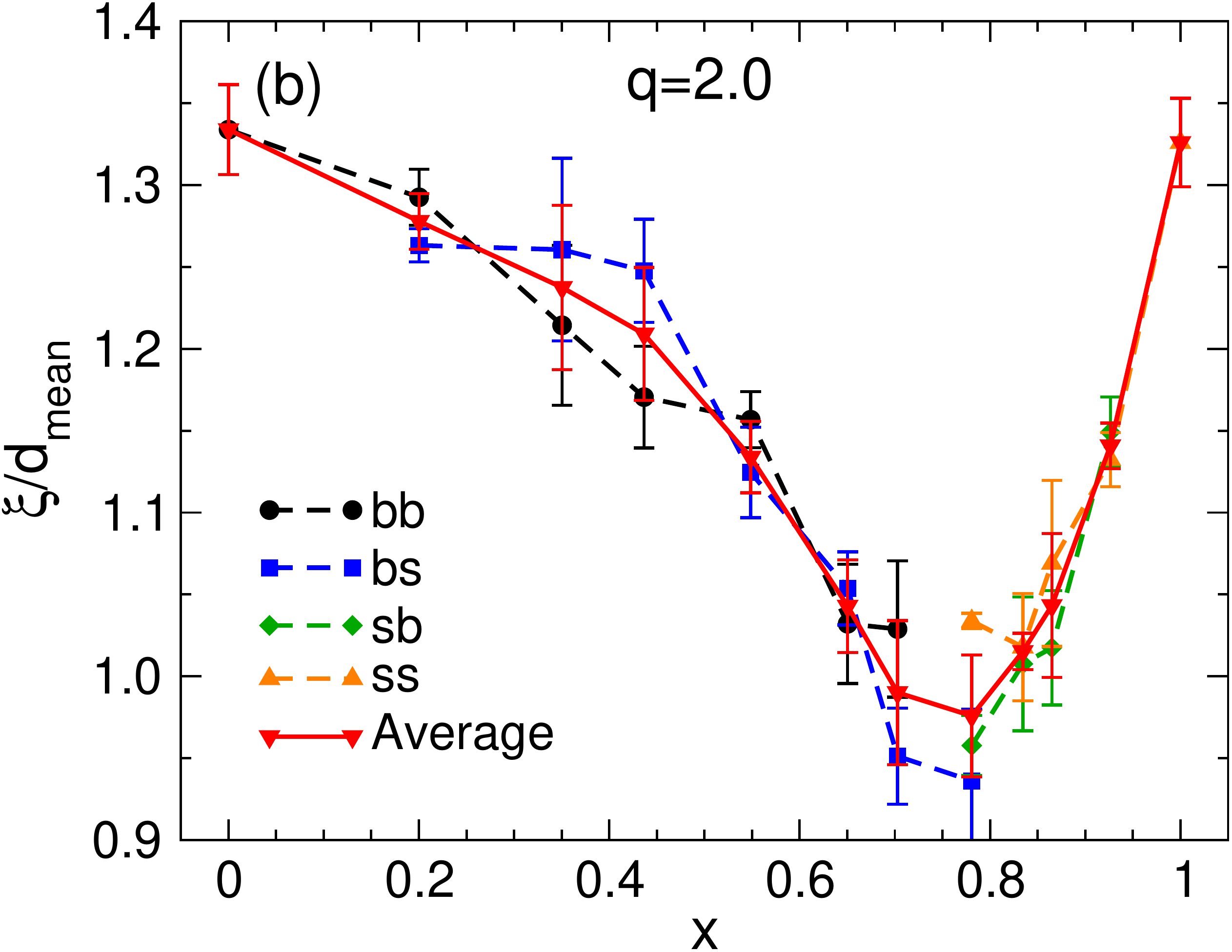}
}
\caption{The correlation lengths $\xi$ for the $bb$, $bs$, $sb$, and $ss$ correlation. (a) The $q=1.33$ systems. (b) The $q=2.0$ systems.}
\label{fig_s7}
\end{figure}

The $x$-dependence of the length scales $\xi$ obtained from Fig.~\ref{fig_s6} is shown in Fig.~\ref{fig_s7}. We see that the length scales obtained from the different correlation functions agree with each other within the error bars which is evidence that this length scale can be determined in a robust manner.

\section{4. Static structure factor}
\vspace*{-5mm}
For simple one-component systems the width of the first peak of the static structure factor $S(\mathbf{k})$ can be used to define a coherence length of the structure~\cite{binder_11,hansen_13}. Using the positions of the particles we have calculated $S(\mathbf{k})$ from its definition

\begin{equation}
S(\mathbf{k}) = \frac{1}{N}\sum_{j=1}^N \sum_{l=1}^N \exp[-i\mathbf{k} \cdot (\mathbf{r}_j-\mathbf{r}_l)]\quad .
\end{equation}

\begin{figure}[thbp]
\centering
\subfigure{
\includegraphics[width=7cm]{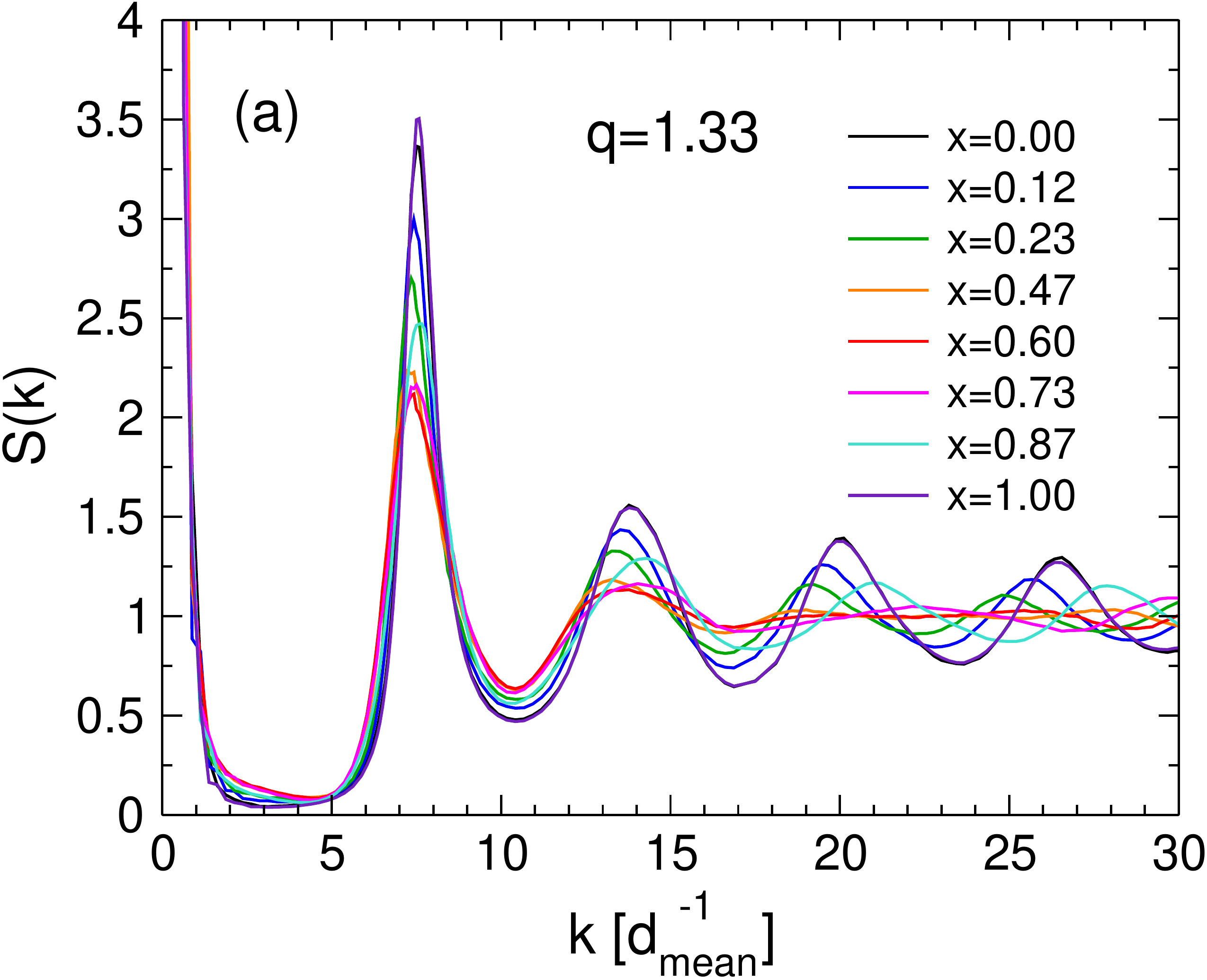}
}
\quad
\subfigure{
\includegraphics[width=7cm]{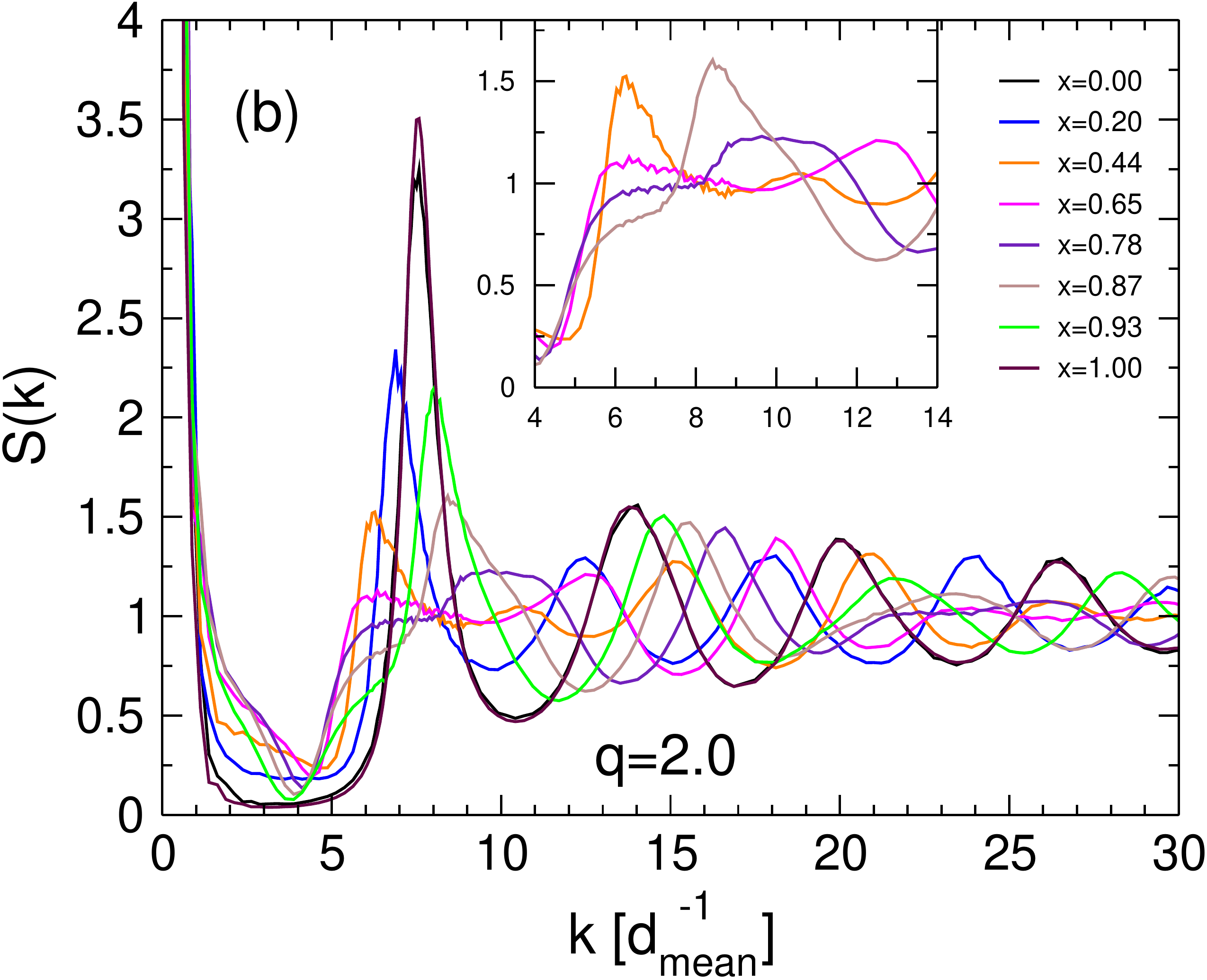}
}
\caption{Static structure factor $S(k)$ for the $q=1.33$ and the $q=2.0$ systems, panels (a) and (b), respectively. The inset in panel (b) is a zoom on the main peak of the main figure.
}
\label{fig_s8}
\end{figure}

Since the system is isotropic we can take the angular average and thus we consider $S(k)$, with $k=|\mathbf{k}|$. The $q$-dependence of $S(k)$ is shown in Fig.~\ref{fig_s8} for the different concentrations $x$. Note that here we express $k$ in units of the inverse mean particle diameter $d_{\rm mean}=(1-x)d_b + xd_s$ in order to remove the trivial dependence on $x$ due to the variation of the mean particle size. This makes that the $S(k)$ for the $x=0$ system coincides with the one for $x=1.0$. Note that here we calculate the total static structure factor, i.e.~we are not considering the partial ones~\cite{binder_11,hansen_13}, since we have found that the conclusions made below are not affected by this choice.

Figure~\ref{fig_s8}(a) shows that for the $q=1.33$ system the $x$-dependence of $S(k)$ is mild, as expected. The height and width of the main peak at around $k=7$ changes smoothly with $x$ and both quantities show a non-monotonic dependence on $x$. This $x$-dependence of the width $\sigma_S$ is included in Fig.~\ref{fig_3}(a) of the main text.

For the case of the $q=2.0$ systems, Fig.~\ref{fig_s8}(b), we see that for intermediate values of $x$ the main peak of $S(k)$ becomes flat since the presence of two competing length scales, $d_s$ and $d_b$, give rise to a destructive interference. As a consequence it is not possible to use  the width of the peak as a definition of the coherence length for such systems.

\section{5. Coordination numbers}
\begin{figure}[htbp]
\centering
\subfigure{
\includegraphics[width=7cm]{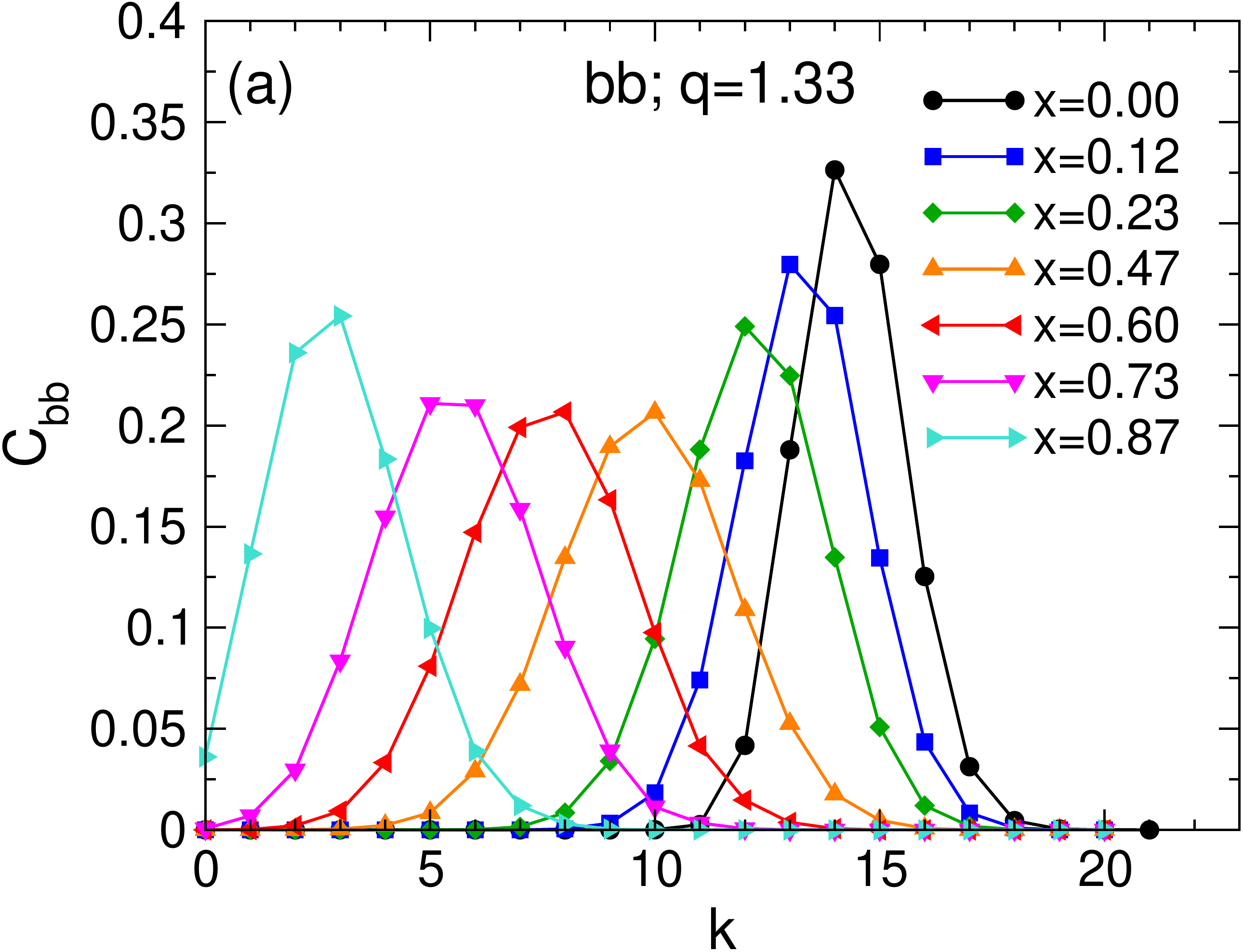}
}
\quad
\subfigure{
\includegraphics[width=7cm]{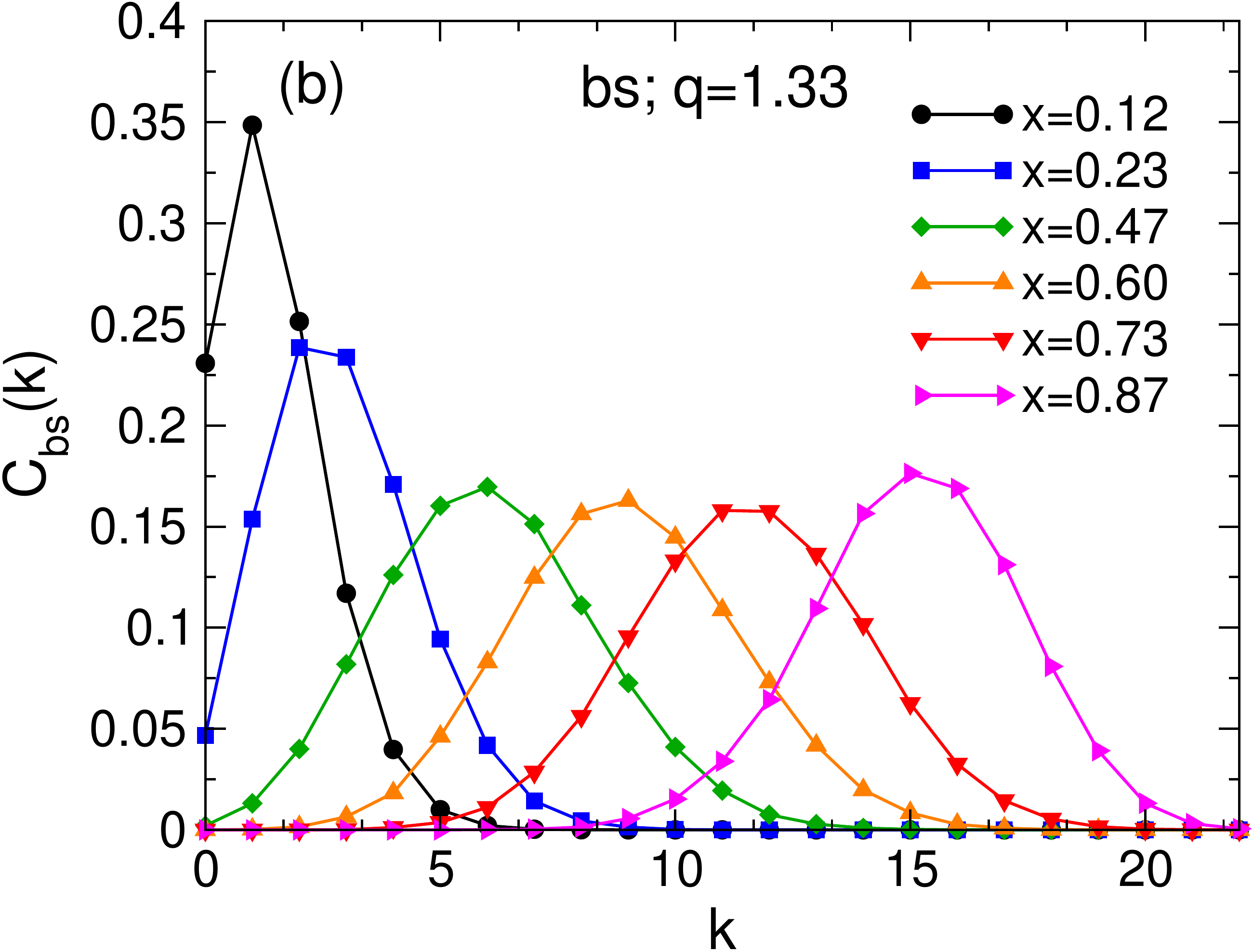}
}
 
\centering
\subfigure{
\includegraphics[width=7cm]{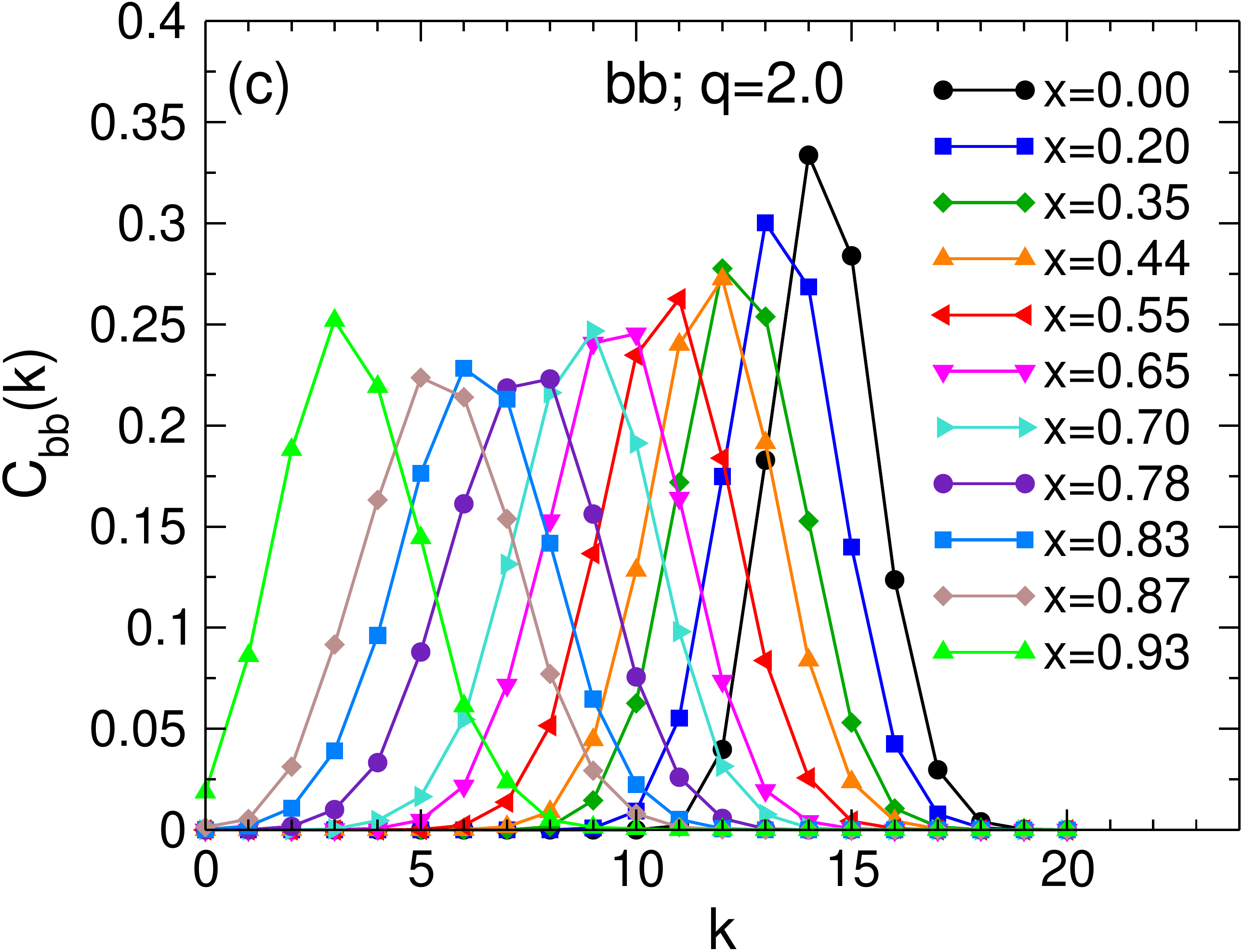}
}
\quad
\subfigure{
\includegraphics[width=7cm]{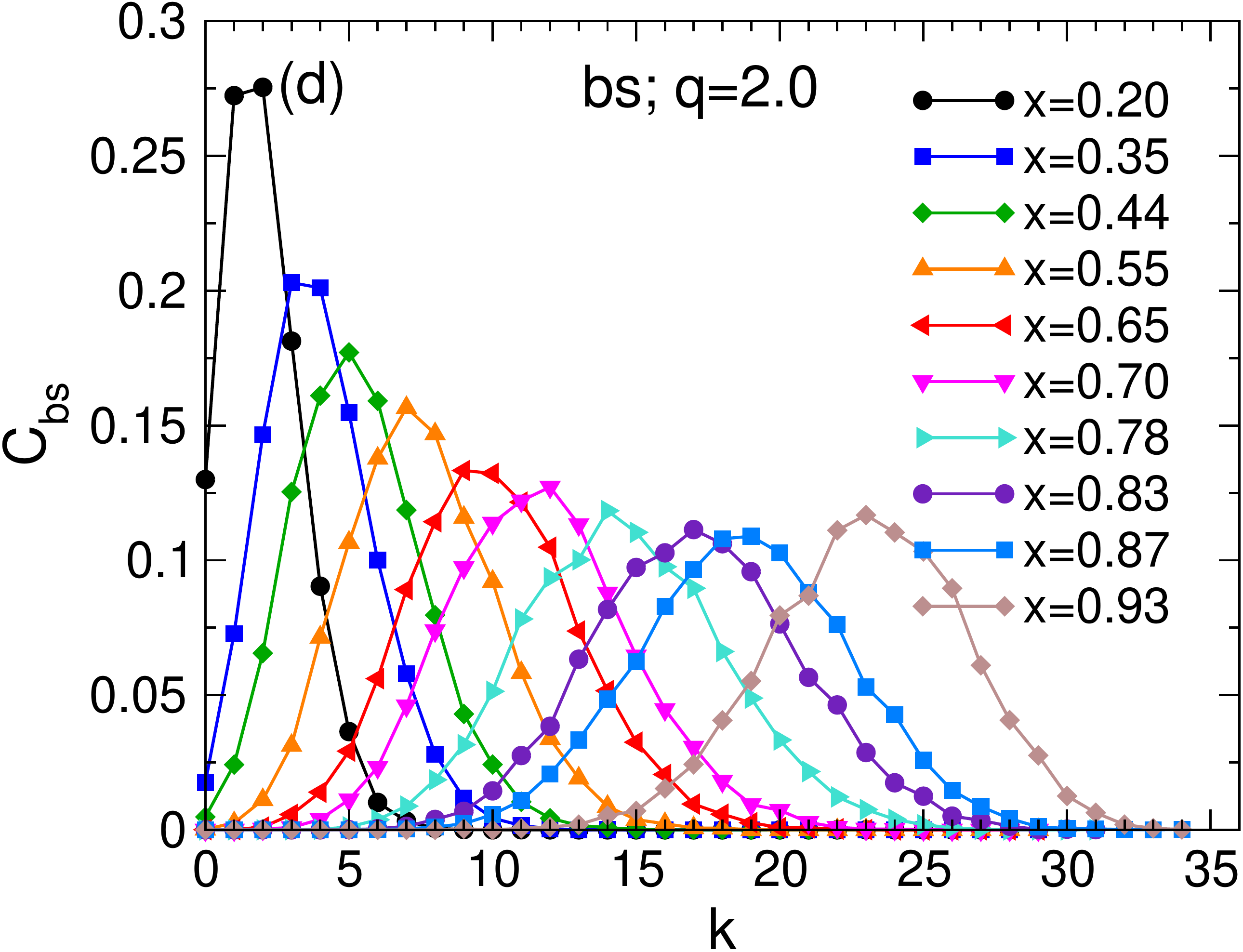}
}

\caption{The distribution of the partial coordination numbers $C_{\alpha\beta}(k)$ for different compositions. Panels (a), $C_{bb}$, and (b), $C_{bs}$, are for the $q=1.33$ systems. Panels (c), $C_{bb}$, and (d), $C_{bs}$, are for the $q=2.0$ systems.
}
\label{fig_s9}
\end{figure}

\vspace*{-5mm}
In the main text we have argued that the packing efficiency is related to the presence of a large variety of different neighborhoods for the particles. One possibility to characterize such a neighborhood is by means of the partial coordination number $C_{\alpha\beta}(k)$, i.e.~the probability that a particle of type $\alpha$ has exactly $k$ neighbors of type $\beta$. Here we define the neighbors as the particles whose Voronoi cell share a face with the Voronoi cell of the central particle. In Fig.~\ref{fig_s9}(a) we show for the $q=1.33$ system the probability distribution of $C_{bb}$ for different concentration $x$ and one sees that as a function of $x$ the peak shifts to smaller values of $k$, as expected. At the same time its height first decreases and then rises again if $x$ becomes large, i.e., the standard deviation $\omega_{bb}$ shows a maximum. Qualitatively the same behavior is observed for the coordination number $C_{bs}$, shown in Fig.~\ref{fig_s9}(b), and the two other combinations of $\alpha$ and $\beta$ (not shown).

\begin{figure}[htbp]
\centering
\subfigure{
\includegraphics[width=7cm]{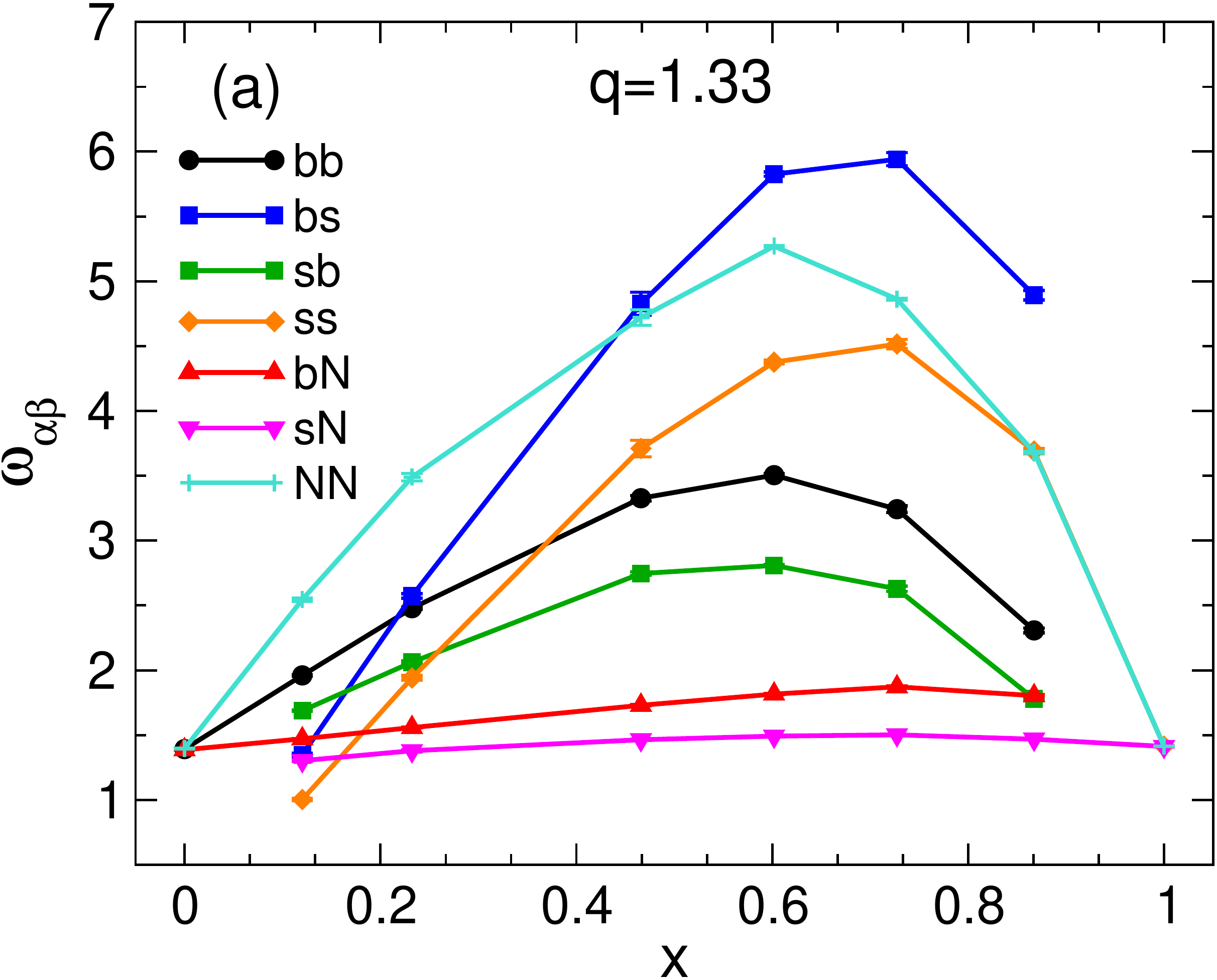}
}
\quad
\subfigure{
\includegraphics[width=7cm]{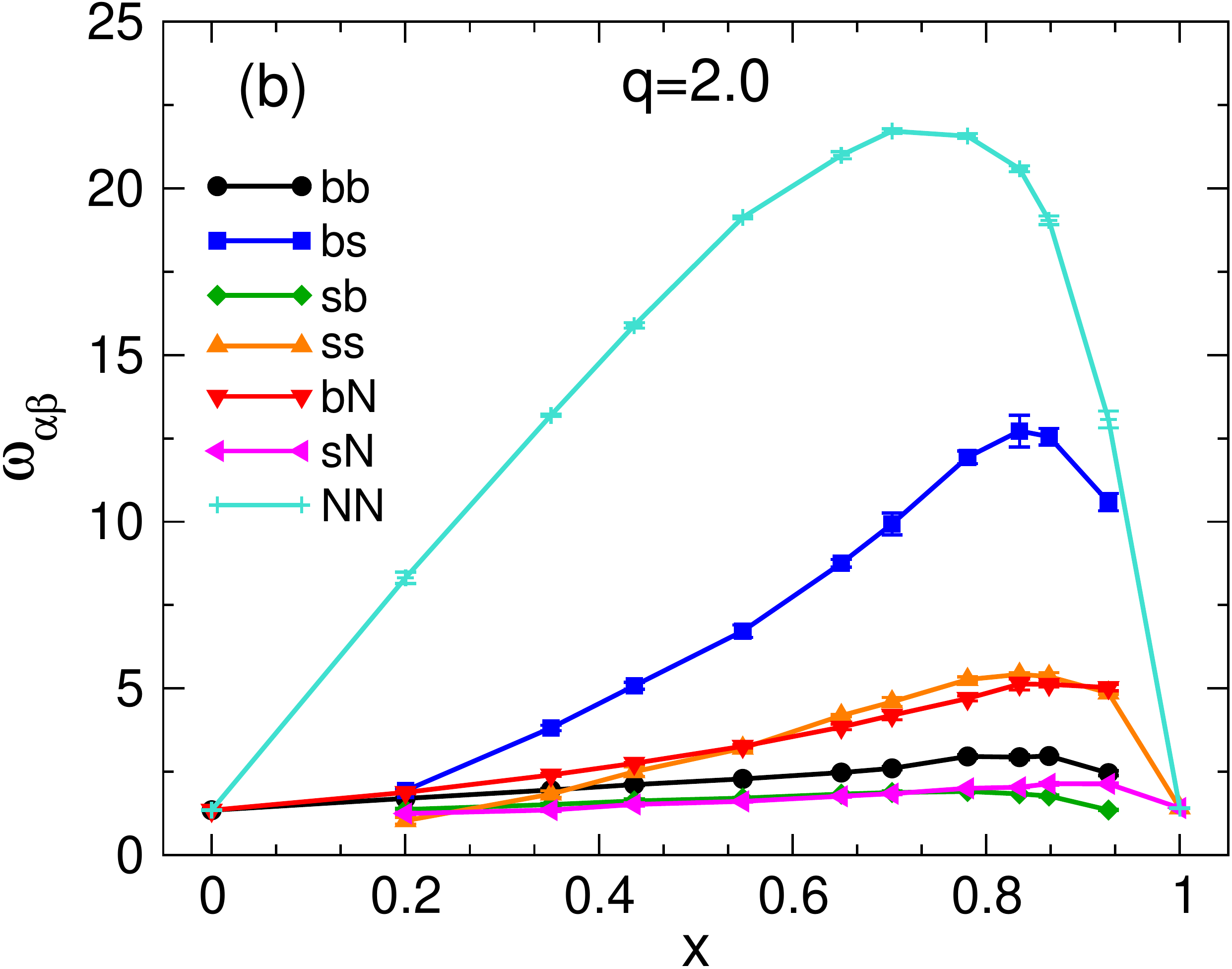}
}

\caption{The standard deviation of the coordination number $C_{\alpha\beta}$ for different choices of $\alpha$ and $\beta$. Panels (a) and (b) are, respectively, for the $q=1.33$ and the $q=2.0$ systems. $N$ stands for ``number'', i.e. ``$b$''+``$s$''.
} 
\label{fig_s10}
\end{figure}

For the case of the $q=2.0$ system we find that $C_{\alpha\beta}$ shows qualitatively the same $x$-dependence as the $q=1.33$ system, but now the trends are more pronounced, see panels (c) and (d) of Fig.~\ref{fig_s9}. This can be recognized directly from Fig.~\ref{fig_s10}, where we plot $\omega_{\alpha\beta}$, the standard deviation of $C_{\alpha\beta}$, as a function of $x$, and one clearly sees that this $x$-dependence is stronger for the $q=2.0$ system than for the $q=1.33$ system. Importantly we find that the location of the maximum in $\omega_{\alpha\beta}$ does not depend on the choice of $\alpha$ or $\beta$, i.e., the variety of possible local structure has a maximum at the same concentration $x_{\rm max}$.

\clearpage
\section{6. Local packing fraction}
\begin{figure}[htbp]
\centering
\subfigure{
\includegraphics[width=7cm]{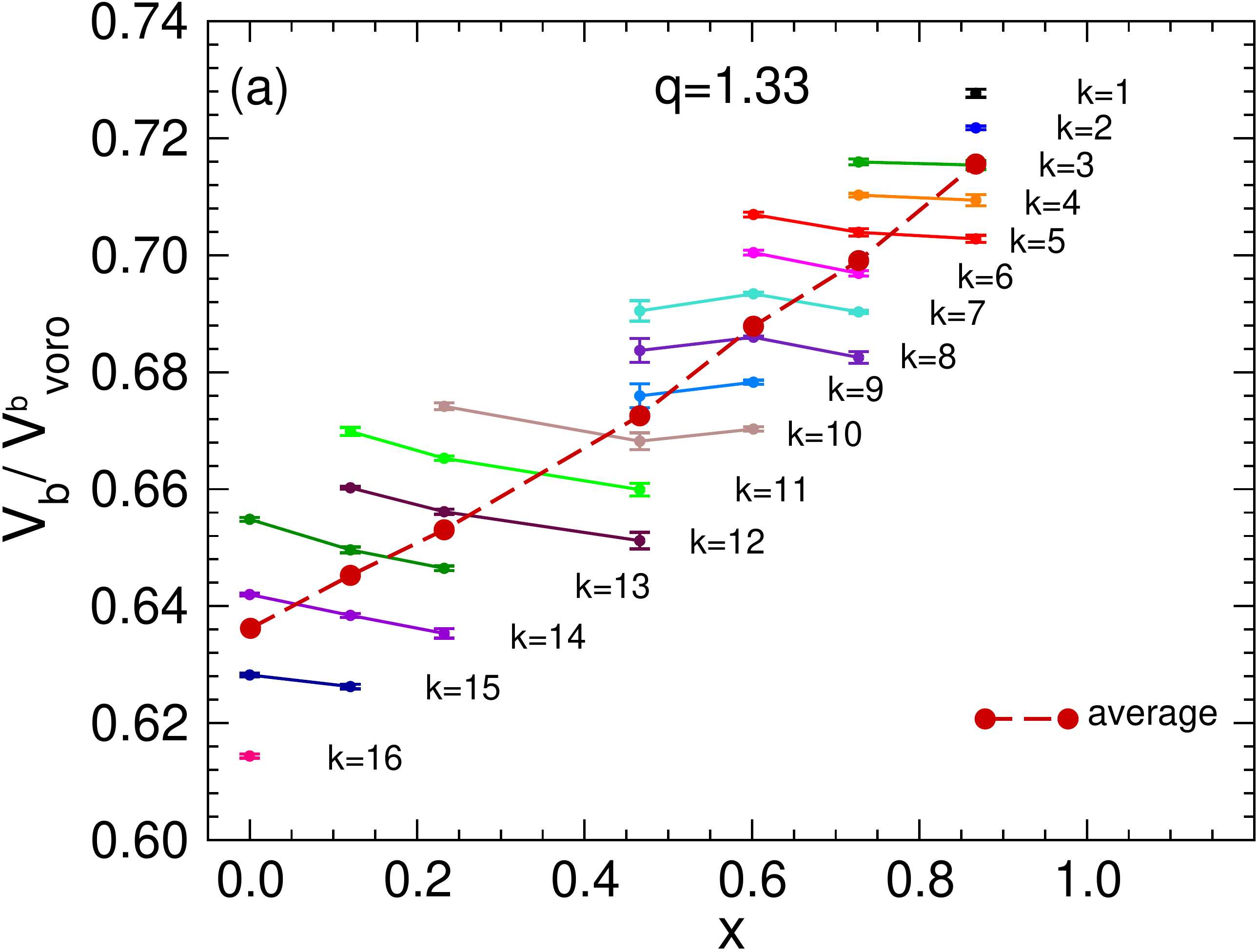}
}
\quad
\subfigure{
\includegraphics[width=7cm]{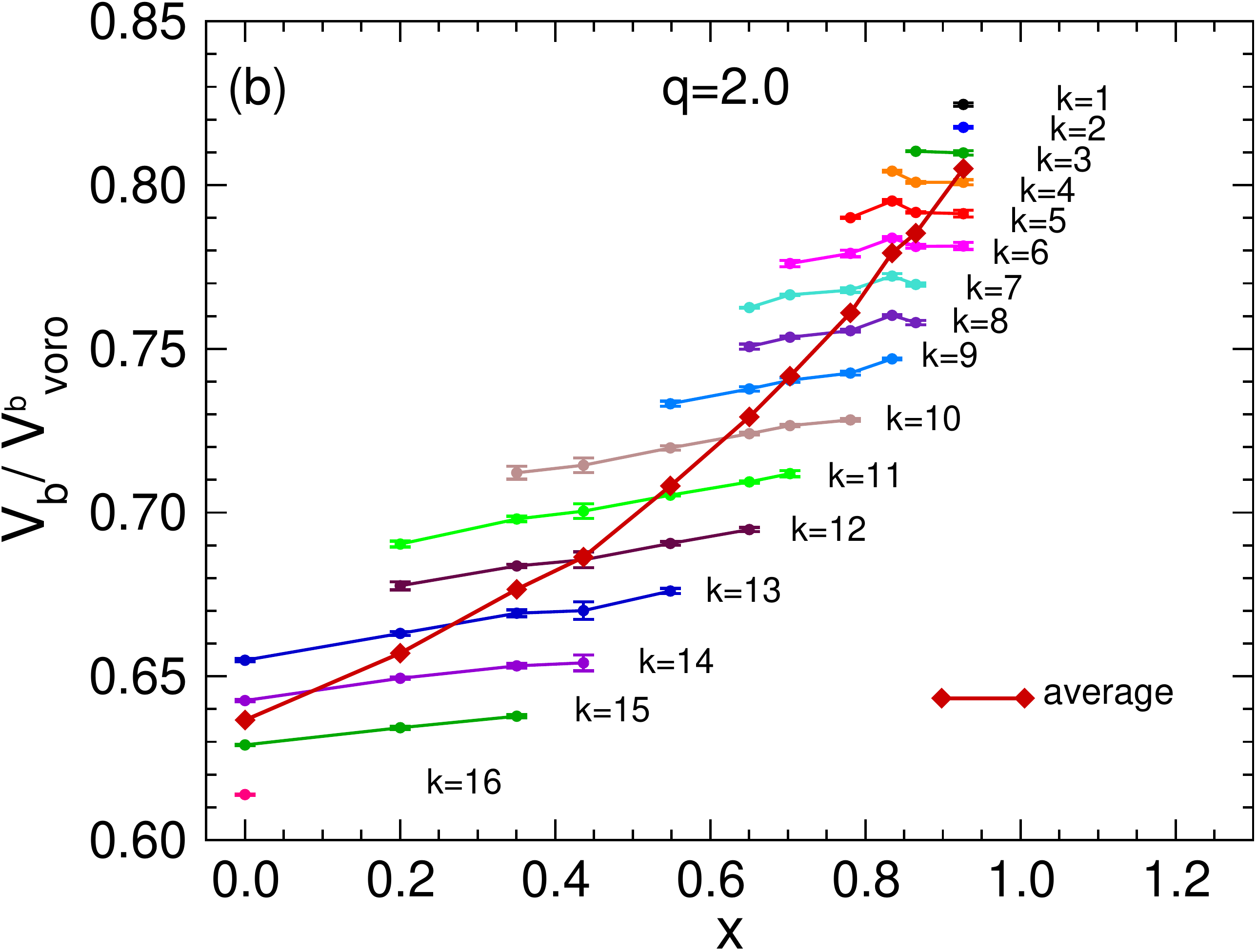}
}
\caption{The local packing fraction for different types of neighborhoods. Only data points whose probability is larger than 5\% are shown. a) $q=1.33$ system. b) $q=2.0$ system.
}
\label{fig_s11}
\end{figure}

\vspace*{-5mm}
In the context of Fig.~\ref{fig_4} of the main text we have demonstrated that the maximum in the global packing fraction $\varphi(x)$ is due to a competition between an increase with $x$ of the local packing fraction for the $b$-particles, $\varphi_b$, and a decrease of $\varphi_s$ for the $s$-particles with $1-x$, see Eq.~(\ref{eq_2}). The presence of this maximum was argued to be related to the existence of a large variety of local environments which allow to have an efficient global packing. However, this argument rests on the hypothesis that all these different environments do have a relatively high local packing. That this is indeed the case is shown in Fig.~\ref{fig_s11} where we present the local packing density of the different local environments of a $b$-particle, i.e.~a $b$-particle that has exactly $k$ other $b$-particles as its nearest neighbor. From the graph we see that, for any given $x$, the packing fractions of the different neighborhoods, i.e.~different values of $k$, are quite similar. Thus we can conclude that the maximum in the global packing fraction at $\varphi_{\rm}$ is not due to a the presence of extreme packing fractions in the particle environment but instead due to the distribution of
the frequency of these environments, i.e., the standard deviation $\omega_{\alpha\beta}$ shown in Fig.~\ref{fig_s10}.

\vspace*{-5mm}

\section{7. Packing structure for the $\mathbf{q=2}$ system.}

\vspace*{-5mm}
In Movie~1 we show the density field of the big particles around a $s$-particle (grey sphere). $q=2.0$ and $x=0.55$, i.e., the same system as shown in Fig.~1c. Only regions with high density (covering 10\% area of the sphere) are depicted. The different colors correspond to different shells.

\end{document}